\documentclass[amsmath,
amssymb,showpacs,aps,
pra,twocolumn,
superscriptaddress,nobibnotes]{revtex4-2}
\usepackage[utf8]{inputenc}
\usepackage[dvipsnames]{xcolor}
\usepackage{amsmath}
\usepackage{mathtools}
\usepackage{adjustbox}
\usepackage{lipsum}
\usepackage[font=small,labelfont=bf,
   justification=Justified,
   format=plain]{caption}
\usepackage{comment}
\usepackage{enumitem}
\usepackage[normalem]{ulem}
\usepackage[clock]{ifsym}
\usepackage{nicefrac}
\usepackage{eurosym}
\usepackage{graphicx}
\usepackage{subcaption}
\usepackage{nomencl}
\usepackage{amsfonts}
\usepackage{tabto}
\numberwithin{equation}{section}
\usepackage{amssymb}
\usepackage{fancyhdr}
\usepackage{amscd}
\usepackage{quantikz}
\usepackage{braket}
\usepackage{booktabs}
\usepackage[english]{babel}
\usepackage{listings}
\usepackage{epstopdf}
\usepackage{multirow}
\usepackage[colorlinks=true,linkcolor=blue,citecolor=blue,urlcolor=blue]{hyperref}
\usepackage[capitalise]{cleveref}

\usetikzlibrary{bending, shapes.geometric}
\usetikzlibrary{calc}
\usetikzlibrary{spy}
\usetikzlibrary{arrows.meta,bending}
\usetikzlibrary{shapes.geometric}
\usetikzlibrary{decorations.markings,calc}

\newcommand{\halfcoloredcircle}[4]{%
  \begin{scope}
    \fill[#3] (#1,#2) -- ++(180:3pt) arc (180:360:3pt) -- cycle; 
    \fill[#4] (#1,#2) -- ++(0:3pt) arc (0:180:3pt) -- cycle; 
    \draw[thick] (#1,#2) circle (3pt); 
  \end{scope}
}

\newcommand{\red}{{\color{red} R}}
\newcommand{\blue}{{\color{blue} B}}
\newcommand{\zero}{\boldsymbol{\emptyset}}

\newcommand{\circleopen}{\tikz[baseline=-0.5ex]{\draw[black, thick] (0,0) circle (0.5ex);}}

\newcommand{\circlefill}{\tikz[baseline=-0.5ex]{\fill[black] (0,0) circle (0.5ex);}}

\newcommand{\redcircleopen}{\tikz[baseline=-0.5ex]{\draw[red, thick] (0,0) circle (0.5ex);}}

\newcommand{\bluecircleopen}{\tikz[baseline=-0.5ex]{\draw[blue, thick] (0,0) circle (0.5ex);}}

\newcommand{\redcirclefill}{\tikz[baseline=-0.5ex]{\fill[red] (0,0) circle (0.5ex);}}

\newcommand{\bluecirclefill}{\tikz[baseline=-0.5ex]{\fill[blue] (0,0) circle (0.5ex);}}

\newcounter{parrow}
\tikzset{
  record path/.style={
    /utils/exec=\tikzset{parrow/.cd,#1},
    decorate,
    decoration={
      markings,
      mark=at position 0 with {
        \setcounter{parrow}{1}
        \path (0,\pgfkeysvalueof{/tikz/parrow/dist}/2)
              coordinate (parrowt-\pgfkeysvalueof{/tikz/parrow/name}-\number\value{parrow})
               (0,-\pgfkeysvalueof{/tikz/parrow/dist}/2)
              coordinate (parrowb-\pgfkeysvalueof{/tikz/parrow/name}-\number\value{parrow});
        \pgfmathsetmacro{\mystep}{(\pgfdecoratedpathlength-4pt)/int(1+(\pgfdecoratedpathlength-4pt)/2pt)}
        \xdef\mystep{\mystep}
      },
      mark=between positions 2pt and 1 step \mystep pt with {
        \stepcounter{parrow}
        \path (0,\pgfkeysvalueof{/tikz/parrow/dist}/2)
              coordinate (parrowt-\pgfkeysvalueof{/tikz/parrow/name}-\number\value{parrow})
              (0,-\pgfkeysvalueof{/tikz/parrow/dist}/2)
              coordinate (parrowb-\pgfkeysvalueof{/tikz/parrow/name}-\number\value{parrow})
              (0,0)
              coordinate (parrowm-\pgfkeysvalueof{/tikz/parrow/name}-\number\value{parrow});
      }
    }
  },
  reconstruct top/.style={
    insert path={
      plot[variable=\t,samples at={1,...,\number\value{parrow}},smooth]
        (parrowt-#1-\t)
    }
  },
  reconstruct bottom/.style={
    insert path={
      plot[variable=\t,samples at={\number\value{parrow},\the\numexpr\value{parrow}-1,...,1},smooth]
        (parrowb-#1-\t)
    }
  },
  parrow/.cd,
  dist/.initial=4pt,
  step/.initial=2pt,
  name/.initial={}
}

\renewcommand{\theequation}{\arabic{equation}}

\makeatletter
\secnums@arabic
\makeatother

\begin{document}

\title{A resource-efficient quantum-walker Quantum RAM}


\author{Giuseppe De Riso}
\affiliation{Scuola Normale Superiore, Piazza dei Cavalieri 7, I-56126, Pisa, Italy}
\email{giuseppe.deriso@sns.it}

\author{Giuseppe Catalano}
\affiliation{NEST-CNR Scuola Normale Superiore, Piazza dei Cavalieri 7, I-56126, Pisa, Italy}

\author{Seth Lloyd}
\affiliation{Dept. of Mech. Eng., Massachusetts Institute of Technology, 77 Mass. Av., Cambridge, MA 02139, USA}

\author{Vittorio Giovannetti}
\affiliation{NEST-CNR Scuola Normale Superiore, Piazza dei Cavalieri 7, I-56126, Pisa, Italy}

\author{Dario De Santis}
\affiliation{Scuola Normale Superiore, Piazza dei Cavalieri 7, I-56126, Pisa, Italy}

\begin{abstract}
\noindent Efficient and coherent data retrieval and storage are essential for harnessing quantum algorithms' speedup. Such a fundamental task is addressed by a quantum Random Access Memory (qRAM). Despite their promising scaling properties, current qRAM proposals demand excessive resources and rely on operations beyond the capabilities of current hardware requirements, rendering their practical realization inefficient. We introduce a novel architecture that significantly reduces resource requirements while preserving optimal complexity scaling for quantum queries. Moreover, unlike previous proposals, our algorithm design leverages a simple, repeated operational block based exclusively on local unitary operations and short-range interactions between a limited number of quantum walkers traveling over a single binary tree. This novel approach not only simplifies experimental requirements by reducing the complexity of necessary operations but also enhances the architecture’s scalability by ensuring a resource-efficient, modular design that maintains optimal quantum query performance. 
\end{abstract}
\maketitle

Computational tasks require fast and efficient data storage and retrieval. Classically, this is achieved by a Random Access Memory (RAM), which stores data in memory cells, each identified by a unique binary address, allowing for deterministic access to them. In a quantum computer, however, it is essential to process superpositions of addresses while preserving coherence. This ability is provided by a quantum Random Access Memory (qRAM)~\cite{nielsen_chuang}.
A qRAM works as follows. Consider a database $M$ with $N = 2^n$ memory cells, where each cell $M^{(\boldsymbol{a})}$ is identified 
by a unique $n$-bit binary address 
$\boldsymbol{a} := a_1, a_2, \dots, a_n$, and 
stores $m$ bits of classical information $\boldsymbol{b}^{(\boldsymbol{a})}:=
b^{(\boldsymbol{a})}_1, b^{(\boldsymbol{a})}_2, \dots, b^{(\boldsymbol{a})}_m$. 
When a specific address $\boldsymbol{a}$ is given as the state
$\ket{\boldsymbol{a}}_A$ of a $n$-qubit register $A$, the qRAM must access the corresponding memory cell $M^{(\boldsymbol{a})}$ and retrieve the bit string $\boldsymbol{b}^{(\boldsymbol{a})}$ it contains
in the form of a state $\ket{\boldsymbol{b}^{(\boldsymbol{a})}}_D$ of a second quantum register~$D$.  
 Importantly, due to the linearity of the procedure, if the address register 
$A$ is prepared in a superposition $\sum_{\boldsymbol{a}}\alpha_{\boldsymbol{a}}\ket{\boldsymbol{a}}_A$,
the qRAM implements  the following (unitary) transformation: 
\begin{equation}\label{generalqRAM}
\sum_{\boldsymbol{a}}\alpha_{\boldsymbol{a}}\ket{\boldsymbol{a}}_A\ket{0}_D\xrightarrow{\text{qRAM}} \sum_a\alpha_{\boldsymbol{a}}\ket{\boldsymbol{a}}_A\ket{\boldsymbol{b}^{(\boldsymbol{a})}}_D\;,
\end{equation}
where $\ket{0}_D$ is a fiduciary state of register $D$.

The qRAM is a crucial component for achieving a quantum advantage over classical computation for many algorithms strictly requiring its use to enable speedups for specific computational tasks. Notable examples include famous algorithms such as Grover's classical database search~\cite{Grover_search} and Shor's factoring~\cite {Shor_factoring}. Additionally, it has applications in linear algebra~\cite{Harrow_Linear_algebra,Low_Linear_algebra, Clader_Linear_algebra, Giovannetti_Linear_algebra, Childs_Linear_algebra}, quantum chemistry \cite{Kassal_quantum_chemistry, Whitfield_Quantum_chemistry, Babbush_Quantum_chemistry, Cao_Quantum_chemistry}, quantum machine learning~\cite{Lloyd_machine_learning, Poggiali_machine_learning, Rebentrost_machine_learning,Lloyd_machine_learning_2, Zhao_Machine_learning, Kerenidis_Machine_learning} and cryptography~\cite{Giovannetti_Cryptography,Kuperberg_cryptography}.
It is important to stress that these speedups are guaranteed only if qRAM enables fast data access. More precisely, the computational cost should scale at most as $\mathcal{O}(\text{poly} \log N)$; otherwise, an exponential scaling would invalidate any advantage offered by the quantum algorithm.

Even though theoretical models achieving fast query times exist \cite{Bucket_Brigade,BB_robustness,Asaka_breve,Cesa_qRAM, Park_qRAM,Weiss_qRAM}, their hardware implementation faces significant challenges. In addition to ensuring rapid data access, scaling issues such as the exponential increase in qubits, the need for extensive gate operations, and reliance on long-range interactions pose major obstacles to constructing a viable qRAM architecture \cite{Jaques_QRAM_survey, Xu_QRAM_critique}. Despite significant efforts in recent years to overcome these problems \cite{Wang_QRAM_design, Mukhopadhyay_QRAM_design} and pave the way for experimental implementations \cite{Jiang_QRAM_implementation, Chen_QRAM_implementation}, a practical realization remains elusive.
Therefore, efficient qRAM design is crucial, as inefficiencies can lead to excessive time and energy overheads and high decoherence rates, ultimately undermining the performance benefits of quantum algorithms.

A notable architecture is the Bucket Brigade (BB)~\cite{Bucket_Brigade}, the first resource-efficient qRAM proposal. Here, the authors use a perfect binary tree of depth $n$  with $N=2^n$ leaves representing the memory cells. Each bifurcation of the three, or node, hosts a qutrit, whose internal state, updated through sequential interactions with the information carriers, dynamically opens a path corresponding to the input address. This approach significantly reduces node activation and tree depth to 
$\mathcal{O}(\log N)$, making it resilient to decoherence~\cite{Hann_noise_resilience}. However, its main drawbacks are the need for an exponential number of active memory elements (qutrits) located at each bifurcation of the binary tree that define the paths the address carriers have to follow to reach the memory cells of the database. 
A recent approach is based on discrete-time quantum walkers~\cite{Quantum_Walker_intro}. {In this context, R.~Asaka, K.~Sakai and R.~Yahagi (ASY) propose} a BB model that avoids qutrits and eliminates the need for entanglement between address walkers and the active elements of the tree~\cite{Asaka_breve}. In this model, the address state is encoded in the position of the walkers, which move simultaneously along multiple binary trees. 
This enables all information carriers to reach their target memory cells in $\mathcal{O}(\log N)$ computational steps, resulting in
a coherent and reversible addressing mechanism that avoids the need for active routing elements, such as the qutrit-based components required in the BB model. This passive retrieval mechanism simplifies error handling
and improves robustness. However, the approach of~\cite{Asaka_breve} requires $2(m+\log N)$ distinct binary trees and long-range interactions among the walkers that propagate along such trees, leading to a substantial overhead in spatial resources and posing greater challenges for physical implementation.

In this work, we introduce a novel qRAM model based on discrete time quantum walks on graphs that preserves the logarithmic runtime cost for a single query while significantly reducing the experimental complexity, making it a more feasible candidate for near-term hardware implementations. Unlike~\cite{Asaka_breve}, our approach uses a single binary tree and can be implemented entirely on local and short-range unitary operations, simplifying hardware requirements without compromising efficiency. We begin by outlining the core elements of our architecture in its general form before introducing a modified variant that enhances coherence and further simplifies hardware requirements at the cost of a minor, constant overhead in walker count.

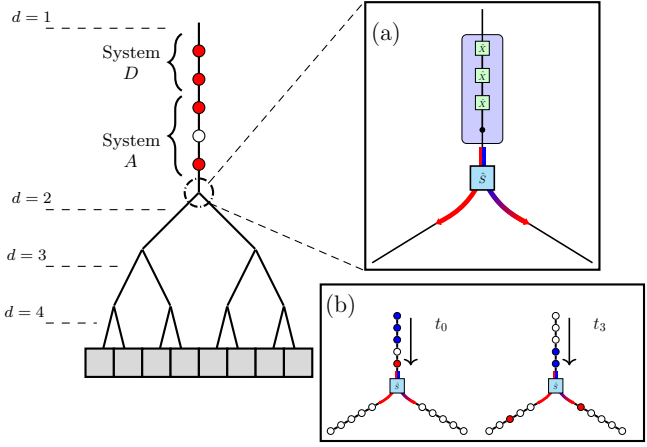
\begin{figure}[t]
\begin{tikzpicture}[scale=0.75, transform shape,
  grow=south,
  level 1/.style={level distance=1cm, sibling distance=2cm},
  level 2/.style={level distance=1cm, sibling distance=1cm},
  level 3/.style={level distance=1cm, sibling distance=0.5cm},
  edge from parent/.style={draw, thick},
  leaf/.style={rectangle, draw, fill=gray!30, minimum size=5mm}
  ]

\node (root) at (0,-1) [draw=none, minimum size=0pt, inner sep=0pt] {}
  child {node  [draw=none, minimum size=0pt, inner sep=0pt]( A0) {}
    child  {node  [draw=none, minimum size=0pt, inner sep=0pt] (A00) {}
      child {
        node[leaf] {}
        edge from parent {}
      }
      child  { 
        node[leaf] {}
        edge from parent {}
      }
      edge from parent {}
    }
    child  {node  [draw=none, minimum size=0pt, inner sep=0pt] (A01) {}
      child {
        node[leaf] {}
        edge from parent {}
      }
      child {
        node[leaf] {}
        edge from parent {}
      }
      edge from parent {}
    }
    edge from parent {}
  }
  child {node  [draw, minimum size=0pt, inner sep=0pt](A1) {}
    child {node [draw=none, minimum size=0pt, inner sep=0pt] (A10) {}
      child {
        node[leaf] {}
        edge from parent {}
      }
      child {
        node[leaf] {}
        edge from parent {}
      }
      edge from parent {}
    }
    child {node  [draw=none, minimum size=0pt, inner sep=0pt] (A11) {}
      child {
        node[leaf] {}
        edge from parent {}
      }
      child {
        node[leaf] {}
        edge from parent {}
      }
      edge from parent {}
    }
    edge from parent {}
  };

\def\n{5}
\def\spacing{0.5} 
\draw[thick] (0,-1) -- (0,{\spacing*(\n-1)});
\foreach \i/\col in {0/red, 1/white, 2/red, 3/red, 4/red} {
\edef\colorname{\col} 
\filldraw[black, fill=\colorname] (0, \i*\spacing-0.5) circle (3pt);}

\draw[decorate,decoration={brace,amplitude=5pt},thick] (-0.3,-0.7) -- (-0.3,0.7);

\draw[decorate,decoration={brace,amplitude=5pt},thick] (-0.3,0.8) -- (-0.3,1.8);

\node[align=center, text width=1cm] at (-1.2, -0.25) {System $A$};

\node[align=center, text width=1cm] at (-1.2, 1.3) { System $D$};

\node (inset1) [draw=black, thick, fill=white, rectangle, minimum width=40mm,minimum height=40mm] at (5,0) {
\begin{tikzpicture}[scale=0.6, transform shape,
  grow=south,
  level 1/.style={level distance=2cm, sibling distance=6.5cm},
  level 2/.style={level distance=1cm, sibling distance=1cm},
  level 3/.style={level distance=1cm, sibling distance=0.5cm},
  edge from parent/.style={draw, thick},
  every node/.style={circle, draw, minimum size=15mm, thick},
  leaf/.style={rectangle, draw, fill=gray!30, minimum size=5mm},
  block/.style={rectangle, draw, fill=blue!20, minimum width=10mm, minimum height=1.3cm},
  gateX/.style={rectangle, draw, fill=green!20, minimum size=10mm},
  ]
    \node[draw=none, fill=none] (rbin) at (0,0){}
  child { 
    node [draw=none, minimum size=0pt, inner sep=0pt] {} 
    edge from parent{
    }
  }
  child { 
    node [draw=none, minimum size=0pt, inner sep=0pt] {}
    edge from parent{}
  };
 \draw[draw, thick] (rbin.north) --++ (0,4.7)
  node[draw=none,midway, xshift=0cm, yshift=0cm] {
    \begin{tikzpicture}[scale=0.8]
      \node [rectangle,draw, rounded corners, fill=blue!20, inner sep=5pt, minimum width=15mm, minimum height=40mm] at (0,0.5) {};
      \def\n{4}
      \def\spacing{1}
      \draw[very thick] (0,-1.5) -- (0,{(\spacing*(\n-1))-0.5 });
      \foreach \i/\col in { 1/red, 2/white, 3/red, 4/white} {
        \filldraw[black, fill=\col] (0, \i*\spacing-2) circle (2pt);}
      \draw[very thick] (0,-1)--(0,2);
      \node[rectangle, draw, fill=green!20, minimum size=5mm] at (0,0) {$\hat{X}$};
      \node[rectangle, draw, fill=green!20, minimum size=5mm] at (0,1) {$\hat{X}$};
      \node[rectangle, draw, fill=green!20, minimum size=5mm] at (0,2) {$\hat{X}$};
      \fill (0,-1) circle (3pt);
    \end{tikzpicture}
  };
\node[draw=none, scale=1] (newrb) at (0,0.25){
\begin{tikzpicture}[line join=round]

  \def\stemthick{5pt}
  \def\branchthick{4pt}
  \def\branchlen{1.4}
  \def\bendangle{20}

  \coordinate (O) at (0,0);
  \coordinate (B) at (0,-0.8); 
  \coordinate (L) at ($ (B) + (-\branchlen,-\branchlen) $);
  \coordinate (R) at ($ (B) + (\branchlen,-\branchlen) $);

  \draw[red,line width=\stemthick] (O) -- (B);

  \draw[red,line width=\branchthick,-{Latex[length=5pt, width=7pt]}]
    (B) to[bend left=\bendangle] (L);

  \path[record path={name=ramo,dist=4pt}] (B) to[bend right=\bendangle] (R);

  \draw[opacity=0, line width=0pt] (B) to[bend right=\bendangle] (R); 
  \draw[line width=\branchthick, -{Latex[length=8pt,width=8pt]}, draw=none] (B) to[bend right=\bendangle] (R);
  \path[draw=red, postaction={draw=red, -{Latex[length=5pt,width=7pt]}, line width=\branchthick}]
    (B) to[bend right=\bendangle] (R);

  \addtocounter{parrow}{-2}
  \draw let \p1=($(R)-(B)$),\n1={atan2(\y1,\x1)} in
    [left color=blue, right color=red, shading angle=\n1+90, draw=none]
    [reconstruct top=ramo] -- (parrowm-ramo-\the\numexpr\value{parrow}+2)
    -- (parrowb-ramo-\number\value{parrow})
    [reconstruct bottom=ramo] -- (parrowm-ramo-3)
    -- (parrowt-ramo-1);

\node[rectangle, draw=none, fill=red, inner sep=0pt, outer sep=0pt, minimum width=3pt, minimum height=10mm
] at (-0.05,-0.45){}; 

\node[rectangle, draw=none, fill=blue, inner sep=0pt, outer sep=0pt, minimum width=3pt, minimum height=10mm
] at (0.05,-0.45){}; 

\end{tikzpicture}
};

\node[rectangle, fill=cyan!30, inner sep=0pt, outer sep=0pt, minimum width=20pt, minimum height=20pt
] at (0,0.5){$\hat{S}$}; 

\end{tikzpicture}
};

\node (inset2) [draw=black, thick, fill=white, rectangle, minimum width=35mm,minimum height=20mm] at (5,-4) {
\begin{minipage}{0.15\textwidth}
\begin{tikzpicture}[scale=0.6, transform shape]
\node[circle, draw=none, fill=none, minimum size=10mm] (O) at (0,0) {};

\node at (1.3,2){\Large$t_0$};
  \def\offset{5mm}
  \def\shortenval{2pt}

  \newcommand{\drawbranch}[3]{%
      \coordinate (prev) at (O.#1);
      \foreach [count=\i] \col in {#3} {%
         \coordinate (curr) at (#1:{\offset+\i*#2});
         \ifnum\i=1
            \draw[line width=1pt, shorten >=\shortenval] (prev) -- (curr);
         \else
            \draw[line width=1pt, shorten <=\shortenval, shorten >=\shortenval] (prev) -- (curr);
         \fi
         \ifthenelse{\equal{\col}{}}{%
           \node[circle, draw, fill=none, inner sep=1.5pt,minimum size=2mm] at (curr) {};%
         }{%
           \node[circle, draw, fill=\col, inner sep=1.5pt,minimum size=2mm] at (curr) {};%
         }%
         \coordinate (prev) at (curr);
      }%
  }

  \drawbranch{90}{10}{red,,blue,blue,blue}
  \drawbranch{210}{10}{,,,,}
  \drawbranch{330}{10}{,,,,}
\def\lunghezza{1.3}

\draw[->, thick] (0.4,2.2) -- (0.4,0.8);

\node[draw=none, scale=0.4] (rbnew1) at (0,0.13){\begin{tikzpicture}[line join=round]

  \def\stemthick{5pt}
  \def\branchthick{6pt}
  \def\branchlen{1.4}
  \def\bendangle{20}

  \coordinate (O) at (0,0);
  \coordinate (B) at (0,-0.8); 
  \coordinate (L) at ($ (B) + (-\branchlen,-\branchlen) $);
  \coordinate (R) at ($ (B) + (\branchlen,-\branchlen) $);

  \draw[red,line width=\stemthick] (O) -- (B);

  \draw[red,line width=\branchthick,-{Latex[length=5pt, width=7pt]}]
    (B) to[bend left=\bendangle] (L);

  \path[record path={name=ramo,dist=6pt}] (B) to[bend right=\bendangle] (R);

  \draw[opacity=0, line width=0pt] (B) to[bend right=\bendangle] (R); 
  \draw[line width=\branchthick, -{Latex[length=8pt,width=8pt]}, draw=none] (B) to[bend right=\bendangle] (R);
  \path[draw=red, postaction={draw=red, -{Latex[length=5pt,width=7pt]}, line width=\branchthick}]
    (B) to[bend right=\bendangle] (R);

  \addtocounter{parrow}{-2}
  \draw let \p1=($(R)-(B)$),\n1={atan2(\y1,\x1)} in
    [left color=blue, right color=red, shading angle=\n1+90, draw=none]
    [reconstruct top=ramo] -- (parrowm-ramo-\the\numexpr\value{parrow}+2)
    -- (parrowb-ramo-\number\value{parrow})
    [reconstruct bottom=ramo] -- (parrowm-ramo-3)
    -- (parrowt-ramo-1);

\node[rectangle, draw=none, fill=red, inner sep=0pt, outer sep=0pt, minimum width=5pt, minimum height=12mm
] at (-0.08,-0.45){}; 

\node[rectangle, draw=none, fill=blue, inner sep=0pt, outer sep=0pt, minimum width=5pt, minimum height=12mm
] at (0.08,-0.45){};

\end{tikzpicture}};

\node[rectangle,draw,scale=0.6, fill=cyan!30, inner sep=0pt, outer sep=0pt, minimum width=20pt, minimum height=20pt
] at (0,0.2){$\hat{S}$}; 

\end{tikzpicture}

\end{minipage}
\hfill
\begin{minipage}{0.15\textwidth}
    \begin{tikzpicture}[scale=0.6, transform shape]

  \node[circle, draw=none, fill=none, minimum size=10mm] (O) at (0,0) {};

  \node at (1.3,2){\Large$t_3$};
  \def\offset{5mm}
  \def\shortenval{2pt}

  \newcommand{\drawbranch}[3]{%
      \coordinate (prev) at (O.#1);
      \foreach [count=\i] \col in {#3} {%
         \coordinate (curr) at (#1:{\offset+\i*#2});
         \ifnum\i=1
            \draw[line width=1pt, shorten >=\shortenval] (prev) -- (curr);
         \else
            \draw[line width=1pt, shorten <=\shortenval, shorten >=\shortenval] (prev) -- (curr);
         \fi
         \ifthenelse{\equal{\col}{}}{%
           \node[circle, draw, fill=none, inner sep=1.5pt,minimum size=2mm] at (curr) {};%
         }{%
           \node[circle, draw, fill=\col, inner sep=1.5pt,minimum size=2mm] at (curr) {};%
         }%
         \coordinate (prev) at (curr);
      }%
  }

  \drawbranch{90}{10}{blue,blue,,,}
  \drawbranch{210}{10}{,,red,,}
  \drawbranch{330}{10}{red,,,,}
\def\lunghezza{1.3}

\draw[->, thick] (0.4,2.2) -- (0.4,0.8);


\node[draw=none, scale=0.4] (rbnew2) at (0,0.13){\begin{tikzpicture}[line join=round]

  \def\stemthick{5pt}
  \def\branchthick{6pt}
  \def\branchlen{1.4}
  \def\bendangle{20}

  \coordinate (O) at (0,0);
  \coordinate (B) at (0,-0.8); 
  \coordinate (L) at ($ (B) + (-\branchlen,-\branchlen) $);
  \coordinate (R) at ($ (B) + (\branchlen,-\branchlen) $);

  \draw[red,line width=\stemthick] (O) -- (B);

  \draw[red,line width=\branchthick,-{Latex[length=5pt, width=7pt]}]
    (B) to[bend left=\bendangle] (L);

  \path[record path={name=ramo,dist=6pt}] (B) to[bend right=\bendangle] (R);

  \draw[opacity=0, line width=0pt] (B) to[bend right=\bendangle] (R); 
  \draw[line width=\branchthick, -{Latex[length=8pt,width=8pt]}, draw=none] (B) to[bend right=\bendangle] (R);
  \path[draw=red, postaction={draw=red, -{Latex[length=5pt,width=7pt]}, line width=\branchthick}]
    (B) to[bend right=\bendangle] (R);

  \addtocounter{parrow}{-2}
  \draw let \p1=($(R)-(B)$),\n1={atan2(\y1,\x1)} in
    [left color=blue, right color=red, shading angle=\n1+90, draw=none]
    [reconstruct top=ramo] -- (parrowm-ramo-\the\numexpr\value{parrow}+2)
    -- (parrowb-ramo-\number\value{parrow})
    [reconstruct bottom=ramo] -- (parrowm-ramo-3)
    -- (parrowt-ramo-1);

\node[rectangle, draw=none, fill=red, inner sep=0pt, outer sep=0pt, minimum width=5pt, minimum height=12mm
] at (-0.08,-0.45){}; 

\node[rectangle, draw=none, fill=blue, inner sep=0pt, outer sep=0pt, minimum width=5pt, minimum height=12mm
] at (0.08,-0.45){}; 

\end{tikzpicture}};

\node[rectangle,draw,scale=0.6, fill=cyan!30, inner sep=0pt, outer sep=0pt, minimum width=20pt, minimum height=20pt
] at (0,0.2){$\hat{S}$}; 

\end{tikzpicture}
\end{minipage}
};

\node(zoom1)[densely dashdotted, draw=black, thick, circle, minimum size=0.5cm] at (root) {};

\draw[densely dashed, draw=black] (zoom1.south east) -- (inset1.south west);
\draw[densely dashed, draw=black] (zoom1.north east) -- (inset1.north west);







\draw[dashed] (-2.7,1.9) -- (-0.5,1.9) node [pos=0.1,above left] {\footnotesize$d=1$};

\draw[dashed] (-2.7,-1.3) -- (-0.5,-1.3) node [above left, pos=0.1] {\footnotesize$d=2$};

\draw[dashed] (-2.7,-2.3) -- (-1.4,-2.3) node [pos=0.1,above left] {\footnotesize$d=3$};

\draw[dashed] (-2.7,-3.3) -- (-1.8,-3.3) node [pos=0.1,above left] {\footnotesize$d=4$};



\node[draw=none] at (3.3,1.8) {\large(a)};

\node[draw=none] at (2.5,-3) {\large(b)};

\end{tikzpicture}
\caption{Schematic depiction of our qRAM model for a binary tree of depth $d=3$ with $N=2^3$ memory cells. At the root node, we inject $n=3$ quantum walkers encoding the address state and $m+1$ walkers for message retrieval. (a) Walkers moving from top to bottom are routed left/right according to their internal degree of freedom by the gate $\hat{S}$. Information about the subsequent route is transmitted by the $i$-th address walker to all the walkers by the unitary gate $\hat{U}^{(i)}$ (purple box in the figure).
(b) Workflow of the scattering gate $\hat{S}$: a walker in the red state $\ket{R}$ is routed left and remains $\ket{R}$, while a walker in the blue state $\ket{B}$ is routed right and is flipped to $\ket{R}$.}
\label{binarytree}
\end{figure}

\textit{Routing architecture:--}
{In our qRAM model, the registers $A$ and $D$ of Eq.~(\ref{generalqRAM}) are implemented via ordered sequences of 
quantum  walkers. As we explain in detail below, the encoding used in our scheme allows each subsystem of $A$ and $D$ -- namely, the elements in $\{A_1, A_2, ..., A_n\}$ and $\{D_0, D_1, D_2, ..., D_m\}$ -- to be implemented either by a quantum walker or by no walker at all.
In the simplest version of our architecture, each walker possesses an internal degree of freedom -- referred to as color -- spanned by two orthogonal states, $\ket{R}$ 
(red) and $\ket{B}$ (blue). 
Instead, when any subsystem is not implemented by a walker, we formally represent it in the state $\ket{\emptyset}$. We emphasize that this empty state is to be understood as the presence of a gap between colored walkers in the input sequence of the algorithm, not representing a third proper internal state.
The quantum walkers considered in our scheme propagate ballistically and with constant speed along the branches of a 
  binary-tree graph of depth $n$  whose emerging paths are connected with 
the $N$ cells of the database}.
As in a Stern-Gerlach apparatus or a polarizing beam splitter, these internal states determine the walker’s path at the bifurcations of the tree. 
{Each bifurcation is realized through a unitary gate $\hat S$ which routes walkers in the state $\ket{R}$ via the left output port, without any change of color, while walkers in the state $\ket{B}$ exit through the right port emerging as $\ket{R}$.} {The main features of our qRAM architecture are depicted in Fig.~\ref{binarytree}.}

The protocol begins by sequentially injecting the $n$ address walkers $A$ into the root node of the tree -- first  $A_1$, then $A_2$, and so on -- followed by the $m+1$ data walkers  $D$ initialized in the fiduciary state
$|R
\rangle_{D}^{\otimes (m+1)}:= |R\rangle_{D_0}|R\rangle_{D_1} \cdots |R\rangle_{D_{m}}$ where each of them are in the red state 
$\ket{R}$. This encoding ensures that data walkers remain present during the entire routing procedure, enabling the activation of memory operations only when the addressed memory cell is reached.
 To encode the $i$-th bit  $a_i$ of the address binary sequence $\boldsymbol{a}$, we adopt the mapping:
 \begin{equation}\label{infoencoding}
a_i=0 \rightarrow \ket{\emptyset}_{A_i},\quad a_i=1 \rightarrow \ket{R}_{A_i}.
\end{equation}
Thus, the all-zero address ${00\dots 0}$ corresponds to the vacuum state 
$\ket{\emptyset \emptyset\dots \emptyset}_A$, while the all-one address ${11\dots 1}$ is 
encoded by the state $\ket{RR\dots R}_A$ of $n$ red walkers. Since Eq.~(\ref{generalqRAM}) necessitates coherent superpositions of address states, the encoding in Eq.~(\ref{infoencoding}) necessitates the use of bosonic carriers (e.g., photons). However, we emphasize that neither the use of bosons nor the use of the vacuum state of the walkers as a logical element is a fundamental constraint. These assumptions can be relaxed, for example,  by employing a dual-rail encoding or by increasing the number of internal states available to each walker, see App.~\ref{SecAlternative}.

The routing strategy we adopt relies on the fact that the $i$-th address walker determines the direction taken by all the subsequent walkers in the initialization sequence -- namely, the address walkers from $A_{i+1}$ to $A_n$, as well as the data walkers $D$ -- when they reach a bifurcation at the $i$-th level of the binary tree. 
This scheme is realized through two unitary operations applied at each node: (i)~the gate $\hat{U}^{(i)}$, which, depending on the internal state of the $i$-th address walker, either changes or preserves the color of all its following walkers, and {(ii) the scattering gate $\hat S$ described above.}
{The unitary gate $\hat U^{(i)}$} is activated by an external clock that, when $A_i$ reaches a bifurcation point at level $i$ of the tree, triggers the following control-unitary operation:
\begin{eqnarray}\label{Ugate}
&& \hat{U}^{(i)}\ket{R}_D^{\otimes (m+1)}\ket{a_n, \dots, a_{i+1}, R}_A  \\ \nonumber 
&&\qquad \qquad  = \ket{B}_D^{\otimes (m+1)}\ket{\bar{a}_n, \dots, \bar{a}_{i+1}, R}_A,
\end{eqnarray}
where $\bar{a}$ denotes the negation of $a$, defined by  the rules
$\bar{R} = B$, $\bar{B} = R$, $\bar{\emptyset} = \emptyset$. If the 
$i$-th address walker is not present (i.e. $\ket{\emptyset}_{A_i}$), then $\hat{U}^{(i)}$ leaves the state of all the other walkers unchanged.
In essence, the operation $\hat{U}^{(i)}$ uses the color state of the $i$-th  address walker to conditionally
invert or preserve the color of all the subsequent walkers: the inversion occurs if the $i$-th walker is red; otherwise, their states are preserved. Hence due to the combined action of $\hat U^{(i)}$ and $\hat S$, when $A_i$ is in the red state, all the walkers $A_{i+1}$, $\cdots$, $A_n$, $D_0$, $\cdots$, $D_{m}$ 
acquire a blue state (action of $\hat{U}^{(i)}$), and turn  right at the following  bifurcation and change back their color to red (action of $\hat S$). If instead 
$A_i$ is in the empty state, all its following walkers retain the red internal (or empty) state and turn left as required by the routing protocol. A more detailed explanation of this routing strategy can
be found in App.~\ref{SecRout}.

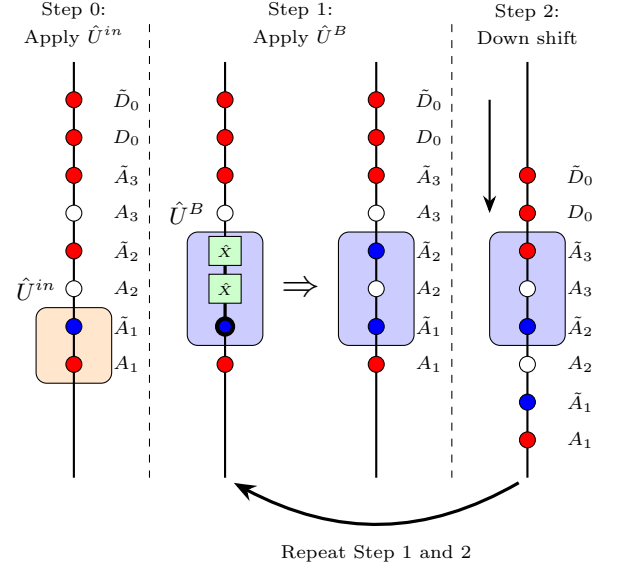
\begin{figure}[t]
\centering
\begin{tikzpicture}[gateX/.style={rectangle, draw, fill=green!20, minimum size=2mm}, scale=1, transform shape]


\node [draw, rounded corners, fill=orange!20, inner sep=5pt, minimum width=10mm, minimum height=10mm] 
        at (2,-1.75) {};

\def\n{8}
\def\spacing{0.5} 
\draw[thick] (2,-3.5) -- (2,{(\spacing*(\n-1))-1.5 });


\foreach \i/\col in {0/red, 1/blue, 2/white, 3/red, 4/white,5/red,6/red,7/red} {
\edef\colorname{\col} 
\filldraw[black, fill=\colorname] (2, \i*\spacing-2) circle (3pt);}


\node at (1.5,-1){$\hat{U}^{in}$};

\node at (2.7,1.5){\scriptsize$\tilde{D}_0$};

\node at (2.7,1){\scriptsize$D_0$};

\node at (2.7,0.5){\scriptsize$\tilde{A}_3$};

\node at (2.7,0){\scriptsize$A_3$};

\node at (2.7,-0.5){\scriptsize$\tilde{A}_2$};

\node at (2.7,-1){\scriptsize$A_2$};

\node at (2.7,-1.5){\scriptsize$\tilde{A}_1$};

\node at (2.7,-2){\scriptsize$A_1$};


\node [draw, rounded corners, fill=blue!20, inner sep=5pt, minimum width=10mm, minimum height=15mm] 
        at (4,-1) {};

\def\n{8}
\def\spacing{0.5} 
\draw[thick] (4,-3.5) -- (4,{(\spacing*(\n-1))-1.5 });

\draw[very thick] (4,-1.5)--(4,-0.5);

\foreach \i/\col in {0/red, 1/blue, 2/white, 3/blue, 4/white,5/red,6/red,7/red} {
\edef\colorname{\col} 
\filldraw[black, fill=\colorname] (4, \i*\spacing-2) circle (3pt);}

\node[midway,yshift=-0.5cm,xshift=4cm,gateX] {\tiny$\hat{X}$};

\node[midway,yshift=-1cm,xshift=4cm,gateX] {\tiny$\hat{X}$};

\draw[ultra thick] (4,-1.5) circle (3pt);









\node at (5,-1){\Large $\Rightarrow$};

\node at (3.5,0){$\hat{U}^{B}$};


\node [draw, rounded corners, fill=blue!20, inner sep=5pt, minimum width=10mm, minimum height=15mm] 
        at (6,-1) {};

\def\n{8}
\def\spacing{0.5} 
\draw[thick] (6,-3.5) -- (6,{(\spacing*(\n-1))-1.5 });



\foreach \i/\col in {0/red, 1/blue, 2/white, 3/blue, 4/white,5/red,6/red,7/red} {
\edef\colorname{\col} 
\filldraw[black, fill=\colorname] (6, \i*\spacing-2) circle (3pt);}

\node at (6.7,1.5){\scriptsize$\tilde{D}_0$};

\node at (6.7,1){\scriptsize$D_0$};

\node at (6.7,0.5){\scriptsize$\tilde{A}_3$};

\node at (6.7,0){\scriptsize$A_3$};

\node at (6.7,-0.5){\scriptsize$\tilde{A}_2$};

\node at (6.7,-1){\scriptsize$A_2$};

\node at (6.7,-1.5){\scriptsize$\tilde{A}_1$};

\node at (6.7,-2){\scriptsize$A_1$};


\node [draw, rounded corners, fill=blue!20, inner sep=5pt, minimum width=10mm, minimum height=15mm] 
        at (8,-1) {};

\def\n{8}
\def\spacing{0.5} 
\draw[thick] (8,-3.5) -- (8,{(\spacing*(\n-1))-1.5 });



\foreach \i/\col in {0/white, 1/blue, 2/white, 3/red, 4/red,5/red} {
\edef\colorname{\col} 
\filldraw[black, fill=\colorname] (8, \i*\spacing-2) circle (3pt);}

\filldraw[color=black, fill=blue](8,-2.5) circle (3pt);

\filldraw[color=black, fill=red](8,-3) circle (3pt);

\node at (8.7,0.5){\scriptsize$\tilde{D}_0$};

\node at (8.7,0){\scriptsize$D_0$};

\node at (8.7,-0.5){\scriptsize$\tilde{A}_3$};

\node at (8.7,-1){\scriptsize$A_3$};

\node at (8.7,-1.5){\scriptsize$\tilde{A}_2$};

\node at (8.7,-2){\scriptsize$A_2$};

\node at (8.7,-2.5){\scriptsize$\tilde{A}_1$};

\node at (8.7,-3){\scriptsize$A_1$};

\draw[->, -Stealth,thick] (7.5,1.5)--(7.5,0);

\draw[dashed] (3,2.5)--(3,-3.5);

\draw[dashed] (7,2.5)--(7,-3.5);

\node(0) at (2,2.5){};

\node(1a) at (4,2.5){};

\node(1b) at (6,2.5){};

\node(2) at (8,2.5){};

\node(in) at (4,-3.5){};

\node(out) at (8,-3.5){};

\node[align=center] at (2,2.5){\scriptsize Step 0: \\ \scriptsize Apply $\hat{U}^{in}$};


\node[align=center] at (5,2.5){\scriptsize Step 1: \\ \scriptsize Apply $\hat{U}^B$ };


\node[align=center] at (8,2.5){\scriptsize Step 2: \\ \scriptsize Down shift};

\draw[->, >={Stealth[bend]}, very thick, out=-150, in=-30] (out) to (in);

\node at (6,-4.5){\scriptsize Repeat Step 1 and 2 };

\end{tikzpicture}
\caption{ Example of the  backup variant implementation of the unitary evolution 
$\hat{U}^{(1)}$ for an input state with address bits $a_1=1$, $a_2=0$, $a_3=0$.
Step 0: the first address bit transfers its path information to its corresponding backup walker through an initial operation $\hat{U}^{in}$. Subsequently, we enter the main body of the iterative procedure, split into two main steps. 
Step 1: the backup, the next address walker, and its associated backup enter the gate $\hat{U}^B$. This gate operates via a C-NOT-NOT mechanism, enabling the backup to transmit its path information to the other two particles. Step 2: after the interaction, all three particles are shifted forward by two positions. As a result, the second backup now holds the same path information originally carried by the first. 
This procedure is repeated $n+m-(d+1)$ times.
}
\label{backupscheme}
\end{figure}

We notice that, even if the application of $\hat U^{(i)}$ appears to require long-range interactions between the $i$-th walker and all the following ones (see Eq.~(\ref{Ugate})), it can be implemented using only short-range couplings with a slight modification of the scheme (see next section).
It is also important to note that in the transformation (\ref{Ugate}), the walker that triggers the operation retains its original color and follows a distinct path. Consequently, at each level of the tree, one address walker is effectively removed from the sequence to be later recollected during the output phase of the protocol -- see below.
On the contrary, all the red walkers of  register $D$  associated with the input state 
$\ket{\boldsymbol{a}}_A\ket{R}_D^{\otimes (m+1)}$ 
reach the $M^{(\boldsymbol{a})}$ memory cell specified by the address 
walkers $A$ via the encoding~(\ref{infoencoding}).

\begin{table*}[ht]
\centering
\begin{adjustbox}{width=\textwidth,center}
\begin{tabular}{cccc|ccc}
\hline
             & \multicolumn{1}{c}{\begin{tabular}{c}
              \textbf{Particles}
         \end{tabular}} & \multicolumn{1}{c}{\begin{tabular}{c}
              \textbf{Circuit depth}
         \end{tabular}} & \multicolumn{1}{c|}{\begin{tabular}{c}
              \textbf{Binary trees}
         \end{tabular}} & \textbf{Call type}  & \multicolumn{1}{c}{\begin{tabular}{c}
              \textbf{Two-qubit gates}
         \end{tabular}} & \textbf{Node operations} \\\hline\hline
\multicolumn{1}{c}{\begin{tabular}{c}           \textbf{BB}~\cite{Bucket_Brigade}
         \end{tabular}} & \multicolumn{1}{c}{\begin{tabular}{c}
             $2^n-1$ qutrits,  \\
             $ n + m$ qubits
         \end{tabular} }
         & $\mathcal{O}(n + m)$~\cite{Hann_noise_resilience} & $1$& \multicolumn{1}{c}{\begin{tabular}{l}
              Classical address  \\
              Superposition
         \end{tabular}} & \multicolumn{1}{c}{\begin{tabular}{c}
              $\mathcal{O}(n^2+nm)$  \\
              $\mathcal{O}((n+m)2^n)$
         \end{tabular}} & \multicolumn{1}{c}{\begin{tabular}{c}
              \textbackslash
         \end{tabular}}\\\hline
\multicolumn{1}{c}{\begin{tabular}{c}
              \textbf{ASY}~\cite{Asaka_breve}
         \end{tabular}} & 
        $n+m$ qubits & $\mathcal{O}(n^2 + nm)$ & $2(n+m)$& \multicolumn{1}{c}{\begin{tabular}{l}
              Classical address  \\
              Superposition
         \end{tabular}} & \multicolumn{1}{c}{\begin{tabular}{c}
              $\mathcal{O}(n^2+nm)$  \\
              $\mathcal{O}((n+m)2^n)$
         \end{tabular}} & \multicolumn{1}{c}{\begin{tabular}{c}
              $\mathcal{O}(n^2+nm)$  \\
              $\mathcal{O}((n+m)2^n)$
         \end{tabular}}\\\hline
\textbf{Backup model} & $\mathcal{O}(n+m)$ qubits & $\mathcal{O}(n + m)$  &$1$ & \multicolumn{1}{c}{\begin{tabular}{l}
              Classical address  \\
              Superposition
         \end{tabular}} & \multicolumn{1}{c}{\begin{tabular}{c}
              $\mathcal{O}(n^2+nm)$  \\
              $\mathcal{O}((n+m)2^n)$
         \end{tabular}} & \multicolumn{1}{c}{\begin{tabular}{c}
              $\mathcal{O}(n^2 + nm)$  \\
              $\mathcal{O}((n+m)2^n)$
         \end{tabular}}\\\hline\hline

\end{tabular}
\end{adjustbox}
\caption{Resources scaling comparison between qRAM architecture. Left side: particles required, circuit depth obtained allowing short-range couplings, and binary trees employed. Here, our backup qRAM achieves the most favorable scalings. For the ASY model, the depth scaling is computed assuming that long-range gates are not allowed; when such gates are permitted, the routing depth reduces to $\mathcal{O}(n\mathrm{log}(n+m))$. Right side: gate activation scalings for both classical and quantum queries, which are comparable among the three architectures.} 
\label{Tab: scaling}
\end{table*}

{\it{Coping and retrieving:--}} Once the routing process is concluded, 
the transfer of the $m$ bits of information  $\boldsymbol{b}^{(\boldsymbol{a})}$, contained in the memory cell $M^{(\boldsymbol{a})}$, 
to $D$ 
is implemented using a control-$COPY$ unitary gate triggered by the walker $D_0$.
Specifically, if $D_0$ is in the red state, then for each $j\in\{ 1,\cdots, n\}$, a coupling is activated  between the $D_j$  and the 
$j$-th bit entries contained in the cell $M^{(\boldsymbol{a})}$, inducing the following mapping 
\begin{equation} \label{defcopy} 
  |R\rangle_{D_j} \xrightarrow{ b^{(\boldsymbol{a})}_j=0} |\emptyset \rangle_{D_j},\qquad|R\rangle_{D_j}\xrightarrow{ b^{(\boldsymbol{a})}_j=1}    |R \rangle_{D_j}.
\end{equation} 
In particular, when the memory entry $M_j^{(\boldsymbol{a})}$ is empty, the copy operation resets the data walker $D_j$ from $\ket{R}$ to $\ket{\emptyset}$, while if $M_j^{(\boldsymbol{a})}$ is in $\ket{R}$ the data walker remains in $\ket{R}$.
If  $D_0$ is in the empty state, no interaction occurs.  As discussed in App.~\ref{SecMessCop}, this transformation can be implemented either by considering an internal degree of freedom of the memory cell and a collection of ${\cal O}(m)$ two-body couplings among the $D$ walkers and the qubits of the memory cell $M^{(\boldsymbol{a})}$ or through long-range interactions. We anticipate that, in what follows, we present a modified version of our qRAM scheme that neither requires long-range interactions nor any internal degrees of freedom in the memory cells.
The recovery of the information in the form specified by Eq.~(\ref{generalqRAM}) is then executed by reversing the routing operations.
A crucial aspect of this process is the coherent recollection of all particles, including those that were diverted from the main path by the action of the gates $\hat{U}^{(i)}$.
Specifically we employ a $n$-level inverted binary tree, where $N$ input branches collapse to a single output line through $N-1$ inverted bifurcations which implement the time-reversed mapping introduced in the routing stage: a red walker entering from the right emerges as a blue walker $|B\rangle$, while a red walker entering from the left input port emerges red $|R\rangle$ (see Fig.~\ref{binarytree}b).

\textit{Backup variant:--}
The previously described model relies on the application of a large number of many-body controlled gates~(\ref{Ugate}). These transformations are challenging to implement due to hardware constraints, error accumulation, and the difficulty of precise control over many-body interactions \cite{BB_robustness,Hann_noise_resilience,Xu_QRAM_critique}. To address these issues, we propose a variation of the model that, at the cost of a linear overhead in the number of walkers, allows for a reformulation entirely in terms of simpler operations acting on blocks of three walkers at a time. 
 We modify our protocol by roughly doubling the total number of walkers: we now use $2n$ walkers for routing and 
{$2m-1$ data} walkers for message retrieval. In this modified scheme, half of the walkers are initialized as in the previously shown protocol, while the other half, our “backup walkers,” are all initialized in the state $\ket{R}$, {where we require no backup for the last data walker $D_m$}. These backup walkers allow us to implement a simplified short-range form for $\hat U^{(i)}$.

The initial state of the qRAM is prepared as in the standard protocol, but with a red backup walker following each primary address and memory walker.  More precisely, after each walker $A_i$ of the address, we place its corresponding backup $\tilde{A}_i$ in the state $\ket{R}_{\tilde{A}_i}$. Similarly, after each data walker $D_j$, apart from $D_m$, we place its corresponding backup $\tilde{D_j}$ in the state $\ket{R}_{\tilde{D_j}}$.
Thus, the overall initial state, for a classical address, is now initialized in the vector {
$\ket{R}_{D_m}\!\!\otimes(\otimes_{j=1}^{m-1}\ket{R}_{\tilde{D}_j}\ket{R}_{D_j})\otimes(\otimes_{i=1}^n\ket{R}_{\tilde{A}_i}\ket{a_i}_{A_i})$.}
In this modified setting, $\hat U^{(i)}$ is replaced by (i) the application of a unitary $\hat U_{in}$ acting on the $i$-th walker $A_i$ and its backup $\tilde A_i$ and (ii) the repeated application of a unitary block $\hat{U}_B$ that acts on three walkers at a time (see Fig.~\ref{backupscheme}).  The unitary transformation $\hat{U}_{in}$ is a C-NOT between $A_i$ and $\tilde A_i$, which is activated if $A_i$ is a red walker, and has the role to transfer the information contained in $A_i$ to $\tilde A_i$. Hence,
if the address state is $\ket{\emptyset}_{A_i}$, $\hat U_{in}$ leaves the state of the backup walker $\tilde A_i$ in the state $\ket{R}_{\tilde A_i}$. Instead, if the $i$-th address walker is in the state $\ket{R}_{A_i}$, $\hat U_{in}$ changes the color of the backup walker into blue, see App.~\ref{BackupGates}. The transformation $\hat U_{B}$ is a C-NOT-NOT gate which at its first application acts on $\tilde A_i$, $A_{i+1}$ and $\tilde A_{i+1}$, which is activated if the control backup walker $\tilde A_i$ is blue. Later the particles are moved forward by two sites and $\hat U_{B}$ is applied again, now with $\tilde A_{i+1}$ in the control site and $A_{i+2}$ and $\tilde A_{i+2}$ on the target sites. This scheme is repeated until all the walkers following the $i$-th walker have gone through $\hat U_B$. Hence, $\hat{U}_B$ has to be repeated $n+m-(d+1)$ times on level $d$. It is straightforward to verify that the action of $\hat U_{in}$ followed by the described repeated application of $U_B$ coincides with the unitary $U^{(i)}$ applied on the standard address and data register walkers $A_i$ and~$D_j$. 
Furthermore, in this scheme the extra flag walker $D_0$ becomes unnecessary: the message copy mechanism trigger can be handled by the final address backup $\tilde{A}_n$, see App.~\ref{BackupMessCopy}.

We finally point out that this backup architecture admits a natural parallelization of the
routing operations, which improves the scalings of the circuit depth and the physical length 
from $\mathcal{O}(n^2+nm)$, as in the case of our standard qRAM, to $\mathcal{O}(n+m)$. See App.~\ref{DepthAnalysis} for more details.

\textit{Discussion:--}
This work introduces a qRAM architecture in which the internal and spatial degrees of freedom of quantum walkers are jointly exploited to encode and propagate the routing information required to access memory cells efficiently.

We first propose a qRAM scheme requiring
long-range controlled operations, which employs a minimal number of quantum walkers and binary trees.
To further bring the architecture closer to near-term experimental realization, we introduce a backup variant that relies exclusively on local and short-range two-body interactions between information carriers during the routing phase. This scheme reduces the physical overhead associated with qRAM implementations and permits an advantageous parallelization of the controlled gates applied along the binary tree. As a consequence, the circuit depth of the architecture scales linearly with $n$, providing a quadratic improvement with respect to the long-range protocol.

We stress that both the standard and backup variants of the qRAM protocol do not inherently rely on vacuum states or bosonic information carriers. Indeed, the underlying architecture is fully compatible with alternative implementations employing fermionic carriers with two internal states (i.e., qubits), or more generally $d$-level systems (qudits). These generalizations require only minor adjustments to the encoding scheme and interaction rules, and can be adapted to a broad class of physical platforms without altering the essential functioning of the protocol. A detailed explanation of such modifications to adapt the protocol can be found in App.~\ref{SecAlternative}.

Overall, our architecture combines the passive routing advantages of the ASY qRAM with the structured efficiency of the BB scheme, while implementing a novel routing strategy that allows our scheme to achieve an optimal balance between robustness and experimental feasibility.
We benchmark our model against the BB architecture~\cite{Bucket_Brigade,BB_robustness} and the {ASY qRAM}~\cite{Asaka_breve}. We summarize this comparison in Table~\ref{Tab: scaling}, where we display how our backup proposal guarantees the best scalings of different resources. In terms of particles employed, our model and the ASY proposal utilize $\mathcal{O}(n+m)$ qubits, while the BB scheme includes an additional exponential number of qutrits, i.e., $\mathcal{O}(2^n)$, that compose the active structure of the qRAM binary tree. Our model achieves the best scaling in terms of circuit depth together with the BB model ~\cite{Hann_noise_resilience}, which is $\mathcal O(n+m)$. Notice that the ASY model has a circuit depth that scales quadratically with~$n$. Finally, our routing procedure requires a single binary tree, as in the BB scheme, differently from the ASY scheme, which requires $2(n+m)$ of them.

Future works will explore the suitability for integration with existing quantum hardware platforms such as photonic processors, trapped ions, or superconducting qubits, where connectivity constraints and coherence times impose strict limits on circuit depth and gate non-locality. Furthermore, we plan to analyze the noise-resilience of the protocol via simulations based on realistic error models to explore the possibilities for error‑mitigation strategies and full error‑correction schemes.

\textit{Acknowledgement:--}We acknowledge financial support
by MUR (Ministero dell’ Università e della
Ricerca) through the PNRR MUR project
PE0000023-NQSTI. D. De Santis acknowledges support from Scuola Normale Superiore di Pisa.

\bibliography{biblio}

@book{nielsen_chuang,
  author = {Nielsen, Michael A. and Chuang, Isaac L.},
  publisher = {Cambridge University Press},
  title = {Quantum Computation and Quantum Information},
  year = 2000
}

@article{Bucket_Brigade,
   title={Quantum Random Access Memory},
   volume={100},
   ISSN={1079-7114},
   url={http://dx.doi.org/10.1103/PhysRevLett.100.160501},
   DOI={10.1103/physrevlett.100.160501},
   number={16},
   journal={Physical Review Letters},
   publisher={American Physical Society (APS)},
   author={Giovannetti, Vittorio and Lloyd, Seth and Maccone, Lorenzo},
   year={2008},
   month=apr }

@article{BB_robustness,
   title={On the robustness of bucket brigade quantum RAM},
   volume={17},
   ISSN={1367-2630},
   url={http://dx.doi.org/10.1088/1367-2630/17/12/123010},
   DOI={10.1088/1367-2630/17/12/123010},
   number={12},
   journal={New Journal of Physics},
   publisher={IOP Publishing},
   author={Arunachalam, Srinivasan and Gheorghiu, Vlad and Jochym-O’Connor, Tomas and Mosca, Michele and Srinivasan, Priyaa Varshinee},
   year={2015},
   month=dec, pages={123010} }

@article{Quantum_Walker_intro,
   title={Quantum random walks: An introductory overview},
   volume={44},
   ISSN={1366-5812},
   url={http://dx.doi.org/10.1080/00107151031000110776},
   DOI={10.1080/00107151031000110776},
   number={4},
   journal={Contemporary Physics},
   publisher={Informa UK Limited},
   author={Kempe, J},
   year={2003},
   month=jul, pages={307–327} }

@article{Asaka_breve,
   title={Quantum random access memory via quantum walk},
   volume={6},
   ISSN={2058-9565},
   url={http://dx.doi.org/10.1088/2058-9565/abf484},
   DOI={10.1088/2058-9565/abf484},
   number={3},
   journal={Quantum Science and Technology},
   publisher={IOP Publishing},
   author={Asaka, Ryo and Sakai, Kazumitsu and Yahagi, Ryoko},
   year={2021},
   month=may, pages={035004} }

@article{Shor_factoring,
   title={Polynomial-Time Algorithms for Prime Factorization and Discrete Logarithms on a Quantum Computer},
   volume={26},
   ISSN={1095-7111},
   url={http://dx.doi.org/10.1137/S0097539795293172},
   DOI={10.1137/s0097539795293172},
   number={5},
   journal={SIAM Journal on Computing},
   publisher={Society for Industrial & Applied Mathematics (SIAM)},
   author={Shor, Peter W.},
   year={1997},
   month=oct, pages={1484–1509} }

@misc{Grover_search,
      title={A fast quantum mechanical algorithm for database search}, 
      author={Lov K. Grover},
      year={1996},
      eprint={quant-ph/9605043},
      archivePrefix={arXiv},
      primaryClass={quant-ph},
      url={https://arxiv.org/abs/quant-ph/9605043}, 
}

@article{Low_Linear_algebra,
  doi = {10.22331/q-2026-03-23-2041},
  url = {https://doi.org/10.22331/q-2026-03-23-2041},
  title = {Quantum linear system algorithm with optimal queries to initial state preparation},
  author = {Low, Guang Hao and Su, Yuan},
  journal = {{Quantum}},
  issn = {2521-327X},
  publisher = {{Verein zur F{\"{o}}rderung des Open Access Publizierens in den Quantenwissenschaften}},
  volume = {10},
  pages = {2041},
  month = mar,
  year = {2026}
}

@article{Harrow_Linear_algebra,
   title={Quantum Algorithm for Linear Systems of Equations},
   volume={103},
   ISSN={1079-7114},
   url={http://dx.doi.org/10.1103/PhysRevLett.103.150502},
   DOI={10.1103/physrevlett.103.150502},
   number={15},
   journal={Physical Review Letters},
   publisher={American Physical Society (APS)},
   author={Harrow, Aram W. and Hassidim, Avinatan and Lloyd, Seth},
   year={2009},
   month=oct }

@article{Childs_Linear_algebra,
   title={Quantum Algorithm for Systems of Linear Equations with Exponentially Improved Dependence on Precision},
   volume={46},
   ISSN={1095-7111},
   url={http://dx.doi.org/10.1137/16M1087072},
   DOI={10.1137/16m1087072},
   number={6},
   journal={SIAM Journal on Computing},
   publisher={Society for Industrial & Applied Mathematics (SIAM)},
   author={Childs, Andrew M. and Kothari, Robin and Somma, Rolando D.},
   year={2017},
   month=jan, pages={1920–1950} }

@article{Clader_Linear_algebra,
   title={Preconditioned Quantum Linear System Algorithm},
   volume={110},
   ISSN={1079-7114},
   url={http://dx.doi.org/10.1103/PhysRevLett.110.250504},
   DOI={10.1103/physrevlett.110.250504},
   number={25},
   journal={Physical Review Letters},
   publisher={American Physical Society (APS)},
   author={Clader, B. D. and Jacobs, B. C. and Sprouse, C. R.},
   year={2013},
   month=jun }

@misc{Giovannetti_Linear_algebra,
      title={A quantum algorithm for estimating the determinant}, 
      author={Vittorio Giovannetti and Seth Lloyd and Lorenzo Maccone},
      year={2025},
      eprint={2504.11049},
      archivePrefix={arXiv},
      primaryClass={quant-ph},
      url={https://arxiv.org/abs/2504.11049}, 
}

@article{Rebentrost_machine_learning,
   title={Quantum Support Vector Machine for Big Data Classification},
   volume={113},
   ISSN={1079-7114},
   url={http://dx.doi.org/10.1103/PhysRevLett.113.130503},
   DOI={10.1103/physrevlett.113.130503},
   number={13},
   journal={Physical Review Letters},
   publisher={American Physical Society (APS)},
   author={Rebentrost, Patrick and Mohseni, Masoud and Lloyd, Seth},
   year={2014},
   month=sep }

@article{Lloyd_machine_learning,
   title={Quantum principal component analysis},
   volume={10},
   ISSN={1745-2481},
   url={http://dx.doi.org/10.1038/nphys3029},
   DOI={10.1038/nphys3029},
   number={9},
   journal={Nature Physics},
   publisher={Springer Science and Business Media LLC},
   author={Lloyd, Seth and Mohseni, Masoud and Rebentrost, Patrick},
   year={2014},
   month=jul, pages={631–633} }

@article{Poggiali_machine_learning,
   title={Quantum clustering with k-Means: A hybrid approach},
   volume={992},
   ISSN={0304-3975},
   url={http://dx.doi.org/10.1016/j.tcs.2024.114466},
   DOI={10.1016/j.tcs.2024.114466},
   journal={Theoretical Computer Science},
   publisher={Elsevier BV},
   author={Poggiali, Alessandro and Berti, Alessandro and Bernasconi, Anna and Del Corso, Gianna M. and Guidotti, Riccardo},
   year={2024},
   month=apr, pages={114466} }

@misc{Lloyd_machine_learning_2,
      title={Quantum algorithms for supervised and unsupervised machine learning}, 
      author={Seth Lloyd and Masoud Mohseni and Patrick Rebentrost},
      year={2013},
      eprint={1307.0411},
      archivePrefix={arXiv},
      primaryClass={quant-ph},
      url={https://arxiv.org/abs/1307.0411}, 
}

@article{Zhao_Machine_learning,
   title={Quantum-assisted Gaussian process regression},
   volume={99},
   ISSN={2469-9934},
   url={http://dx.doi.org/10.1103/PhysRevA.99.052331},
   DOI={10.1103/physreva.99.052331},
   number={5},
   journal={Physical Review A},
   publisher={American Physical Society (APS)},
   author={Zhao, Zhikuan and Fitzsimons, Jack K. and Fitzsimons, Joseph F.},
   year={2019},
   month=may }

@inproceedings{Kerenidis_Machine_learning,
 author = {Kerenidis, Iordanis and Landman, Jonas and Luongo, Alessandro and Prakash, Anupam},
 booktitle = {Advances in Neural Information Processing Systems},
 editor = {H. Wallach and H. Larochelle and A. Beygelzimer and F. d\textquotesingle Alch\'{e}-Buc and E. Fox and R. Garnett},
 pages = {},
 publisher = {Curran Associates, Inc.},
 title = {q-means: A quantum algorithm for unsupervised machine learning},
 url = {https://proceedings.neurips.cc/paper_files/paper/2019/file/16026d60ff9b54410b3435b403afd226-Paper.pdf},
 volume = {32},
 year = {2019}
}

@article{Cao_Quantum_chemistry,
title = "Quantum chemistry in the age of quantum computing",
author = "Yudong Cao and Jonathan Romero and Olson, {Jonathan P.} and Matthias Degroote and Johnson, {Peter D.} and M{\'a}ria Kieferov{\'a} and Kivlichan, {Ian D.} and Tim Menke and Borja Peropadre and Sawaya, {Nicolas P.D.} and Sukin Sim and Libor Veis and Al{\'a}n Aspuru-Guzik",
year = "2019",
month = oct,
day = "9",
doi = "10.1021/acs.chemrev.8b00803",
language = "English",
volume = "119",
pages = "10856--10915",
journal = "Chemical Reviews",
issn = "0009-2665",
publisher = "American Chemical Society",
number = "19",
}

@article{Whitfield_Quantum_chemistry,
   title={Simulation of electronic structure Hamiltonians using quantum computers},
   volume={109},
   ISSN={1362-3028},
   url={http://dx.doi.org/10.1080/00268976.2011.552441},
   DOI={10.1080/00268976.2011.552441},
   number={5},
   journal={Molecular Physics},
   publisher={Informa UK Limited},
   author={Whitfield, James D. and Biamonte, Jacob and Aspuru-Guzik, Alán},
   year={2011},
   month=mar, pages={735–750} }

@article{Kassal_quantum_chemistry,
   title={Polynomial-time quantum algorithm for the simulation of chemical dynamics},
   volume={105},
   ISSN={1091-6490},
   url={http://dx.doi.org/10.1073/pnas.0808245105},
   DOI={10.1073/pnas.0808245105},
   number={48},
   journal={Proceedings of the National Academy of Sciences},
   publisher={Proceedings of the National Academy of Sciences},
   author={Kassal, Ivan and Jordan, Stephen P. and Love, Peter J. and Mohseni, Masoud and Aspuru-Guzik, Alán},
   year={2008},
}

@article{Babbush_Quantum_chemistry,
  title = {Encoding Electronic Spectra in Quantum Circuits with Linear T Complexity},
  author = {Babbush, Ryan and Gidney, Craig and Berry, Dominic W. and Wiebe, Nathan and McClean, Jarrod and Paler, Alexandru and Fowler, Austin and Neven, Hartmut},
  journal = {Phys. Rev. X},
  volume = {8},
  issue = {4},
  pages = {041015},
  numpages = {36},
  year = {2018},
  month = {Oct},
  publisher = {American Physical Society},
  doi = {10.1103/PhysRevX.8.041015},
  url = {https://link.aps.org/doi/10.1103/PhysRevX.8.041015}
}

@article{Kuperberg_cryptography,
author = {Kuperberg, Greg},
title = {A Subexponential-Time Quantum Algorithm for the Dihedral Hidden Subgroup Problem},
journal = {SIAM Journal on Computing},
volume = {35},
number = {1},
pages = {170-188},
year = {2005},
doi = {10.1137/S0097539703436345},
URL = {https://doi.org/10.1137/S0097539703436345},
eprint = {https://doi.org/10.1137/S0097539703436345}
}

@article{Giovannetti_Cryptography,
  title = {Quantum Private Queries},
  author = {Giovannetti, Vittorio and Lloyd, Seth and Maccone, Lorenzo},
  journal = {Physical Review Letters},
  volume = {100},
  issue = {23},
  pages = {230502},
  numpages = {4},
  year = {2008},
  month = {Jun},
  publisher = {American Physical Society (APS)},
  doi = {10.1103/PhysRevLett.100.230502},
  url = {https://link.aps.org/doi/10.1103/PhysRevLett.100.230502}
}

@article{Jiang_QRAM_implementation,
   title={Experimental realization of 105-qubit random access quantum memory},
   volume={5},
   ISSN={2056-6387},
   url={http://dx.doi.org/10.1038/s41534-019-0144-0},
   DOI={10.1038/s41534-019-0144-0},
   number={1},
   journal={npj Quantum Information},
   publisher={Springer Science and Business Media LLC},
   author={Jiang, N. and Pu, Y.-F. and Chang, W. and Li, C. and Zhang, S. and Duan, L.-M.},
   year={2019},
   month=apr }

@article{Wang_QRAM_design,
   title={Hardware‐Efficient Quantum Random Access Memory Design with a Native Gate Set on Superconducting Platforms},
   volume={8},
   ISSN={2511-9044},
   url={http://dx.doi.org/10.1002/qute.202400519},
   DOI={10.1002/qute.202400519},
   number={5},
   journal={Advanced Quantum Technologies},
   publisher={Wiley},
   author={Wang, Yun‐Jie and Zhang, Sheng and Sun, Tai‐Ping and Zhao, Ze‐An and Xu, Xiao‐Fan and Zhuang, Xi‐Ning and Liu, Huan‐Yu and Xue, Cheng and Duan, Peng and Wu, Yu‐Chun and Chen, Zhao‐Yun and Guo, Guo‐Ping},
   year={2025},
   month=jan }

@article{Jaques_QRAM_survey,
  doi = {10.22331/q-2025-12-02-1922},
  url = {https://doi.org/10.22331/q-2025-12-02-1922},
  title = {{QRAM}: {A} {S}urvey and {C}ritique},
  author = {Jaques, Samuel and Rattew, Arthur G.},
  journal = {{Quantum}},
  issn = {2521-327X},
  publisher = {{Verein zur F{\"{o}}rderung des Open Access Publizierens in den Quantenwissenschaften}},
  volume = {9},
  pages = {1922},
  month = dec,
  year = {2025}
}

@article{Chen_QRAM_implementation,
  title = {Scalable and High-Fidelity Quantum Random Access Memory in Spin-Photon Networks},
  author = {Chen, K. C. and Dai, W. and Errando-Herranz, C. and Lloyd, S. and Englund, D.},
  journal = {PRX Quantum},
  volume = {2},
  issue = {3},
  pages = {030319},
  numpages = {19},
  year = {2021},
  month = {Aug},
  publisher = {American Physical Society},
  doi = {10.1103/PRXQuantum.2.030319},
  url = {https://link.aps.org/doi/10.1103/PRXQuantum.2.030319}
}

@article{Mukhopadhyay_QRAM_design,
   title={A quantum random access memory (QRAM) using a polynomial encoding of binary strings},
   volume={15},
   ISSN={2045-2322},
   url={http://dx.doi.org/10.1038/s41598-025-95283-5},
   DOI={10.1038/s41598-025-95283-5},
   number={1},
   journal={Scientific Reports},
   publisher={Springer Science and Business Media LLC},
   author={Mukhopadhyay, Priyanka},
   year={2025},
   month=mar }

@inproceedings{Xu_QRAM_critique, series={MICRO ’23},
   title={Systems Architecture for Quantum Random Access Memory},
   url={http://dx.doi.org/10.1145/3613424.3614270},
   DOI={10.1145/3613424.3614270},
   booktitle={56th Annual IEEE/ACM International Symposium on Microarchitecture},
   publisher={ACM},
   author={Xu, Shifan and Hann, Connor T. and Foxman, Ben and Girvin, Steven M. and Ding, Yongshan},
   year={2023},
   month=oct, pages={526–538},
   collection={MICRO ’23} }

@article{Weiss_qRAM,
  title = {Quantum Random Access Memory Architectures Using 3D Superconducting Cavities},
  author = {Weiss, D.K. and Puri, Shruti and Girvin, S.M.},
  journal = {PRX Quantum},
  volume = {5},
  issue = {2},
  pages = {020312},
  numpages = {23},
  year = {2024},
  month = {Apr},
  publisher = {American Physical Society},
  doi = {10.1103/PRXQuantum.5.020312},
  url = {https://link.aps.org/doi/10.1103/PRXQuantum.5.020312}
}

@article{Park_qRAM,
   title={Circuit-Based Quantum Random Access Memory for Classical Data},
   volume={9},
   ISSN={2045-2322},
   url={http://dx.doi.org/10.1038/s41598-019-40439-3},
   DOI={10.1038/s41598-019-40439-3},
   number={1},
   journal={Scientific Reports},
   publisher={Springer Science and Business Media LLC},
   author={Park, Daniel K. and Petruccione, Francesco and Rhee, June-Koo Kevin},
   year={2019},
   month=mar }

@misc{Cesa_qRAM,
      title={Fast and Error-Correctable Quantum RAM}, 
      author={Francesco Cesa and Hannes Bernien and Hannes Pichler},
      year={2025},
      eprint={2503.19172},
      archivePrefix={arXiv},
      primaryClass={quant-ph},
      url={https://arxiv.org/abs/2503.19172}, 
}

@article{Hann_noise_resilience,
   title={Resilience of Quantum Random Access Memory to Generic Noise},
   volume={2},
   ISSN={2691-3399},
   url={http://dx.doi.org/10.1103/PRXQuantum.2.020311},
   DOI={10.1103/prxquantum.2.020311},
   number={2},
   journal={PRX Quantum},
   publisher={American Physical Society (APS)},
   author={Hann, Connor T. and Lee, Gideon and Girvin, S.M. and Jiang, Liang},
   year={2021},
   month=apr }



\widetext
\appendix
\renewcommand{\theequation}{\Alph{section}\arabic{equation}}

\section{Introduction}

In this work we present several possible qRAM implementations having different required resources, e.g. more/less walkers, short/long range interactions, bosonic/fermionic walkers and using single/double binary trees. Nonetheless, there is a unique main idea behind all these implementations. We have a collection of quantum walkers ordered in line where the first ones contain the information of the first bits of the address register $A=\{A_1,A_2,\dots,A_n\}$. Hence, the first (sometimes depicted as the rightmost) particle contains the information of $A_1$, then it follows $A_2$, $A_3$, up to $A_n$. Then the data register $D=\{D_0,D_1,\dots,D_m\}$ particles follow, where the first walker represents $D_0$, then it follow $D_2$, $D_3$, up to the last data particle $D_m$. We remind that, given the encoding of Eq.~(2), some address or data subsystem may be represented by a no-walker state. 

The particles travel along a binary tree, where, before each bifurcation, a gate $\hat U^{(i)}$ is applied. In particular, at the $i$-th bifurcation $\hat U^{(i)}$ effectively transfers the information contained in $A_i$, which may be represented by a red walker $\ket{R}_{A_i}$ or by no particle $\ket{\emptyset}_{A_i}$, to all the following ones, namely $A_{i+1},\dots, A_n, D_0,\dots, D_m$. The gate  $\hat U^{(i)}$ effectively coincides with a controlled operation that either changes or not the color of the following walkers depending on the state of $A_i$. If it is in the state $\ket{R}_{A_i}$, the gate $\hat U^{(i)}$ changes state of the following walkers from red to blue (if present). Otherwise, if $A_i$ in not present, the controlled operation is not activated and no change of color occurs. 

Immediately after the application of  $\hat U^{(i)}$, we apply the scattering gate $\hat S$, which routes to the left/right the red/blue-colored walkers, respectively (in this Supplemental Material red walkers go down and blue walkers go up). Moreover, this gate turns back to red the blue walkers after they have been routed. As a consequence, the quantum walkers exploited in our algorithm can be blue-colored solely when traveling between $\hat U^{(i)}$ and $\hat S$. During all the other phases, our quantum walkers are either red or not present.

The main target of this Supplemental Material is to describe more in detail the technicalities of the proposed qRAM schemes. We separately describe the standard and the backup variants, where the former makes use of long-range interactions with fewer quantum walkers, namely $ n+m+1$, while the latter adopts short-range gates only and exploits $ 2(n+m)-1$ quantum walkers. The main differences between the two approaches rely on: (i) the realization of the gate $\hat U^{(i)}$, and (ii) the process to copy the information from the target memory cell to the data register $D$.

We start by describing the standard approach (Section~\ref{SecStandard}). Later, we discuss the main features of the backup variant (Section~\ref{SecBackup}) and finally we propose some possible physical systems that can be adopted to implement our protocol (Section~\ref{SecAlternative}).

\subsection{Ket notation}
We represent the systems $A$ and $D$ through the following notation. Each quantum walker has an internal degree of freedom $c$ called color that can either be red $c=R$ or blue $c=B$. Moreover, its position along the three is represented by the parameters $d$ and $l$, where $d=1,\dots,n+1$ coincides with the depth and $l=1,\dots,2^{d-1}$ with branches present at the level $d$. 
Hence, the color $c$ and the position $(d,l)$ of a generic walker $W$ is represented by the ket  $\ket{c^{d,l}}_W$.
For instance, if the last data walker $D_m$ is in the red state at the position $(d,l)=(3,1)$, we report its state as $\ket{R^{3,1}}_{D_m}$. Instead, if the first address walker is not present and all the other walkers are at the level $d=2$, we report the state of $A_1$ as $\ket{\emptyset^{2}}_{A_1}$. 

During the inverse routing, namely after that the particles crossed the memory cells, we keep using the same $(d,l)$ notation, where now the value of the depth is reported with an apex in order to distinguish this position from the one reached during the routing from the initial input port to the memory cells (see Fig.~\ref{rout_qram_cl}).

\section{Technical description of the qRAM}\label{SecStandard}

In this section we describe the functioning of our qRAM scheme, which is divided into the routing phase, namely the phase where the data walkers of qRAM are routed from the input node to the memory cells, the writing of the information contained into the memory cells into the data register and the inverse routing, where the data particles are routed to the exit port of the qRAM. 
In Fig.~\ref{rout_qram_cl} we show these phases in case of a classic address of $n=2$ subsystems and a data register made of $m=1$ quantum walkers.

The technical ingredients of the routing phase, namely the description of the gates $\hat U^{(i)}$ and the scattering gate $\hat S$, are explained in Section~\ref{SecRout}. Similarly, we introduce the tools necessary for the writing of the message contained in the memory cells in Section~\ref{SecMessCop} and the inverse routing in Section~\ref{SecInvRout}. Finally, in Section~\ref{example} we describe step-by-step the whole qRAM algorithm in case of a classical address, namely the one represented in Fig.~\ref{rout_qram_cl}.

\subsection{Routing}\label{SecRout}

In this section we describe the routing phase of our qRAM scheme, namely how the quantum walkers are routed from the root input node to the memory cells. This is achieved thanks to three main ingredients: (i) how the address and data walkers are initialized (see the encoding procedure described in the main text of this work), (ii) how the address particles transfer their information to the other particles through the gate $\hat U^{(i)}$ (see Section~\ref{Ud}) and (iii) how particles are physically routed along the different branches of the binary tree through the scattering gate $\hat S$ (see Section~\ref{scattS}). 

\subsubsection{Information transmission: $\hat U^{(d)}$}\label{Ud}

The gate $\hat{U}^{(i)}$ is used to propagate the information carried by the $i$-th address particle to all the walkers following it. 
More precisely, given a sequence of information carriers, the gate is a multi-target controlled-NOT operation taking $i$-th address particle as control and all the following ones as targets, where the gate is activated if $A_i$ is in the red state.
Recall that at level $d$ of the tree we have $m+1$ data particles $D$ and $n-d$ address particles $A=\{A_{d+1},\dots,A_n\}$ following $A_d$. Thus, the gate 
$\hat{U}^{(d)}$ uses the 
$d$-th address walker as a control: if that particle is in the state $\ket{R}_{A_{d}}$ it applies Pauli
$\hat{X}$ to all of the subsequent data and address walkers, otherwise it does nothing

\begin{equation*}
    \hat{U}^{(d)}=\hat{X}_{D_m,D_1}\otimes \hat{X}_{A_{n},{A_{d+1}}} \otimes \ket{R^{d,l}}_{A_{d}}\bra{R^{d,l}}\otimes \hat{I}_{A_{d-1}, A_1} +\sum_{i\in\{\emptyset,B\}} \hat{I}_{D_m,D_1}\otimes \hat{I}_{A_{n},A_{d+1}}\otimes \ket{i^{d,l}}_{A_{d}}\bra{i^{d,l}}\otimes \hat{I}_{A_{d-1},A_1},
\end{equation*}
where $\hat{O}_{i,j}\equiv \hat{O}_i\otimes \hat{O}_{i+1}\otimes\dots\otimes \hat{O}_j$ means the tensor product of operator $\hat{O}$ for all indexes between $i$ and $j$. Notice that the same gate $\hat U^{(d)}$ is adopted for each branch $l=1,\dots,2^{d-1}$ at the depth $d$. 
Thus, given a state of the form $\ket{\delta_m^{d,l}}_{D_m}\dots\ket{R^{d,l}}_{A_d}\dots\ket{a_1^{d,l}}_{A_1}$, the action of $\hat U^{(d)}$ on it is (we omit the position $(d,l)$ as it is the same for all the walkers): 
\begin{align*}
\hat{U}^{(d)}\ket{\delta_m}_{D_m}\dots\ket{R}_{A_d}\dots\ket{a_1}_{A_1} =\left(\bigotimes_{j=0}^{m+1}\hat{X}_{D_j}\ket{\delta_j}_{D_j}\right) \otimes \left(\bigotimes_{i=d+1}^{n}\hat{X}_{A_i}\ket{a_i}_{A_i}\right)\otimes \ket{R}_{A_{d}}\otimes \dots\otimes\ket{a_1}_{A_1}
\end{align*}
where $\delta_j\in\{R,B,\emptyset\}$ represents the color/absence of the $j$-th data walker and $\hat{X}\ket{R}=\ket{B},\; \hat{X}\ket{B}=\ket{R},\; \hat{X}\ket{\emptyset}=\ket{\emptyset}$.
For instance, consider a qRAM with an address register made of $n=3$ subsystems, and a data register made of $m=2$ (red) particles, where the initial address is $\boldsymbol{a}=a_1a_2a_3=110$. Therefore, the initial state would be $\ket{\psi_{in}}=\ket{R^{1,1}}_{D_1} \ket{R^{1,1}}_{D_2} \ket{\emptyset^1}_{A_3}\ket{R^{1,1}}_{A_2}\ket{R^{1,1}}_{A_1}$ and the action of $\hat U^{(1)}$ on $\ket{\psi_{in}}$ would be:
\begin{eqnarray} \nonumber
\hat U^{(1)} \ket{\psi_{in}} = \hat U^{(1)}\ket{\red^{1,1}}_{D_1} \ket{\red^{1,1}}_{D_2} \ket{\emptyset^1}_{A_3}\ket{\red^{1,1}}_{A_2}\ket{\red^{1,1}}_{A_1}&& \\ \label{explU1}
= \ket{\blue^{1,1}}_{D_1} \ket{\blue^{1,1}}_{D_2} \ket{\emptyset^1}_{A_3}\ket{\blue^{1,1}}_{A_2}\ket{\red^{1,1}}_{A_1}&&.
\end{eqnarray}

\begin{figure}[t]
\begin{tikzpicture}[scale=0.7, transform shape,
  level 1/.style={level distance=4cm, sibling distance=8cm},
  level 2/.style={level distance=4cm, sibling distance=4cm},
  grow=east, 
  edge from parent/.style={draw, thick}, 
  every node/.style={rectangle, draw, minimum size=6mm,fill=cyan!30}, 
  leaf/.style={rectangle, draw, fill=gray!30, minimum size=6mm}, 
  block/.style={rectangle, draw, fill=blue!20, minimum width=0.7cm, minimum height=5mm, rotate around={0:(0,0)}}, 
  gateX/.style={rectangle, draw, fill=green!20, minimum size=5mm, rotate around={0:(0,0)}} 
]

\node (root1) {$\hat{S}$}
  child {
    node {$\hat{S}$}
    child { 
      node[leaf] {$\boldsymbol{x}_3$} 
      edge from parent {}
    }
    child { 
      node[leaf] {$\boldsymbol{x}_2$} 
      edge from parent{
      }
    }
    edge from parent {
      node[midway,xshift=0.4cm, yshift=-0.4cm,  block, rotate=-45] {$\hat{U}^{(2)}$} 
    }
  }
  child {
    node {$\hat{S}$}
    child { 
      node[leaf] {$\boldsymbol{x}_1$} 
      edge from parent {}
    }
    child { 
      node[leaf] {$\boldsymbol{x}_0$} 
      edge from parent{ 
      }
    }
    edge from parent {
      node[midway, xshift=0.4cm,yshift=0.4cm, block, rotate=45] {$\hat{U}^{(2)}$} 
    }
  };
\node (root2) at (16,0) {$\hat{S}^{\dagger}$}
[grow=left]
  child {
    node {$\hat{S}^{\dagger}$}
    child { 
      node[leaf] {$\boldsymbol{x}_{11}$} 
      edge from parent {
      }
    }
    child { 
      node[leaf] {$\boldsymbol{x}_{10}$} 
      edge from parent {}
    }
    edge from parent {
      node(Ugate)[midway,xshift=-0.4cm,yshift=0.4cm, block, rotate=-45] {$\hat{U}^{(2)}$} 
    }
  }
  child {
    node {$\hat{S}^{\dagger}$}
    child { 
      node[leaf] {$\boldsymbol{x}_{01}$} 
      edge from parent {
      }
    }
    child { 
      node[leaf] {$\boldsymbol{x}_{00}$} 
      edge from parent {}
    }
    edge from parent {
      node[midway, xshift=-0.4cm, yshift=-0.4cm, block, rotate=45] {$\hat{U}^{(2)}$} 
    }
  };

\draw[thick] (-4, 0) -- (root1) node(Ugateroot)[midway, block, rotate=0] {$\hat{U}^{(1)}$};
\draw[thick] (root2) -- (20,0) node[midway, block] {$\hat{U}^{(1)}$};


\draw[thick, fill=red] (-3.6,0) circle (3pt);
\draw[thick, fill=white] (-3.3,0) circle (3pt);
\halfcoloredcircle{-3.3}{0}{white}{white}
\draw[thick, fill=red] (-3,0) circle (3pt);
\halfcoloredcircle{-3}{0}{red}{red}

\draw[thick, fill=blue] (-1.4,0) circle (3pt);
\halfcoloredcircle{-1.4}{0}{blue}{blue}
\draw[thick,fill=white] (-1.1,0) circle (3pt);
\halfcoloredcircle{-1.1}{0}{white}{white}
\draw[thick, fill=red] (-0.8,0) circle (3pt);
\halfcoloredcircle{-0.8}{0}{red}{red}


\draw[thick, fill=white] (1.7,1.7) circle (3pt);
\draw[thick,fill=white] (1.5,1.5) circle (3pt);
\draw[thick,fill=red] (1.3,1.3) circle (3pt);

\draw[thick,fill=white] (3.4,3.4) circle (3pt);
\draw[thick,fill=white] (3.2,3.2) circle (3pt);
\draw[thick,fill=red] (3,3) circle (3pt);


\draw[thick, fill=red] (1.7,-1.7) circle (3pt);
\draw[thick,fill=white] (1.5,-1.5) circle (3pt);
\draw[thick,fill=white] (1.3,-1.3) circle (3pt);

\draw[thick,fill=red] (3.4,-3.4) circle (3pt);
\draw[thick,fill=white] (3.2,-3.2) circle (3pt);
\draw[thick,fill=white] (3,-3) circle (3pt);




\draw[thick, fill=red] (6,3) circle (3pt);
\draw[thick, fill=white] (6.25,2.875) circle (3pt);
\draw[thick,fill=white] (6.5,2.745) circle (3pt);




\draw[thick,fill=white] (6,-5) circle (3pt);
\draw[thick,fill=white] (6.25,-5.125) circle (3pt);
\draw[thick, fill=red] (6.5,-5.25) circle (3pt);




\draw[thick, fill=red] (9.5,2.745) circle (3pt);
\draw[thick, fill=white] (9.75,2.875) circle (3pt);
\draw[thick,fill=white] (10,3) circle (3pt);




\draw[thick,fill=white] (9.5,-5.25) circle (3pt);
\draw[thick,fill=white] (9.75,-5.125) circle (3pt);
\draw[thick, fill=red] (10,-5) circle (3pt);


\draw[thick, fill=red] (14.3,1.7) circle (3pt);
\draw[thick,fill=white] (14.5,1.5) circle (3pt);
\draw[thick,fill=white] (14.7,1.3) circle (3pt);

\draw[thick, fill=red] (12.6,3.4) circle (3pt);
\draw[thick,fill=white] (12.8,3.2) circle (3pt);
\draw[thick,fill=white] (13,3) circle (3pt);


\draw[thick,fill=white] (14.3,-1.7) circle (3pt);
\draw[thick,fill=white] (14.5,-1.5) circle (3pt);
\draw[thick, fill=red] (14.7,-1.3) circle (3pt);

\draw[thick,fill=white] (12.6,-3.4) circle (3pt);
\draw[thick,fill=white] (12.8,-3.2) circle (3pt);
\draw[thick, fill=red] (13,-3) circle (3pt);


\draw[thick, fill=red] (19.6,0) circle (3pt);
\halfcoloredcircle{19.6}{0}{red}{red}
\draw[thick,fill=white] (19.3,0) circle (3pt);
\halfcoloredcircle{19.3}{0}{white}{white}
\draw[thick, fill=red] (19,0) circle (3pt);

\draw[thick, fill=red] (17.4,0) circle (3pt);
\halfcoloredcircle{17.4}{0}{red}{red}
\draw[thick,fill=white] (17.1,0) circle (3pt);
\halfcoloredcircle{17.1}{0}{white}{white}
\draw[thick, fill=blue] (16.8,0) circle (3pt);
\halfcoloredcircle{16.8}{0}{blue}{blue}


\draw[dashed] (-2.7,6.5)--(-2.7,-6.5) node[fill=none,draw=none,left,anchor=east]
{};
\draw[dashed] (-1.65,6.5)--(-1.65,-6.5) node[fill=none,draw=none,left,anchor=east]
{};
\fill[gray!20, opacity=0.5] (-2.7,6.5) -- (-1.65,6.5)  -- (-1.65,-6.5) -- (-2.7,-6.5)  -- cycle;
\node[draw=none, fill=none] at (-2.1,-6.7){ Step 1};

\draw[->,-Stealth] (-1,7)-- (-1,6){};

\node[fill=none,draw=none, fill=none] at (-1,7.5){\Large$\ket{\psi_1}$};

\draw[dashed] (-0.5,6.5)--(-0.5,-6.5) node[fill=none,draw=none,left,anchor=east]
{};
\draw[dashed] (1.15,6.5)--(1.15,-6.5) node[fill=none,draw=none,left,anchor=east]
{};
\fill[gray!20, opacity=0.5] (-0.5,6.5)  --  (1.15,6.5)--(1.15,-6.5) -- (-0.5,-6.5) -- cycle;
\node[fill=none,draw=none] at (0.4,-6.7){Step 2};

\draw[->,-Stealth] (1.5,7)-- (1.5,6){};

\node[fill=none,draw=none] at (1.5,7.5){\Large$\ket{\psi_2}$};

\draw[dashed] (1.9,6.5)--(1.9,-6.5) node[fill=none,draw=none,left,anchor=east]{};
\draw[dashed] (2.85,6.5)--(2.85,-6.5) node[fill=none,draw=none,left,anchor= east]{};
\fill[gray!20, opacity=0.5] (1.9,6.5)  --  (2.85,6.5)--(2.85,-6.5) -- (1.9,-6.5) -- cycle;
\node[fill=none,draw=none] at (2.4,-6.7){Step 3};

\draw[->,-Stealth] (3.2,7)-- (3.2,6){};

\node[fill=none,draw=none] at (3.2,7.5){\Large$\ket{\psi_3}$};

\draw[dashed] (3.6,6.5)--(3.6,-6.5) node[fill=none,draw=none,left,anchor=east]
{};
\draw[dashed] (5.7,6.5)--(5.7,-6.5) node[fill=none,draw=none,left,anchor=east]
{};
\fill[gray!20, opacity=0.5] (3.6,6.5)  --  (5.7,6.5)--(5.7,-6.5) -- (3.6,-6.5) -- cycle;
\node[fill=none,draw=none] at (4.7,-6.7){Step 4};

\draw[->,-Stealth] (6.2,7)-- (6.2,6){};

\node[fill=none,draw=none] at (6.2,7.5){\Large$\ket{\psi_4}$};

\draw[dashed,] (6.8,6.5)--(6.8,-6.5) node[fill=none,draw=none,left,anchor=east]
{};
\draw[dashed,] (9.2,6.5)--(9.2,-6.5) node[fill=none,draw=none,left,anchor=east]
{};
\fill[gray!20, opacity=0.5] (6.8,6.5)  --  (9.2,6.5)--(9.2,-6.5) -- (6.8,-6.5) -- cycle;
\node[fill=none,draw=none] at (8,-6.7){Querying};

\draw[->,-Stealth] (9.8,7)-- (9.8,6){};

\node[fill=none,draw=none] at (9.8,7.5){\Large$\ket{\psi_Q}$};

\draw[dashed] (10.3,6.5)--(10.3,-6.5) node[fill=none,draw=none,left,anchor=east]
{};
\draw[dashed] (12.4,6.5)--(12.4,-6.5) node[fill=none,draw=none,left,anchor=east]
{};
\fill[gray!20, opacity=0.5] (10.3,6.5)  --  (12.4,6.5)--(12.4,-6.5) -- (10.3,-6.5) -- cycle;
\node[fill=none,draw=none] at (11.4,-6.7){Step 5};

\draw[->,-Stealth] (12.7,7)-- (12.7,6){};

\node[fill=none,draw=none] at (12.7,7.5){\Large$\ket{\psi_5}$};

\draw[dashed] (13.15,6.5)--(13.15,-6.5) node[fill=none,draw=none,left,anchor=east]{};
\draw[dashed] (14.1,6.5)--(14.1,-6.5) node[fill=none,draw=none,left,anchor=east]{};
\fill[gray!20, opacity=0.5] (13.15,6.5)  --  (14.1,6.5)--(14.1,-6.5) -- (13.15,-6.5) -- cycle;

\node[fill=none,draw=none] at (13.7,-6.7){Step 6};

\draw[->,-Stealth] (14.5,7)-- (14.5,6){};

\node[fill=none,draw=none] at (14.5,7.5){\Large$\ket{\psi_6}$};

\draw[dashed] (14.85,6.5)--(14.85,-6.5) node[fill=none,draw=none,left,anchor=east]
{};
\draw[dashed] (16.5,6.5)--(16.5,-6.5) node[fill=none,draw=none,left,anchor=east]
{};
\fill[gray!20, opacity=0.5] (14.85,6.5)  --  (16.5,6.5)--(16.5,-6.5) -- (14.85,-6.5) -- cycle;
\node[fill=none,draw=none] at (15.7,-6.7){Step 7};

\draw[->,-Stealth] (17,7)-- (17,6){};

\node[fill=none,draw=none] at (17,7.5){\Large$\ket{\psi_7}$};

\draw[dashed] (17.65,6.5)--(17.65,-6.5) node[fill=none,draw=none,left,anchor=east]
{};
\draw[dashed] (18.7,6.5)--(18.7,-6.5) node[fill=none,draw=none,left,anchor=east]
{};
\fill[gray!20, opacity=0.5] (17.65,6.5)  --  (18.7,6.5)--(18.7,-6.5) -- (17.65,-6.5) -- cycle;
\node[fill=none,draw=none] at (18.2,-6.7){Step 8};


\node[draw=none, fill=none] at (-3,-0.8){\scalebox{1.2}{$(d,l)=(1,1)$}};

\node[draw=none, fill=none, rotate=-45] at (2,-1){\scalebox{1.2}{$(2,1)$}};

\node[draw=none, fill=none, rotate=45] at (2,1){\scalebox{1.2}{$(2,2)$}};

\node[draw=none, fill=none,rotate=-22.5] at (5,-5){\scalebox{1.2}{$(3,1)$}};

\node[draw=none, fill=none, rotate=22.5] at (5,-2.9){\scalebox{1.2}{$(3,2)$}};

\node[draw=none, fill=none, rotate=-22.5] at (5,2.9){\scalebox{1.2}{$(3,3)$}};

\node[draw=none, fill=none, rotate=22.5] at (5,5){\scalebox{1.2}{$(3,4)$}};


\node[draw=none, fill=none,rotate=22.5] at (11,-5){\scalebox{1.2}{$(3',1)$}};

\node[draw=none, fill=none, rotate=-22.5] at (11,-2.9){\scalebox{1.2}{$(3',2)$}};

\node[draw=none, fill=none, rotate=22.5] at (11,2.9){\scalebox{1.2}{$(3',3)$}};

\node[draw=none, fill=none, rotate=-22.5] at (11,5){\scalebox{1.2}{$(3',4)$}};

\node[draw=none, fill=none, rotate=45] at (14,-1){\scalebox{1.2}{$(2',1)$}};

\node[draw=none, fill=none, rotate=-45] at (14,1){\scalebox{1.2}{$(2',2)$}};

\node[draw=none, fill=none] at (19,-0.8){\scalebox{1.2}{$(1',1)$}};


\node (insetRB) [draw=black, fill=white, thick, rectangle, minimum width=40mm,minimum height=40mm, rotate=90, transform shape] at (-1,-4){\begin{tikzpicture}[line join=round]

  \def\stemthick{5pt}
  \def\branchthick{6pt}
  \def\branchlen{1.4}
  \def\bendangle{20}

  \coordinate (O) at (0,0);
  \coordinate (B) at (0,-0.8); 
  \coordinate (L) at ($ (B) + (-\branchlen,-\branchlen) $);
  \coordinate (R) at ($ (B) + (\branchlen,-\branchlen) $);

  \draw[red,line width=\stemthick] (O) -- (B);

  \draw[red,line width=\branchthick,-{Latex[length=5pt, width=7pt]}]
    (B) to[bend left=\bendangle] (L);

  \path[record path={name=ramo,dist=6pt}] (B) to[bend right=\bendangle] (R);

  \draw[opacity=0, line width=0pt] (B) to[bend right=\bendangle] (R); 
  \draw[line width=\branchthick, -{Latex[length=8pt,width=8pt]}, draw=none] (B) to[bend right=\bendangle] (R);
  \path[draw=red, postaction={draw=red, -{Latex[length=5pt,width=7pt]}, line width=\branchthick}]
    (B) to[bend right=\bendangle] (R);

  \addtocounter{parrow}{-2}
  \draw let \p1=($(R)-(B)$),\n1={atan2(\y1,\x1)} in
    [left color=blue, right color=red, shading angle=\n1+90, draw=none]
    [reconstruct top=ramo] -- (parrowm-ramo-\the\numexpr\value{parrow}+2)
    -- (parrowb-ramo-\number\value{parrow})
    [reconstruct bottom=ramo] -- (parrowm-ramo-3)
    -- (parrowt-ramo-1);

\node[rectangle, draw=none, fill=red, inner sep=0pt, outer sep=0pt, minimum width=5pt, minimum height=12mm
] at (-0.08,-0.45){}; 

\node[rectangle, draw=none, fill=blue, inner sep=0pt, outer sep=0pt, minimum width=5pt, minimum height=12mm
] at (0.08,-0.45){}; 

\node[rectangle, fill=cyan!30, inner sep=0pt, outer sep=0pt, minimum width=20pt, minimum height=20pt, rotate=-90
] at (0,-1){$\hat{S}$};
\end{tikzpicture} 
};


\node (insetU) [draw=black, fill=white, thick, rectangle, minimum width=40mm,minimum height=40mm, rotate=90, transform shape] at (-1,3.7){
\begin{tikzpicture}[gateX/.style={rectangle, draw, fill=green!20, minimum size=5mm}, scale=0.7, transform shape]
\node [draw, rounded corners, rectangle, fill=blue!20, inner sep=5pt, minimum width=15mm, minimum height=30mm] 
        at (0,0) {};
\def\n{5}
\def\spacing{1} 
\draw[thick] (0,-2.5) -- (0,{(\spacing*(\n-1))-1.5 });
\foreach \i/\col in {0/white, 1/red, 2/white, 3/red, 4/white} {
\edef\colorname{\col} 
\filldraw[black, fill=\colorname] (0, \i*\spacing-2) circle (2pt);}

\draw[very thick] (0,-1)--(0,1);

\node[midway, rotate=-90, xshift=0cm,yshift=0cm,gateX] {$\hat{X}$};

\node[midway, rotate=-90,xshift=-1cm,yshift=0cm,gateX] {$\hat{X}$};

\fill (0,-1) circle (3pt);

\draw[->,-Stealth, thick] (1.5,1.5)--(1.5,-1.5);

\end{tikzpicture}
};

\node (insetU2) [draw=black, fill=white, thick, rectangle, minimum width=40mm,minimum height=40mm, rotate=90, transform shape] at (17,3.7){
\begin{tikzpicture}[gateX/.style={rectangle, draw, fill=green!20, minimum size=5mm}, scale=0.7, transform shape]
\node [draw, rounded corners, rectangle, fill=blue!20, inner sep=5pt, minimum width=15mm, minimum height=20mm] 
        at (0,-0.5) {};
\def\n{4}
\def\spacing{1} 
\draw[thick] (0,-2.5) -- (0,{(\spacing*(\n-1))-1.5 });
\foreach \i/\col in {0/white, 1/red, 2/white, 3/white} {
\edef\colorname{\col} 
\filldraw[black, fill=\colorname] (0, \i*\spacing-2) circle (2pt);}

\draw[very thick] (0,-1)--(0,0);

\node[midway, rotate=-90,xshift=0cm,yshift=0cm,gateX] {$\hat{X}$};

\fill (0,-1) circle (3pt);

\draw[->, -Stealth, thick] (1.5,1)--(1.5,-2);

\end{tikzpicture}
};

\node (insetRB1) [draw=black, fill=white, thick, rectangle, minimum width=40mm,minimum height=40mm, rotate=90, transform shape, yscale=-1] at (17,-4){\begin{tikzpicture}[line join=round]

  \def\stemthick{5pt}
  \def\branchthick{6pt}
  \def\branchlen{1.4}
  \def\bendangle{20}

  \coordinate (O) at (0,0);
  \coordinate (B) at (0,-0.8); 
  \coordinate (L) at ($ (B) + (-\branchlen,-\branchlen) $);
  \coordinate (R) at ($ (B) + (\branchlen,-\branchlen) $);

  \draw[red,line width=\stemthick] (O) -- (B);

  \draw[red,line width=\branchthick,{Latex[length=5pt, width=7pt]}-]
    (B) to[bend left=\bendangle] (L);

  \path[record path={name=ramo,dist=6pt}] (B) to[bend right=\bendangle] (R);

  \draw[opacity=0, line width=0pt] (B) to[bend right=\bendangle] (R); 
  \draw[line width=\branchthick,{Latex[length=8pt,width=8pt]}-, draw=none] (B) to[bend right=\bendangle] (R);
  \path[draw=red, postaction={draw=red,{Latex[length=5pt,width=7pt]}-, line width=\branchthick}]
    (B) to[bend right=\bendangle] (R);

  \addtocounter{parrow}{-2}
  \draw let \p1=($(R)-(B)$),\n1={atan2(\y1,\x1)} in
    [left color=blue, right color=red, shading angle=\n1+90, draw=none]
    [reconstruct top=ramo] -- (parrowm-ramo-\the\numexpr\value{parrow}+2)
    -- (parrowb-ramo-\number\value{parrow})
    [reconstruct bottom=ramo] -- (parrowm-ramo-3)
    -- (parrowt-ramo-1);

\node[rectangle, draw=none, fill=red, inner sep=0pt, outer sep=0pt, minimum width=5pt, minimum height=12mm
] at (-0.08,-0.45){}; 

\node[rectangle, draw=none, fill=blue, inner sep=0pt, outer sep=0pt, minimum width=5pt, minimum height=12mm
] at (0.08,-0.45){}; 

\node[rectangle, fill=cyan!30, inner sep=0pt, outer sep=0pt, minimum width=20pt, minimum height=20pt, rotate=-90,
xscale=-1, yscale=1] at (0,-1){$\hat{S}^{\dagger}$};

\begin{scope}[shift={(0,0.45)}, scale=1.7]
  \fill[red]  (-0.12,-0.2) -- (0,0) -- (0,-0.2) -- cycle;
  \fill[blue] (0.12,-0.2) -- (0,0) -- (0,-0.2) -- cycle;
\end{scope}

\end{tikzpicture} 
};

\scalebox{0.8}{\node[fill=none,draw=none, align=center] at (-3,-4){Input\\$(d,l)$};}

\scalebox{0.8}{\node[fill=none,draw=none, align=center] at (0,-7.1){Output\\$(d+1,2l-1)$};}

\scalebox{0.8}{\node[fill=none,draw=none, align=center] at (0,-2.9){Output \\$(d+1,2l)$};}


\scalebox{0.8}{
\node[fill=none,draw=none, align=center] at (22.8,-4){Output\\ $(d-1,l/2)$};}

\scalebox{0.8}{\node[fill=none,draw=none, align=center] at (20,-7.1){Input\\$(d,l-1)$};}

\scalebox{0.8}{\node[fill=none,draw=none, align=center] at (20,-2.9){Input\\$(d,l)$};}


\draw[solid, draw=black] (root1.south east) -- (insetRB.south east);
\draw[solid, draw=black] (root1.south west) -- (insetRB.north east);

\draw[solid, draw=black] (Ugateroot.north west) -- (insetU.north west);
\draw[solid, draw=black] (Ugateroot.north east) -- (insetU.south west);

\draw[solid, draw=black] (Ugate.north west) -- (insetU2.north east);
\draw[solid, draw=black] (Ugate.north east) -- (insetU2.north west);


\draw[solid, draw=black] (root2.south west) -- (insetRB1.south east);
\draw[solid, draw=black] (root2.south east) -- (insetRB1.north east);


\draw[->, >={Stealth[length=6pt]}, very thick] (-4,0.5) -- (-3,0.5)node[fill=none,draw=none,midway,above] {\Large$\ket{\psi_{in}}$};
\draw[->, >={Stealth[length=6pt]}, very thick] (19,0.5) -- (20,0.5)node[fill=none,draw=none,midway,above] {\Large$\ket{\psi_{out}}$};

\end{tikzpicture}
\caption{
Step-by-step evolution of the qRAM protocol in the case of a classical address $\boldsymbol{a}= a_1\, a_2 = 1 0$ and $m=1$, namely a single data walker. An ordered sequence of quantum walkers are inserted in the leftmost input port of the qRAM, where their initialization follows a red-particle/no-particle encoding for the address walkers. The first particle to be injected is $A_1$ and then $A_2$ and $D_1$ follows. The positional notation of the walkers follows the $(d,l)$ representation, where $d$ is the depth and $l$ indexes a specific branch. At each depth the quantum walkers undergo the action of two unitary gates, $\hat U^{(d)}$, which has the role to transfer the information contained in the address walkers to all the following ones, and the scattering gate $\hat S$, which spatially routes particles around the tree. The gate $\hat U^{(d)}$ has the cotrol on the $d$-th address walker. In case this walker is red, it flips the color of all the following walkers (if present). Notice how $D_1$, after the application of $\hat U^{(1)}$ (top left inset), receives a color flip. In case no red walker is present in the control site of $\hat U^{(d)}$, no effect occurs on the other particles, as in the case of $\hat U^{(2)}$ (top right inset).
The scattering gate $\hat S$ sends red walkers to the $(d+1,2l-1)$ branch and blue walkers to the $(d+1,2l)$ branch with a color change to red (bottom left inset).
During the querying phase (center of the qRAM), the memory cells write the information on those data walkers reaching the memory cells, in this case $D_1$ in the $x_{10}$ memory cell.  
During the output phase, the action of the gates $\hat U^{(d)}$ and $\hat S^\dagger$ (bottom right inset) routes the particles to the final output port.} 

\label{rout_qram_cl}
\end{figure}
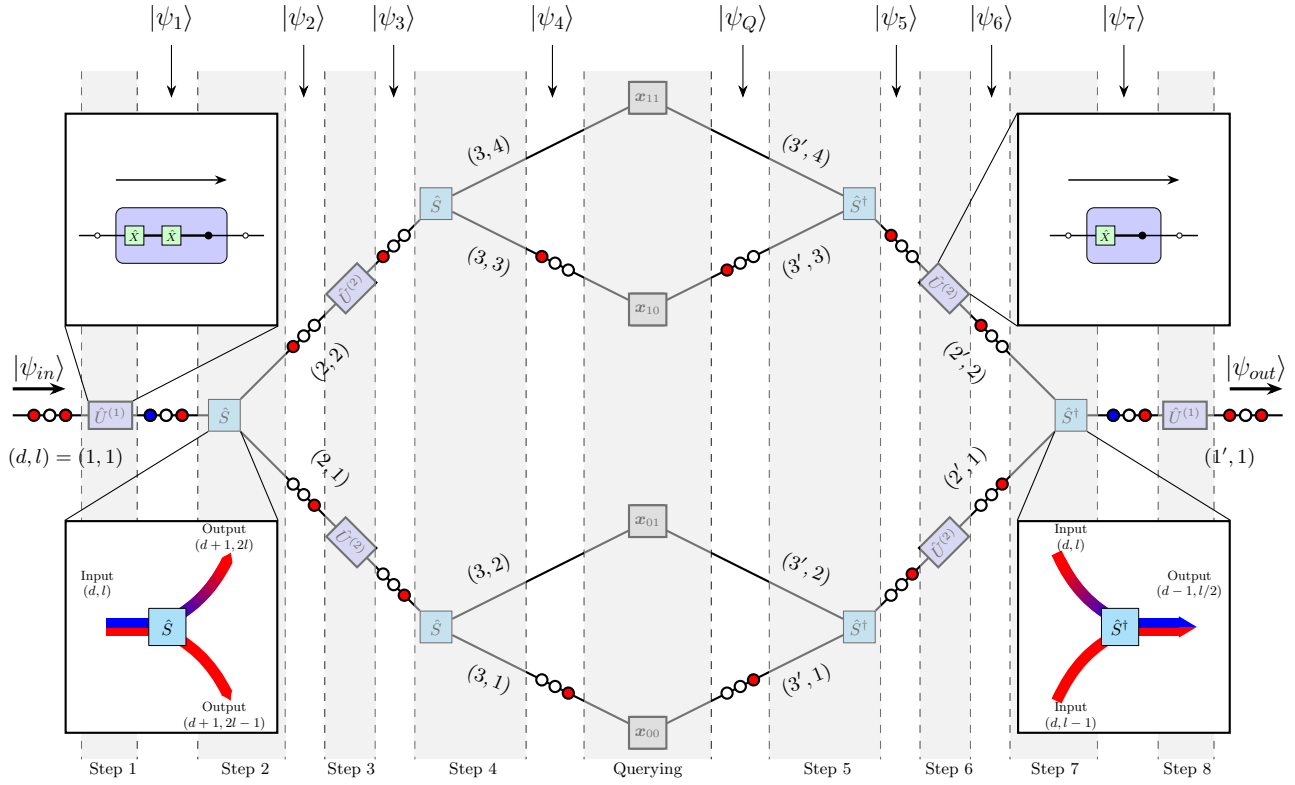

\subsubsection{Scattering gate: $\hat S$}\label{scattS}

We model each internal node of the binary-tree qRAM as a localized scattering region where three possible positions on the binary tree are connected, namely $(d,l)$, $(d+1,2l-1)$ and $(d+1,2l)$ (see Fig.~1).  
These positions correspond to the (parent) branch $(d,l)$ and the two outgoing (child) branches $(d+1,2l-1)$ and $(d+1,2l)$.
For instance, the node $(d,l)=(2,1)$ is connected to the two child branches $(d,l)=(3,1),(3,2)$.
In our model, each walker can be localized only in one of these three sites. Hence, the states that a walker can assume just before and after the application of the scattering gate $\hat S$ belong to:
\begin{equation*}
  \mathcal F^{d,l}=
  \text{Span}\Big\{ \ket{R^{d,l}},\ket{R^{d+1,2l-1}},\ket{R^{d+1,2l}},\ket{B^{d,l}},\ket{B^{d+1,2l-1}},\ket{B^{d+1,2l}},\ket{\emptyset^d},\ket{\emptyset^{d+1}} \Big\},
\end{equation*}

The gate $\hat{S}$ is defined as a unitary on $\mathcal{F}^{d,l}$
that routes and possibly change color of a single carrier in a certain position depending on its internal color, while leaving the no-particle state unchanged. More precisely, each particle arriving at the node in $(d,l)$, is routed to $(d+1,2l-1)$ if in state $\ket{R}$, while it's routed to $(d+1,2l)$ if in state $\ket{B}$. Moreover, in this latter case, the blue walker gets a color change, becoming red. Explicitly the operator $\hat{S}$ is given by 
\begin{eqnarray}\label{hatS}   \hat{S}&=& 
\ket{R^{d+1,2l-1}}\bra{R^{d,l}}+ 
\ket{R^{d+1,2l}}\bra{B^{d,l}}+
\ket{\emptyset^{d+1}}\bra{\emptyset^{d}} \\
&+& \ket{B^{d+1,2l}}\bra{R^{d+1,2l-1}} 
+ \ket{B^{d+1,2l-1}}\bra{R^{d+1,2l}} 
+ \ket{B^{d,l}}\bra{B^{d+1,2l-1}}
+ \ket{R^{d,l}}\bra{B^{d+1,2l}}
+ \ket{\emptyset^{d}}\bra{\emptyset^{d+1}} 
\end{eqnarray}
where the first three terms (first line) are those effectively needed during the routing of the address and data particles from the root node to the memory cells, while the last five (second line) guarantee the unitarity of the operator $\hat S$. Its action on the elements of $\mathcal F^{d,l}$ positioned in $(d,l)$ reads:
\begin{gather*}
\hat S \ket{R^{d,l}} = \ket{R^{d+1,2l-1}}, \,\,\,\,
\hat S \ket{B^{d,l}} = \ket{R^{d+1,2l}},
\,\,\,\,
\hat S \ket{\emptyset^{d}} = \ket{\emptyset^{d+1}}
\end{gather*}
Finally, the action of $\hat{S}$ on any superposition in $\mathcal{F}^{d,l}$ follows by linearity.

\begin{figure}[t]
\centering
\begin{tikzpicture}[x=1cm,y=1cm, scale=0.95]
  \def\w{8}
  \def\h{1.5}
  \def\dr{3pt}
  \def\dist{0.3}

  \begin{scope}[shift={(-2,2)}, scale=0.9, transform shape]
    \draw[rounded corners, thick, fill=gray!30] (0.8,0) rectangle (\w-1.5,\h);

    \foreach \i in {1,...,4} {
      \pgfmathsetmacro{\x}{(\i+0.25)/6*\w}
      \draw[line width=1pt] (\x, \h/2) -- (\x, {-\dist});
    }

      \draw[line width=1pt] (1.7, {\h/2}) -- (5.68, {\h/2});
      \filldraw[fill=black, draw=black] (5.68,{\h/2}) circle (\dr-1);
    
    \foreach[count=\k from 1] \col in {red, white, red} {
        \pgfmathsetmacro{\x}{(\k+0.25)/6*\w}  
        \filldraw[fill=\col, draw=black, thick] (\x, {\h/2}) circle [radius=\dr];
      }

    \draw[thick] (-1,-\dist) -- (\w,-\dist);

    \foreach[count=\k from 1] \col in {red,red,red,red} {
        \pgfmathsetmacro{\x}{(\k-2.1)/7*(\w/2)}
        \pgfmathsetmacro{\y}{-\dist}
        \filldraw[fill=\col, draw=black, thick] (\x, \y) circle [radius=\dr];
      }

    \foreach[count=\k from 1] \col in {gray!30,gray!30,gray!30,black} {
        \pgfmathsetmacro{\x}{(\k+0.25)/6*(\w)}
        \pgfmathsetmacro{\y}{-\dist}
        \filldraw[fill=\col, draw=black, thick] (\x, \y) circle [radius=\dr-1];
      }

    \node at (1.6,1.2) {{\scalebox{0.8}{$M_3^{(\boldsymbol{a})}$}}};
    
    \node at (3,1.2) {{\scalebox{0.8}{$M_2^{(\boldsymbol{a})}$}}};
    
    \node at (4.4,1.2) {{\scalebox{0.8}{$M_1^{(\boldsymbol{a})}$}}};


    \node at (-0.6,-0.9) {{\scalebox{0.8}{$D_{3}$}}};
      
    \node at (0,-0.9) {{\scalebox{0.8}{$D_2$}}};
    
    \node at (0.6,-0.9) {{\scalebox{0.8}{$D_1$}}};
    
    \node at (1.2,-0.9) {{\scalebox{0.8}{$D_{0}$}}};
  \end{scope}

\draw[dashed] (-3,{\h-0.8})--({\w-2.2},{\h-0.8});

  \begin{scope}[shift={(-2,-1)}, scale=0.9, transform shape]
    \draw[rounded corners, thick, fill=gray!30] (0.8,0) rectangle (\w-1.5,\h);

    \foreach \i in {1,...,4} {
      \pgfmathsetmacro{\x}{(\i+0.25)/6*\w}
      \draw[line width=1pt] (\x, \h/2) -- (\x, {-\dist});
    }

      \draw[line width=1pt] (1.7, {\h/2}) -- (5.68, {\h/2});
      \filldraw[fill=black, draw=black] (5.68,{\h/2}) circle (\dr-1);
    
    \foreach[count=\k from 1] \col in {red, white, red} {
        \pgfmathsetmacro{\x}{(\k+0.25)/6*\w}  
        \filldraw[fill=\col, draw=black, thick] (\x, {\h/2}) circle [radius=\dr];
      }

    \draw[thick] (-1,-\dist) -- (\w,-\dist);

    \foreach[count=\k from 1] \col in { red, red,red,red} {
        \pgfmathsetmacro{\x}{(\k+0.25)/6*(\w)}
        \pgfmathsetmacro{\y}{-\dist}
        \filldraw[fill=\col, draw=black, thick] (\x, \y) circle [radius=\dr];
      }

    \node at (1.6,1.2) {{\scalebox{0.8}{$M_3^{(\boldsymbol{a})}$}}};
    
    \node at (3,1.2) {{\scalebox{0.8}{$M_2^{(\boldsymbol{a})}$}}};
    
    \node at (4.4,1.2) {{\scalebox{0.8}{$M_1^{(\boldsymbol{a})}$}}};

    \node at (1.7,-0.9) {{\scalebox{0.8}{$D_3$}}};

    \node at (3,-0.9) {{\scalebox{0.8}{$D_2$}}};
 
    \node at (4.3,-0.9) {{\scalebox{0.8}{$D_1$}}};
    
    \node at (5.7,-0.9) {{\scalebox{0.8}{$D_{0}$}}};

  \end{scope}

\draw[dashed] (-3,{-(2*\h)+0.8})--({\w-2.2},{-(2*\h)+0.8});

     \begin{scope}[shift={(-2,-4)}, scale=0.9, transform shape]
    \draw[rounded corners, thick, fill=gray!30] (0.8,0) rectangle (\w-1.5,\h);

    \foreach \i in {1,...,4} {
      \pgfmathsetmacro{\x}{(\i+0.25)/6*\w}
      \draw[line width=1pt] (\x, \h/2) -- (\x, {-\dist});
    }

      \draw[line width=1pt] (1.7, {\h/2}) -- (5.68, {\h/2});
      \filldraw[fill=black, draw=black] (5.68,{\h/2}) circle (\dr-1);
    
    \foreach[count=\k from 1] \col in {red, white, red} {
        \pgfmathsetmacro{\x}{(\k+0.25)/6*\w}  
        \filldraw[fill=\col, draw=black, thick] (\x, {\h/2}) circle [radius=\dr];
      }

    \draw[thick] (-1,-\dist) -- (\w,-\dist);

    \foreach[count=\k from 1] \col in { red, white,red,red} {
        \pgfmathsetmacro{\x}{(\k+0.25)/6*(\w)}
        \pgfmathsetmacro{\y}{-\dist}
        \filldraw[fill=\col, draw=black, thick] (\x, \y) circle [radius=\dr];
      }

    \node at (1.6,1.2) {{\scalebox{0.8}{$M_3^{(\boldsymbol{a})}$}}};
    
    \node at (3,1.2) {{\scalebox{0.8}{$M_2^{(\boldsymbol{a})}$}}};
    
    \node at (4.4,1.2) {{\scalebox{0.8}{$M_1^{(\boldsymbol{a})}$}}};

    
    \node at (1.7,-0.9) {{\scalebox{0.8}{$D_3$}}};

    \node at (3,-0.9) {{\scalebox{0.8}{$D_2$}}};
 
    \node at (4.3,-0.9) {{\scalebox{0.8}{$D_1$}}};
    
    \node at (5.7,-0.9) {{\scalebox{0.8}{$D_{0}$}}};

  \end{scope}

   \draw[dashed] (-3,{-(3*\h)-0.7})--({\w-2.2},{-(3*\h)-0.7});

 \begin{scope}[shift={(-2,-7)}, scale=0.9, transform shape]
    \draw[rounded corners, thick, fill=gray!30] (0.8,0) rectangle (\w-1.5,\h);

    \foreach \i in {1,...,4} {
      \pgfmathsetmacro{\x}{(\i+0.25)/6*\w}
      \draw[line width=1pt] (\x, \h/2) -- (\x, {-\dist});
    }

      \draw[line width=1pt] (1.7, {\h/2}) -- (5.68, {\h/2});
      \filldraw[fill=black, draw=black] (5.68,{\h/2}) circle (\dr-1);
    
    \foreach[count=\k from 1] \col in {red, white, red} {
        \pgfmathsetmacro{\x}{(\k+0.25)/6*\w}  
        \filldraw[fill=\col, draw=black, thick] (\x, {\h/2}) circle [radius=\dr];
      }

    \draw[thick] (-1,-\dist) -- (\w,-\dist);

    \foreach[count=\k from 1] \col in {red,white,red,red} {
        \pgfmathsetmacro{\x}{(\k-2.1)/7*(\w/2)}
        \pgfmathsetmacro{\y}{-\dist}
        \filldraw[fill=\col, draw=black, thick] (\x+7, \y) circle [radius=\dr];
      }

    \foreach[count=\k from 1] \col in {gray!30,gray!30,gray!30,black} {
        \pgfmathsetmacro{\x}{(\k+0.25)/6*(\w)}
        \pgfmathsetmacro{\y}{-\dist}
        \filldraw[fill=\col, draw=black, thick] (\x, \y) circle [radius=\dr-1];
      }

    \node at (1.6,1.2) {{\scalebox{0.8}{$M_3^{(\boldsymbol{a})}$}}};
    
    \node at (3,1.2) {{\scalebox{0.8}{$M_2^{(\boldsymbol{a})}$}}};
    
    \node at (4.4,1.2) {{\scalebox{0.8}{$M_1^{(\boldsymbol{a})}$}}};


    \node at (6.4,-0.9) {{\scalebox{0.8}{$D_{3}$}}};
      
    \node at (7,-0.9) {{\scalebox{0.8}{$D_2$}}};
    
    \node at (7.6,-0.9) {{\scalebox{0.8}{$D_1$}}};
    
    \node at (8.2,-0.9) {{\scalebox{0.8}{$D_{0}$}}};
  \end{scope}

\draw[thick](6.4,4)--(6.4,-8);

 \begin{scope}[shift={(8,2)}, scale=0.9, transform shape]
    \draw[rounded corners, thick, fill=gray!30] (1,0) rectangle ({\w-3},\h);

    \foreach \i in {1,...,4} {
      \pgfmathsetmacro{\x}{(\i+0.25)/6*\w}
      \draw[line width=1pt] (\x, \h/2) -- (\x, {-\dist});
    }
  
    \draw[line width=1pt] ({\w-2.6},{\h/2})--({\w-2.35},{\h/2});

    \draw[rounded corners, thick, fill=orange!30] ({\w-3.3},0) rectangle ({\w-2.6},\h) node[midway,anchor=center] {{\scalebox{0.8}{$\ket{\text{off}}$}}};

    \foreach[count=\k from 1] \col in {red, white, red} {
        \pgfmathsetmacro{\x}{(\k+0.25)/6*\w}  
        \filldraw[fill=\col, draw=black, thick] (\x, {\h/2}) circle [radius=\dr];
        }

    \draw[thick] (-1,-\dist) -- ({\w+0.5},-\dist);

    \foreach[count=\k from 1] \col in {red,red,red,red,red} {
        \pgfmathsetmacro{\x}{(\k-2.7)/7*(\w/2)}
        \pgfmathsetmacro{\y}{-\dist}
        \filldraw[fill=\col, draw=black, thick] (\x, \y) circle [radius=\dr];
      }

    \foreach[count=\k from 1] \col in {black,black,black,black} {
        \pgfmathsetmacro{\x}{(\k+0.26)/6*(\w)}
        \pgfmathsetmacro{\y}{-\dist}
        \filldraw[fill=\col, draw=black, thick] (\x, \y) circle [radius=\dr-1];
      }

    \node at (1.6,1.2) {{\scalebox{0.8}{$M_3^{(\boldsymbol{a})}$}}};
    
    \node at (3,1.2) {{\scalebox{0.8}{$M_2^{(\boldsymbol{a})}$}}};
    
    \node at (4.4,1.2) {{\scalebox{0.8}{$M_1^{(\boldsymbol{a})}$}}};

    \node at (-1,-0.9) {{\scalebox{0.8}{$D_{4}$}}};

    \node at (-0.4,-0.9) {{\scalebox{0.8}{$D_{3}$}}};
      
    \node at (0.2,-0.9) {{\scalebox{0.8}{$D_2$}}};
    
    \node at (0.75,-0.9) {{\scalebox{0.8}{$D_1$}}};
    
    \node at (1.4,-0.9) {{\scalebox{0.8}{$D_{0}$}}};
  \end{scope}

  \draw[dashed] (7,{\h-0.8})--({\w+7.8},{\h-0.8});

  \begin{scope}[shift={(8,-1)}, scale=0.9, transform shape]
    \draw[rounded corners, thick, fill=gray!30] (1,0) rectangle ({\w-3},\h);

   \node[draw, dashed, rounded corners, fill=green!20, minimum width=12mm, minimum height=16mm, opacity=0.9] at (3.2,0.13) {};

    \foreach \i in {1,...,4} {
      \pgfmathsetmacro{\x}{(\i+0.25)/6*\w}
      \draw[line width=1pt] (\x, \h/2) -- (\x, {-\dist});
    }
  
    \draw[line width=1pt] ({\w-2.6},{\h/2})--({\w-2.35},{\h/2});

    \draw[rounded corners, thick, fill=orange!30] ({\w-3.3},0) rectangle ({\w-2.6},\h) node[midway,anchor=center] {{\scalebox{0.8}{$\ket{\text{on}}$}}};

    \foreach[count=\k from 1] \col in {red, white, red} {
        \pgfmathsetmacro{\x}{(\k+0.25)/6*\w}  
        \filldraw[fill=\col, draw=black, thick] (\x, {\h/2}) circle [radius=\dr];
        }

    \draw[thick] (-1,-\dist) -- (\w+0.5,-\dist);

    \foreach[count=\k from 1] \col in { red, red,red,red,red} {
        \pgfmathsetmacro{\x}{(\k-0.75)/6*(\w)}
        \pgfmathsetmacro{\y}{-\dist}
        \filldraw[fill=\col, draw=black, thick] (\x, \y) circle [radius=\dr];
      }

    \node at (1.6,1.2) {{\scalebox{0.8}{$M_3^{(\boldsymbol{a})}$}}};
    
    \node at (3,1.2) {{\scalebox{0.8}{$M_2^{(\boldsymbol{a})}$}}};
    
    \node at (4.4,1.2) {{\scalebox{0.8}{$M_1^{(\boldsymbol{a})}$}}};

    \node at (0.4,-0.9) {{\scalebox{0.8}{$D_{4}$}}};
       
    \node at (1.7,-0.9) {{\scalebox{0.8}{$D_3$}}};

    \node at (3,-0.9) {{\scalebox{0.8}{$D_2$}}};
 
    \node at (4.3,-0.9) {{\scalebox{0.8}{$D_1$}}};
    
    \node at (5.7,-0.9) {{\scalebox{0.8}{$D_{0}$}}};

    \node at (3.4,0.1) {{\scalebox{0.8}{$\hat{U}_{\text{copy}}^{\text{loc}}$}}};
 \end{scope}

\draw[dashed] (7,{-(2*\h)+0.8})--({\w+7.8},{-(2*\h)+0.8});

   \begin{scope}[shift={(8,-4)}, scale=0.9, transform shape]
    \draw[rounded corners, thick, fill=gray!30] (1,0) rectangle ({\w-3},\h);

    \foreach \i in {1,...,4} {
      \pgfmathsetmacro{\x}{(\i+0.25)/6*\w}
      \draw[line width=1pt] (\x, \h/2) -- (\x, {-\dist});
    }
  
    \draw[line width=1pt] ({\w-2.6},{\h/2})--({\w-2.35},{\h/2});

    \draw[rounded corners, thick, fill=orange!30] ({\w-3.3},0) rectangle ({\w-2.6},\h) node[midway,anchor=center] {{\scalebox{0.8}{$\ket{\text{on}}$}}};

    \foreach[count=\k from 1] \col in {red, white, red} {
        \pgfmathsetmacro{\x}{(\k+0.25)/6*\w}  
        \filldraw[fill=\col, draw=black, thick] (\x, {\h/2}) circle [radius=\dr];
        }

    \draw[thick] (-1,-\dist) -- (\w+0.5,-\dist);

    \foreach[count=\k from 1] \col in { red, red,white,red,red} {
        \pgfmathsetmacro{\x}{(\k-0.75)/6*(\w)}
        \pgfmathsetmacro{\y}{-\dist}
        \filldraw[fill=\col, draw=black, thick] (\x, \y) circle [radius=\dr];
      }

    \node at (1.6,1.2) {{\scalebox{0.8}{$M_3^{(\boldsymbol{a})}$}}};
    
    \node at (3,1.2) {{\scalebox{0.8}{$M_2^{(\boldsymbol{a})}$}}};
    
    \node at (4.4,1.2) {{\scalebox{0.8}{$M_1^{(\boldsymbol{a})}$}}};

    \node at (0.4,-0.9) {{\scalebox{0.8}{$D_{4}$}}};
       
    \node at (1.7,-0.9) {{\scalebox{0.8}{$D_3$}}};

    \node at (3,-0.9) {{\scalebox{0.8}{$D_2$}}};
 
    \node at (4.3,-0.9) {{\scalebox{0.8}{$D_1$}}};
    
    \node at (5.7,-0.9) {{\scalebox{0.8}{$D_{0}$}}};

 \end{scope}

 \draw[dashed] (7,{-(3*\h)-0.7})--({\w+7.8},{-(3*\h)-0.7});

   \begin{scope}[shift={(8,-7)}, scale=0.9, transform shape]
    \draw[rounded corners, thick, fill=gray!30] (1,0) rectangle ({\w-3},\h);

    \foreach \i in {1,...,4} {
      \pgfmathsetmacro{\x}{(\i+0.25)/6*\w}
      \draw[line width=1pt] (\x, \h/2) -- (\x, {-\dist});
    }
  
    \draw[line width=1pt] ({\w-2.6},{\h/2})--({\w-2.35},{\h/2});

    \draw[rounded corners, thick, fill=orange!30] ({\w-3.3},0) rectangle ({\w-2.6},\h) node[midway,anchor=center] {{\scalebox{0.8}{$\ket{\text{off}}$}}};

    \foreach[count=\k from 1] \col in {red, white, red} {
        \pgfmathsetmacro{\x}{(\k+0.25)/6*\w}  
        \filldraw[fill=\col, draw=black, thick] (\x, {\h/2}) circle [radius=\dr];
        }

    \draw[thick] (-1,-\dist) -- ({\w+0.5},-\dist);

    \foreach[count=\k from 1] \col in {black,black,black,black} {
        \pgfmathsetmacro{\x}{(\k+0.26)/6*(\w)}
        \pgfmathsetmacro{\y}{-\dist}
        \filldraw[fill=\col, draw=black, thick] (\x, \y) circle [radius=\dr-1];
      }

    \foreach[count=\k from 1] \col in {red,red,white,red,red} {
        \pgfmathsetmacro{\x}{(\k+8.95)/7*(\w/2)}
        \pgfmathsetmacro{\y}{-\dist}
        \filldraw[fill=\col, draw=black, thick] (\x, \y) circle [radius=\dr];
      }

    \node at (1.6,1.2) {{\scalebox{0.8}{$M_3^{(\boldsymbol{a})}$}}};
    
    \node at (3,1.2) {{\scalebox{0.8}{$M_2^{(\boldsymbol{a})}$}}};
    
    \node at (4.4,1.2) {{\scalebox{0.8}{$M_1^{(\boldsymbol{a})}$}}};

    \node at (5.6,-0.9) {{\scalebox{0.8}{$D_{4}$}}};

    \node at (6.25,-0.9) {{\scalebox{0.8}{$D_{3}$}}};
      
    \node at (6.85,-0.9) {{\scalebox{0.8}{$D_2$}}};
    
    \node at (7.4,-0.9) {{\scalebox{0.8}{$D_1$}}};
    
    \node at (8.05,-0.9) {{\scalebox{0.8}{$D_{0}$}}};
 \end{scope}

\node at (-3,3.3){(a)};
\node at (-3,0.3){(b)};
\node at (-3,-2.7){(c)};
\node at (-3,-5.7){(d)};

\node at (7,3.3){(e)};
\node at (7,0.3){(f)};
\node at (7,-2.7){(g)};
\node at (7,-5.7){(h)};

\end{tikzpicture}

\caption{Schematic representation of the control-copy schemes introduced in Section~\ref{SecMessCop}, for a message of length $m=3$. The left column illustrates the scheme based on a global unitary acting on all walkers: (a) after traversing the binary tree, the data walkers reach the targeted memory cell; (b) the flag walker $D_0$ acts as the control for the global unitary operation; (c) the message is copied to the data walkers according to Eq.~(\ref{locexample}); (d) the particles then exit the cell.
The right column shows the main steps of the cell state switch variant: (e) after the routing process, the particles arrive at the memory cell, which is initially in the state $\ket{\text{off}}_{F^{(\boldsymbol{a})}}$; (f) the flag walker $D_0$ switches the cell state to $\ket{\text{on}}_{F^{(\boldsymbol{a})}}$; (g) the cell activates $m=3$ local copy operations between the memory and data walkers, copying the message according to Eq.~(\ref{writer}); (h) finally, the particles leave the cell, and the last data walker, $D_{4}$, acts as a flag to reset the switch state back to $\ket{\text{off}}_{F^{(\boldsymbol{a})}}$.  }
\label{copy_standard}
\end{figure}
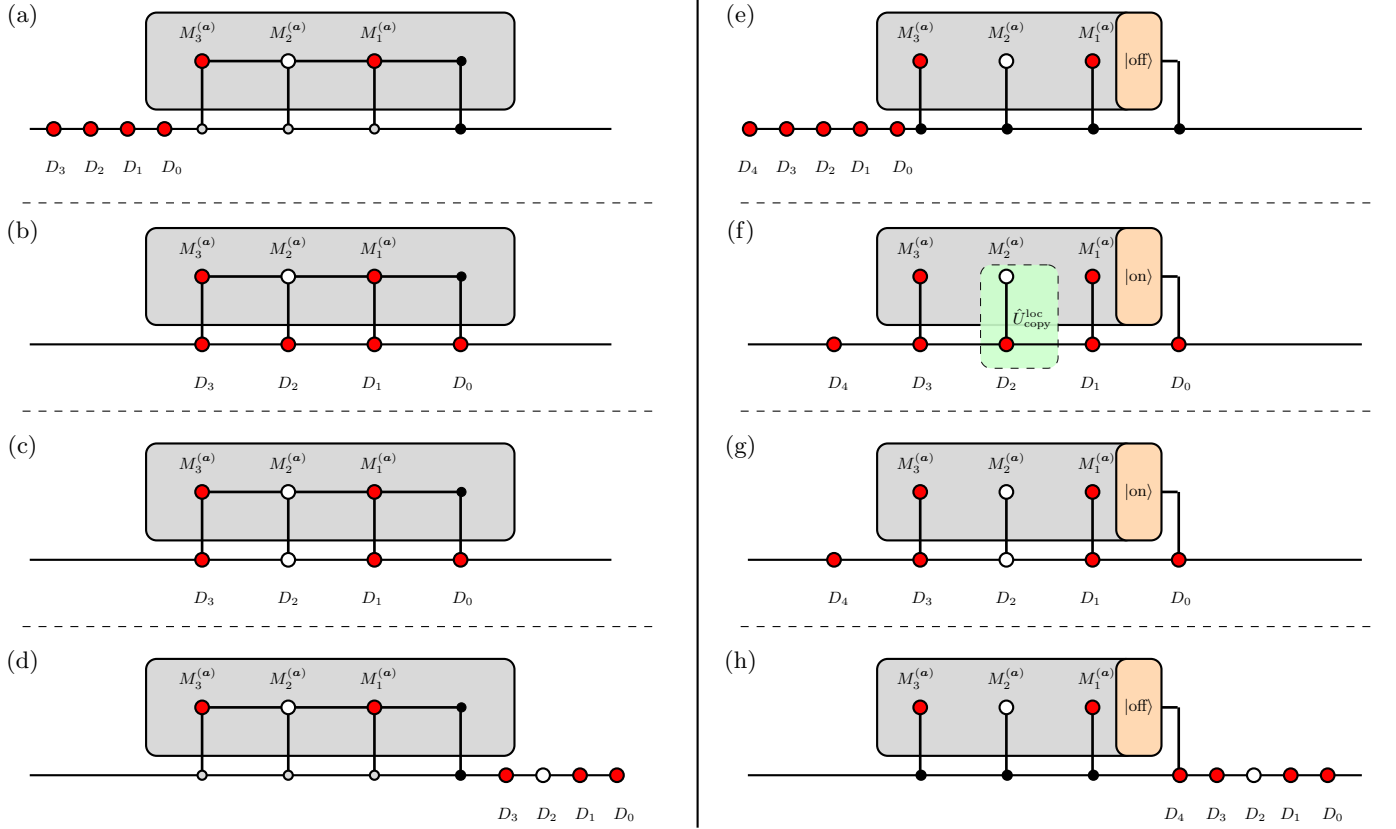

\subsection{Message Copy}\label{SecMessCop}

To retrieve the message stored in the memory cells, a controlled operation must be applied between the memory particles and the data particles traversing the binary tree. However, only those memory cells that are reached by the $m$ data walkers need to be accessed. To ensure that the COPY operation is performed selectively and only at the target locations, we introduce an auxiliary “flag” walker, denoted by $D_0$, that precedes the  $m$ data walkers $D_1, \dots, D_m$. 

A simple way of doing the copying operation is to use the flag walker $D_0$ as a control qubit for global unitary acting on all data and memory walkers. More precisely we can define a unitary $\hat{U}_{copy}^{(\boldsymbol{a})}$ which, depending on the color of the flag walker $D_0$, applies a local control copy operation $\hat{U}_{copy}^{loc}$ between the $i$-th data walker $D_i$ and the corresponding memory particle $M_i^{(\boldsymbol{a})}$. Explicitly, the unitary can be written as
\begin{equation}\label{Ucopya}
\hat{U}_{copy}^{(\boldsymbol{a})}=\ket{R^{\boldsymbol{a}}}_{D_0}\bra{R^{\boldsymbol{a}}}\otimes \left[\bigotimes_{j=1}^m\hat{U}_{copy}^{loc}(M_j^{(\boldsymbol{a})},D_j)\right]+\left(\ket{\emptyset^{\boldsymbol{a}}}_{D_0}\bra{\emptyset^{\boldsymbol{a}}}+\ket{B^{\boldsymbol{a}}}_{D_{0}}\bra{B^{\boldsymbol{a}}}\right)\otimes \hat{I}_{M^{(\boldsymbol{a})},D}
\end{equation}
where here we replaced the positional notation $(d,l)$ with that of memory cells, namely through the corresponding address ${\boldsymbol{a}}$.
The gate $\hat{U}_{\text{copy}}^{\text{loc}}(M_j^{(\boldsymbol{a})}, D_j)$ is a local unitary that couples each message subsystem $M_j^{(\boldsymbol{a})}$ with its corresponding data walker $D_j$. These local gates are defined as:
\begin{equation}\label{loc_copy}
\hat{U}_{\text{copy}}^{\text{loc}}(M_j^{(\boldsymbol{a})}, D_j) =
    \left(    \ket{B}_{M_j^{(\boldsymbol{a})}}\bra{B} + \ket{R}_{M_j^{(\boldsymbol{a})}}\bra{R}
    \right) \otimes \hat{I}_{D_j}
    + \ket{\emptyset}_{M_j^{(\boldsymbol{a})}}\bra{\emptyset} \otimes 
    \left(
  \ket{\emptyset^{\boldsymbol{a}}}_{D_j}\bra{R^{\boldsymbol{a}}} + \ket{R^{\boldsymbol{a}}}_{D_j}\bra{\emptyset^{\boldsymbol{a}}} + \ket{B^{\boldsymbol{a}}}_{D_j}\bra{B^{\boldsymbol{a}}}
    \right).
\end{equation}
We graphically represent this approach to copy the information from a data cell into the data register $D$ in Fig.~\ref{copy_standard}(left).

{A more refined approach to classical data copying is to let the flag walker $D_0$
act as a control mechanism for activating the memory cell.}
As the data walkers approach the memory cell, depending on the internal state of $D_0$, we change the internal state of $F^{(\boldsymbol a)}$, 
namely a switch state of the cell. The two-level system representing this switch is initialized in $\ket{\text{off}}_{F^{(\mathbf a)}}$. We perform a control-NOT operation between $D_0$ and $F^{(\boldsymbol a)}$ so that, if $D_0$ is present (in the state red) then we induce the transformation $\ket{  \text{off}}_{F^{(\mathbf a)}}\rightarrow \ket{\text{on}}_{F^{(\mathbf a)}}$, otherwise, if $D_0$ is not present (in the empty state), the switch keeps being off. This is represented by the following unitary operation:

\begin{equation}\label{switch}
\hat{U}^{(\boldsymbol{a})}_{act}=\ket{R^{\boldsymbol{a}}}_{D_0}\bra{R^{\boldsymbol{a}}}\otimes \left(\ket{\text{on}}_{F^{(\boldsymbol{a})}}\bra{\text{off}}+\ket{\text{off}}_{F^{(\boldsymbol{a})}}\bra{\text{on}}\right)+\left(\ket{\emptyset^{\boldsymbol{a}}}_{D_0}\bra{\emptyset^{\boldsymbol{a}}}+\ket{B^{\boldsymbol{a}}}_{D_0}\bra{B^{\boldsymbol{a}}}\right)\otimes \hat{I}_{F^{(\boldsymbol{a})}}.
\end{equation}

This operation is performed because we want the cell to write the information on the site of the data particles only if the data particles are entering the memory cell. Indeed, we want to avoid the action of the memory cells when they are not reached by any particle.

Hence, if the state of the target memory cells is $\ket{\text{on}}_{F^{(\boldsymbol{a})}}$, we apply the
COPY gates between the data walkers and the message stored in the addressed cell $M^{(\boldsymbol{a})}$ in parallel. The unitary operator implementing this operation can be written as
\begin{equation}\label{writer} \hat{U}^{(\boldsymbol{a})}_{\text{copy}} = 
    \ket{\text{on}}_{F^{(\boldsymbol{a})}}\bra{\text{on}} \otimes 
    \left[
        \bigotimes_{j=1}^m \hat{U}_{\text{copy}}^{\text{loc}}(M_j^{(\boldsymbol{a})}, D_j)
    \right] 
    +  \ket{\text{off}}_{(F^{\boldsymbol{a})}}\bra{\text{off}} 
     \otimes \hat{I}_{M^{(\boldsymbol{a})}, D},
\end{equation}
{where $\hat{U}_{\text{copy}}^{\text{loc}}(M_j^{(\boldsymbol{a})}, D_j)$ is the unitary defined by Eq.~(\ref{loc_copy}).}
The action of this gate is illustrated by the following examples:
\begin{equation}\label{locexample}
\hat{U}_{\text{copy}}^{\text{loc}}(M_j^{(\boldsymbol{a})}, D_j)\ket{R}_{M_j^{(\boldsymbol{a})}}\ket{R^{\boldsymbol{a}}}_{D_j}
    = \ket{R}_{M_j^{(\boldsymbol{a})}}\ket{R^{\boldsymbol{a}}}_{D_j}, \quad
\hat{U}_{\text{copy}}^{\text{loc}}(M_j^{(\boldsymbol{a})}, D_j)\ket{\emptyset}_{M_j^{(\boldsymbol{a})}}\ket{R^{\boldsymbol{a}}}_{D_j}
    = \ket{\emptyset}_{M_j^{(\boldsymbol{a})}}\ket{\emptyset^{\boldsymbol{a}}}_{D_j}.
\end{equation}

In other words, if the $j$-th bit of the message is $b_j^{(\boldsymbol{a})} = 1$, the state of the corresponding data walker $D_j$ remains unchanged. Conversely, if $b_j^{(\boldsymbol{a})} = 0$, the internal state of $D_j$ is flipped from red to empty. The use of the flag walker ensures that the COPY operation is applied only to those memory cells reached during the routing phase.
Since the COPY operation acts independently on each data-message pair, the $m$ local gates can be executed in parallel, resulting in an overall complexity of $\mathcal{O}(m)$ for the readout stage. To signal the conclusion of the COPY phase and synchronize the subsequent steps of the protocol, we introduce an additional data walker, denoted $D_{m+1}$. This walker reaches the memory cell immediately after the $m+1$ data walkers $D_0,\dots,D_m$ and serves as a temporal delimiter for the end of the message retrieval process. 
Indeed, after the action of $\hat{U}^{(\boldsymbol{a})}_{\text{copy}}$, we perform again a unitary operation similar to $\hat{U}^{(\boldsymbol{a})}_{\text{act}}$, where the role of $D_0$ is now played by $D_{m+1}$. 
We graphically represent this approach in Fig.~\ref{copy_standard}(right).

In summary, if the data particles $D_0,...,D_{m+1}$ reaches the target memory cell, our scheme: (i) activates the memory cell switch state $\ket{  \text{off}}_{F^{(\boldsymbol{a}) }}\rightarrow \ket{\text{on}}_{F^{(\boldsymbol{a}) }}$ through the presence of $D_0$ (see Eq.~(\ref{switch})), (ii) writes the information $\boldsymbol{b}^{(\boldsymbol{a})}$ contained in $M^{(\boldsymbol{a})}$ into $D_1,...,D_m$ through $\hat{U}^{(\boldsymbol{a})}_{\text{copy}}$ (see Eq.~(\ref{writer})) and (iii) turns off the memory cell switch state $\ket{  \text{on}}_{F^{(\boldsymbol{a}) }}\rightarrow \ket{\text{off}}_{F^{(\boldsymbol{a}) }}$ through the presence of $D_{m+1}$ (see Eq.~(\ref{switch})). 

\subsection{Inverse routing}\label{SecInvRout}
This phase is realized with the same ingredients explained during the routing phase (see Section~\ref{SecRout}), where just small clarifications are needed. First, we discuss the role of the routing gates $\hat U^{(d)}$ and the scattering gate $\hat S^\dagger$ needed during this last phase (Section \ref{invrout}) and later we clarify the dispersion and recollection of the address quantum walkers around our qRAM binary tree. Indeed, as we can see from Fig.~\ref{rout_qram_cl}, red address walkers, after they have been in control of a $U^{(d)}$ operation, they separate from the rest of the quantum walkers that travel together the data register. We explain how a recollection of these walkers always take place during the inverse routing in Section~\ref{dispersion}.

\subsubsection{Routing gates $\hat U^{(d)}$ and $\hat S^\dagger$}\label{invrout}

Concerning the gate $\hat U^{(d)}$, no modifications are needed during the inverse routing. Notice that, immediately after the particles crossed the memory cells, the first address particle used as control for this gate is the last one $A_n$, namely the $n$-th one, and not the first $A_1$. This is graphically shown in Fig.~\ref{rout_qram_cl} in case of a classical address.

The scattering gate $\hat S^\dagger $ used during the inverse routing is the "left/right" mirrored version of the same gate used during the routing phase, where particles are routed from higher to lower depths $d$, as shown in Fig.~\ref{rout_qram_cl} (bottom right corner).  
The design of our protocol is such that, independently from the address and the message contained into the data walkers, only red walkers enter from the $(d,l)$ and $(d,l-1)$ input ports. This is in full analogy with what we saw for $\hat S$ during the routing to the memory cells, where the output walkers were always colored red. Hence, during our inverse routing phase,  red walkers coming from $(d,l)$ are routed to the depth $d-1$ with a change of color to blue, while red walkers coming from $(d,l-1)$ are routed to the depth $d-1$ without any change of color.

The mathematical description of this gate has a similar structure of Eq.~(\ref{hatS}):
\begin{eqnarray}\label{hatSout}   \hat{S}^\dagger&=& \ket{B^{d-1,l/2}}\bra{R^{d,l}} + \ket{R^{d-1,l/2}}\bra{R^{d,l-1}} +  \ket{\emptyset^{d-1}}\bra{\emptyset^{d}}  \\ 
&+& \nonumber
\ket{R^{d,l-1}}\bra{B^{d,l}} + \ket{R^{d,l}}\bra{B^{d,l-1}} + \ket{B^{d,l}}\bra{R^{d-1,l/2}} + \ket{B^{d,l-1}}\bra{B^{d-1,l/2}} + \ket{\emptyset^{d}}\bra{\emptyset^{d-1}} \, ,
\end{eqnarray}
where the first three terms (first line) are those effectively needed during the inverse routing of the quantum walkers from the memory cells to the final output port, while the last five (second line) guarantee the unitarity of the operator $\hat S^\dagger$. 
The action of $\hat S^\dagger$ on the possible walkers states arriving at the input ports are:
\begin{gather}\label{Sdagga}
 \hat S^\dagger \ket{R^{d,l}} = \ket{B^{d-1,l/2}}, \,\,\,\,
 \hat S^\dagger \ket{R^{d,l-1}} = \ket{R^{d-1,l/2}},
\,\,\,\,
\hat S^\dagger \ket{\emptyset^{d}} = \ket{\emptyset^{d-1}} \, .
\end{gather}

\subsubsection{Dispersion and recollection of the address quantum walkers}\label{dispersion}

A characteristic feature of our qRAM scheme is that, even for a classical address, the walkers do not travel along the binary tree altogether, namely crossing the same $(d,l)$ branch at each time step. During the routing phase from the initial input port to the memory cells, in the case of a red $d$-th address walker, all the particles following it are scattered to one branch after the action of $\hat{S}$, while in contrast, this $d$-th address walker is routed to a different one. 

An example of this phenomenon can be appreciated when the first scattering gate $\hat S$ acts on $A_1, A_2, D$ in Fig.~\ref{rout_qram_cl}. 
At this point of the routing, $A_1$ has transferred the information it contains to the other walkers through $\hat U^{(1)}$ and, from this moment on, it is not subject to the action of any other gate but $\hat S$ and $\hat S^\dagger$, until the last $\hat{U}^{(1)}$ gate at the final output port, where it acts as control. Indeed, $\hat{U}^{(2)}$ takes $A_2$ as control walker and acts on all its following walkers, leaving $A_1$ untouched, no matter the state of $A_2$. Hence, the design of this qRAM ensures that $A_1$ is found again in the red state as the control of $\hat{U}^{(1)}$ at the final output port.

The described dispersion and recollection of the address walker $A_1$ is not specific to the example shown in Fig.~\ref{rout_qram_cl}, but it is general for any address walker $A_i$ and any address chosen. The $i$-th address walker -- being empty or red-colored -- follows the train of walkers containing the data particles up to depth $i$. At this depth, it plays the role of control for $\hat U^{(i)}$ and then, as explained before, it separates from this train, as red or empty. In the latter case, no explanations is needed: an empty $A_i$ shows up at any following stage of the protocol. In case the address walker is red, we have to guarantee the recollection of this walker with the train of particles containing the data ones at the same depth, but during the inverse routing. After a careful look at the binary tree and the unitary gates implemented in our qRAM, we can see that this recollection always takes place. This is because the $i$-th walker separates as red and never changes its color, because it cannot be targeted by any $\hat U^{(d)}$ for $d>i$. Hence, its route is determined solely by the scattering gates $\hat S$ and $\hat S^\dagger$. This means that, in reaching the memory cells, $\hat S$ always routes it through the transformation $(d,l)\rightarrow (d+1,2l-1)$ and after the memory cells $\hat S^\dagger$ always routes it through the transformation $(d,l-1)\rightarrow (d-1,l/2)$: see Fig.~\ref{rout_qram_cl} for the $(d,l)$ index explanations of $\hat S$ and $\hat S^\dagger$.
Thanks to these rules, it is possible to see that indeed $A_i$ reaches again the control of $\hat U^{(i)}$ at the branch where the data particles travel during the inverse routing phase.

\subsection{Example: functioning of the qRAM for a classical address}\label{example}

Step-by-step, we show the functioning of our qRAM scheme in case of a simple classical address $\boldsymbol{a}=a_1\,a_2= 1\,0$, where the leftmost bit is the most significant one, namely $a_1$. The corresponding initial state of the address register is given by $\ket{\boldsymbol{a}}=\ket{\emptyset^{1} }_{A_2}\ket{R^{1,1}}_{A_1}$. Notice that the order of the kets goes from the least significant (left) to the most significant (right), namely as they are physically ordered inside the qRAM.
This state is followed by the data register $D$, which is made of $m=1$ quantum walkers. As explained above, all the particles of the data register are always initialized in the red state.  We graphically represent the steps that are described in this section in Fig.~\ref{rout_qram_cl}. 

\subsubsection*{Routing}


We use the instruments introduced in Section~\ref{SecRout} to explain how our qRAM works in case of the classical address $\boldsymbol{a}=a_1\, a_2=10$.
The overall $A+D$ state at the beginning of the protocol is: 
\begin{equation*}
\ket{\psi_{in}}={\ket{\red^{1,1}}_{D_1}\ket{\zero^{1} }_{A_2}\ket{\red^{1,1}}_{A_1}} \hspace{1cm} d=1,\,\mbox{ pre-}\hat U^{(1)}\mbox{ gate}.
\end{equation*}
This encoding has been explained in the main text of this work and consists in: using a red walker $\ket{R}$ for each 1 in the address $\boldsymbol{a}=a_1\,a_2\,\dots\, a_n$, an empty address subsystem $\ket{\emptyset}$ for each 0 in the address and a data register made of $m$ red walkers, where $m$ is the dimension of the messages $\boldsymbol{b}^{(\boldsymbol{a})}$ written inside the memory cells. 
The order chosen for the kets of $\ket{\psi_{in}}$ is the same of the spatial order of the particles represented in  Fig.~\ref{rout_qram_cl}: from right to left we have $A_1$ (the first to enter), then $A_2$ and finally $D$ (the last to enter). Moreover, to make the role of the address walkers more explicit, as they transfer their routing information to the other walkers through $\hat U^{(d)}$, namely at depth $d$ for the $d$-th walker, we remove the color from their (red) ket state and we write it in black: $\ket{\red }_{A_i}\rightarrow\ket{R}_{A_i}$. The same for empty address subsystems, which go from bold to normal: $\ket{\boldsymbol{\emptyset}}_{A_i}\rightarrow \ket{{\emptyset}}_{A_i}$. In what follows we report the step number written at the bottom of Fig.~\ref{rout_qram_cl}.

{\bf Steps 1 and 2:} We sequentially insert these particles in the root of the binary tree considered by our qRAM scheme. The first gate that acts on the walkers is  $\hat{U}^{(1)}$. The control particle $A_1$ is in the red state. Thus, $\hat U^{(1)}$ will apply a $\hat{X}$ gate, namely a NOT, on all subsequent particles color degree of freedom. The state of the walkers after the application of the gate is 
\begin{equation*}
\ket{\psi_1}= \hat{U}^{(1)}\ket{\psi_{in}}=\ket{\blue^{1,1}}_{D_1}\ket{\zero^{1} }_{A_2}\ket{R^{1,1}}_{A_1} \hspace{1cm} d=1,\,\mbox{ post-}\hat U^{(1)}\mbox{ gate, \,pre-}\hat S\mbox{ gate}.
\end{equation*}
Immediately after, we apply $\hat S$ to this train of particles. We remember that red particles are routed left, while blue particles are routed right together with a change of color. 
Hence, we obtain:
\begin{equation*}
\ket{\psi_2} = \hat S  \ket{\psi_1} = \ket{\red^{2,2}}_{D_1}\ket{\zero^{2} }_{A_2}\ket{R^{2,1}}_{A_1} \hspace{1cm} d=2,\,\mbox{ pre-}\hat U^{(2)}\mbox{ gate} \, .
\end{equation*}

{\bf Steps 3 and 4}: Now, $\hat U^{(2)}$ is applied. Hence, the control is on $A_2$. Since there is no particle on $A_2$, this gate is not activated and therefore the state after $\hat U^{(2)}$ is again
\begin{equation*}
\ket{\psi_3} = \hat U^{(2)}  \ket{\psi_2} = \ket{\psi_2} =  \ket{\red^{2,2}}_{D_1}\ket{\emptyset^{2} }_{A_2}\ket{R^{2,1}}_{A_1} \hspace{1cm} d=2,\,\mbox{ post-}\hat U^{(2)}\mbox{ gate, \,pre-}\hat S\mbox{ gate} \, .
\end{equation*}
Finally, the particles go through the last scattering gate $\hat S$ and reach the depth $d=3$, namely just before the memory cells:
\begin{equation*}
\ket{\psi_4} = \hat S  \ket{\psi_3} = \ket{\red^{3,3}}_{D_1}\ket{\emptyset^{3} }_{A_2}\ket{R^{3,1}}_{A_1} \hspace{1cm} d=3,\,\mbox{ pre-memory cell} \, .
\end{equation*}
As we can see from Fig.~\ref{rout_qram_cl}, the data particle $D$ correctly reaches the memory cell at $(d,l)=(3,3)$, which is the one with the label $\boldsymbol{a}=10$ (see Fig.~\ref{rout_qram_cl}). The crucial ingredients to achieve this result have been to (i) our encoding of the classical address $\boldsymbol a=10$ into the initial state $\ket{\psi_{in}}$ and (ii) the sequential application of $\hat U^{(i)}$ and $\hat S$ to, respectively, transfer the address information encoded in $A=\{A_1,A_2\}$ and route the particles.

In order to make the explanation of the routing clearer and adherent to Fig.~\ref{rout_qram_cl}, we implemented one data walker alone. Nonetheless, as we explained in Section~\ref{rout_qram_cl}, at least an additional data particle $D_0$ is needed for the following phase for the copy of the message from the memory cell. Therefore, we consider, as final state of this first routing phase, the state
\begin{equation}\label{psi4D0D1}
\ket{\psi_4} = \hat S  \ket{\psi_3} = \ket{\red^{3,3}}_{D_1}\ket{\red^{3,3}}_{D_0}\ket{\emptyset^{3} }_{A_2}\ket{R^{3,1}}_{A_1} \hspace{1cm} d=3,\,\mbox{ pre-memory cell} \, .
\end{equation}
Below, we explain how to generalize the routing of $1$ data walker, as we have done here, to $m>1$ walkers.

{\bf Case of ${\boldsymbol{m>1}}$ data walkers:} Consider a scenario characterized by the same classical address $\boldsymbol{a} = a_1\,a_2=10$, where the data register $D$ is now a generic $m$-particle one. In this case, the algorithm works as described above in the case of a single data walker, with the difference of having $m$ walkers receiving the same action of the gates obtained in the example explained above. In particular, the corresponding initial state is:
\begin{equation*}
\ket{\psi_{in}}=\ket{\red^{1,1}}^{\otimes m}_{D}\ket{\zero^{1} }_{A_2}\ket{\red^{1,1}}_{A_1} = \ket{\red^{1,1}}_{D_1}\ket{\red^{1,1}}_{D_2}\dots\ket{\red^{1,1}}_{D_m}\ket{\zero^{1} }_{A_2}\ket{\red^{1,1}}_{A_1} \, .
\end{equation*}
The action of $\hat U^{(i)}$ is similar as before, where now all the data particles are target of its action, while $\hat S$ acts identically. Hence, is this scenario, the state of the walkers arriving at the $d=3$ level of the memory cells is:
\begin{equation*}
\ket{\psi_4} = \hat S  \ket{\psi_3} = \ket{\red^{3,3}}^{\otimes m}_{D}\ket{\emptyset^{3} }_{A_2}\ket{R^{3,1}}_{A_1}  \, ,
\end{equation*}
where all the data particles $D=\{D_1,D_2,\dots,D_m\}$ reach the correct memory cell positioned at $(d,l) = (3,3) $ and labeled by $ \boldsymbol{a}=10$.

\subsubsection*{Message copy}
 
We apply the first method explained in Section~\ref{SecMessCop} to write down the message into the data register of the state $\ket{\psi_4}$ written in Eq.~(\ref{psi4D0D1}). Hence, we imagine that in the memory cell $x_{10}$ is contained the message $\boldsymbol{b}^{\boldsymbol{(10)}}=1$. It follows that, by using the instrument of Section~\ref{SecMessCop}, we obtain:
\begin{eqnarray}\nonumber
\ket{\psi_Q} &=& U^{(10)}_{copy} \, \ket{\psi_4} = \ket{R^{\boldsymbol{3,3}}}_{D_0}\bra{R^{\boldsymbol{3,3}}}\otimes \hat{U}_{copy}^{loc}(M_1^{3,3},D_1) \otimes I_{A_2A_1}
\left(\ket{R^{3,3}}_{D_1}\ket{R^{3,3}}_{D_0}\ket{\emptyset^{3} }_{A_2}\ket{R^{3,1}}_{A_1}\right)
 \\  \label{apex} &=&\ket{R^{3',3}}_{D_1}\ket{R^{3',3}}_{D_0}\ket{\emptyset^{3'} }_{A_2}\ket{R^{3',1}}_{A_1} \simeq \ket{\psi_4} \, ,
\end{eqnarray}
where we used the symbol $\simeq$ to emphasize that, even if the internal state of the walkers is the same for $\ket{\psi_4}$ and $\ket{\psi_Q}$, the positions are different. Indeed, during the inverse routing phase, we write the values of depth $d$ with the symbol~$'$, as shown in Eq.~(\ref{apex}) and on the right side of Fig.~\ref{rout_qram_cl}. We are now ready to explain the inverse routing phase.

\subsubsection*{Inverse routing}

We exploit the tools introduced in Section~\ref{SecInvRout} to route the quantum walkers from the output of the memory cells ($d=3'$) to the final output port $(d,l)=(1',1)$ of the qRAM. 

{\bf Steps 5 and 6:} We apply the gate $\hat S^\dagger$ to $\ket{\psi_Q}$ and we route all the particles from depth $d=3'$ to $d=2'$ without any change of color (see Fig.~\ref{rout_qram_cl}):
\begin{equation*}
\ket{\psi_5} = \hat S^\dagger \ket{\psi_Q} = \ket{\red^{2',2}}_{D_1}\ket{\red^{2',2}}_{D_0}\ket{\zero^{2'}}_{A_2}\ket{\red^{2',1}}_{A_1} \hspace{1cm} d=2',\,\mbox{ post-}\hat S^\dagger \mbox{ gate, \,pre-}\hat U^{(2)}\mbox{ gate} \, .
\end{equation*}
Since $A_2$ is empty, the control-gate $\hat U^{(2)}$ present at this depth has no effect on the walkers state. Hence, the state just before the last scattering gate $\hat S^\dagger$ is:
 \begin{equation*}
\ket{\psi_6} = \hat U^{(2)} \ket{\psi_5} = \ket{\red^{2',2}}_{D_1}\ket{\red^{2',2}}_{D_0}\ket{\emptyset^{2'}}_{A_2}\ket{\red^{2',1}}_{A_1} 
\hspace{1cm} d=2',\,\mbox{ post-}\hat U^{(2)} \mbox{ gate, \,pre-}\hat S^\dagger\mbox{ gate} \, . 
\end{equation*}

{\bf Steps 7 and 8}:
The last $\hat S^\dagger$ gate routes the walkers from the depth $d=2'$ to $d=1'$ and changes the color of the data walkers $D_0 D_1$ from red to blue, while the address walker $A_1$ remains red (see Fig.~\ref{rout_qram_cl} and Eq.~(\ref{Sdagga})). 
\begin{equation*}
\ket{\psi_7} = \hat S^\dagger \ket{\psi_6} = \ket{\blue^{1',1}}_{D_1}\ket{\blue^{1',1}}_{D_0}\ket{\emptyset^{1'}}_{A_2}\ket{\red^{1',1}}_{A_1} \hspace{1cm} d=1',\,\mbox{ post-}\hat S^\dagger \mbox{ gate, \,pre-}\hat U^{(1)}\mbox{ gate} \, .  .
\end{equation*}
Now, the last step of the protocol is to apply $\hat U^{(1)}$, which is a control-gate operation with the control on $A_1$, which in this case is a red particle. Hence, the gate is activated with targets on $D1,\,D_0$ and $A_2$ and we obtain:
\begin{equation*}
\ket{\psi_{out}} = \hat U^{(1)} \ket{\psi_7} = \ket{\red^{1',1}}_{D_1}\ket{\red^{1',1}}_{D_0}\ket{\emptyset^{1'}}_{A_2}\ket{R^{1',1}}_{A_1} \, .
\end{equation*}

\subsection{Example: functioning of the qRAM for an entangled address}
We exploit what was introduced in the previous section to explain how the qRAM works if we send as input a superposition of two addresses, namely $\boldsymbol{a}= a_1 \, a_2=0\,0$ and $\boldsymbol{a}'= a_1'\,a_2'=1\,1$. The initial quantum states corresponding to such addresses are, respectively,
$\ket{\boldsymbol{a}}=\ket{\emptyset^1}_{A_2}\ket{\emptyset^1}_{A_1}$ and $\ket{\boldsymbol{a'}}=\ket{R^{1,1}}_{A_2}\ket{R^{1,1}}_{A_1}$, where again the kets are ordered from the least significant (left) to the most significant (right), to follow the order in which they are physically placed in the qRAM. Notice that the resulting state is a maximally entangled superposition of addresses. Following the superposition of addresses, there is the data register $D$, which is made up of $m=1$ quantum walkers initialized in the red state $\ket{R^{1,1}}_D$. The protocol steps described here are graphically represented in Fig. ~\ref{rout_qram_ent}.
\subsubsection{Routing}
The overall $A+D$ state sent as input for the algorithm is given by
\begin{equation*}
    \ket{\psi_{in}}={\ket{\red^{1,1}}_{D_1}\left(\frac{\ket{\zero^{1} }_{A_2}\ket{\zero^1}_{A_1}+\ket{\red^{1,1}}_{A_2}\ket{\red^{1,1}}_{A_1}}{\sqrt{2}}\right)} \hspace{1cm} d=1,\,\mbox{ pre-}\hat U^{(1)}\mbox{ gate}.
\end{equation*}
{\bf Steps 1 and 2:} We sequentially insert the particles in the binary tree of our qRAM. The first gate acting on them is $\hat{U}^{(1)}$, which takes as control the first particle $A_1$ of the sequence.
Since the address particles are in a superposition, the action of $\hat{U}^{(1)}$ will differ depending on the components of the state. In the first component, the first address particle is in the vacuum state $\ket{\zero^1}_{A_1}$ (no particle is present), so $\hat{U}^{(1)}$ acts trivially and does not affect the subsequent particles. In the second component, $A_1$ is in the state $\ket{\red^{1,1}}_{A_1}$, meaning a particle is present, and therefore a Pauli $\hat{X}$ gate is applied to the states that follow.
Hence, after the application of the gate, we obtain the state
\begin{equation*}  \ket{\psi_1}=\hat{U}^{(1)}\ket{\psi_{in}}=\frac{\ket{\red^{1,1}}_{D_1}\ket{\zero^{1}}_{A_2}\ket{\emptyset^1}_{A_1}+\ket{\blue^{1,1}}_{D_1}\ket{\blue^{1,1}}_{A_2}\ket{R^{1,1}}_{A_1}}{\sqrt{2}}\hspace{1cm} d=1,\,\mbox{ post-}\hat U^{(1)}\mbox{ gate, \,pre-}\hat S\mbox{ gate}\,.
\end{equation*}
After, we apply the gate $\hat{S}$ to the train of particles. Remember that red particles are routed left and blue particles right, together with a change of color. We obtain
\begin{equation*} \ket{\psi_2}=\hat{S}\ket{\psi_{1}}=\frac{\ket{\red^{2,1}}_{D_1}\ket{\zero^{2}}_{A_2}\ket{\emptyset^2}_{A_1}+\ket{\red^{2,2}}_{D_1}\ket{\red^{2,2}}_{A_2}\ket{R^{2,1}}_{A_1}}{\sqrt{2}}\hspace{1cm} d=2,\,\mbox{ pre-}\hat U^{(2)}\mbox{ gate}\,.
\end{equation*}
{\bf Steps 3 and 4}: 
We apply $\hat{U}^{(2)}$, which uses the second address particle $A_2$ as a control. As before, the operation differs between the two components of the superposition: in the first component, no particle is present in $A_2$, so the gate remains inactive and no operation is applied to $D_1$; in the second component, $A_2$ is red, and a Pauli $\hat{X}$ gate is applied to $D_1$.
The state after the application is
\begin{equation*}   \ket{\psi_3}=\hat{U}^{(2)}\ket{\psi_{2}}=\frac{\ket{\red^{2,1}}_{D_1}\ket{\emptyset^{2}}_{A_2}\ket{\emptyset^2}_{A_1}+\ket{\blue^{2,2}}_{D_1}\ket{R^{2,2}}_{A_2}\ket{R^{2,1}}_{A_1}}{\sqrt{2}}\hspace{1cm} d=2,\,\mbox{ post-}\hat U^{(2)}\mbox{ gate, \,pre-}\hat S\mbox{ gate}\,.
\end{equation*}
In the last routing step, the particles pass through the scattering gate $\hat S$, placed before the memory cells:
\begin{equation*}   \ket{\psi_4}=\hat{S}\ket{\psi_{3}}=\frac{\ket{\red^{3,1}}_{D_1}\ket{\emptyset^{3}}_{A_2}\ket{\emptyset^3}_{A_1}+\ket{\red^{3,4}}_{D_1}\ket{R^{3,3}}_{A_2}\ket{R^{3,1}}_{A_1}}{\sqrt{2}}\hspace{1cm} d=3,\,\mbox{ pre-memory cell} \, .
\end{equation*}
We can see in Fig.~\ref{rout_qram_ent}  that the data particle has correctly reached the position of the targeted memory cells associated with the addresses $\boldsymbol{a}=0\,0$, corresponding to position $(d,l)=(3,1)$, and $\boldsymbol{a}'=1\,1$, corresponding to position $(d,l)=(3,4)$.
Recall again that, for simplicity, we considered the case of $m=1$ data walker; however, the procedure can be easily generalized for the case $m>1$. In particular, for the copy procedure, we need at least one additional data walker $D_0$. Thus, we consider as a final state for the routing procedure 
\begin{equation}\label{exsuperposprecopy}\ket{\psi_4}=\frac{\ket{\red^{3,1}}_{D_1}\ket{\red^{3,1}}_{D_0}\ket{\emptyset^{3}}_{A_2}\ket{\emptyset^3}_{A_1}+\ket{\red^{3,4}}_{D_1}\ket{\red^{3,4}}_{D_0}\ket{R^{3,3}}_{A_2}\ket{R^{3,1}}_{A_1}}{\sqrt{2}}\hspace{1cm} d=3,\,\mbox{ pre-memory cell} \, .
\end{equation}

\begin{figure}[t]
\begin{tikzpicture}[scale=0.7, transform shape,
  level 1/.style={level distance=4cm, sibling distance=8cm},
  level 2/.style={level distance=4cm, sibling distance=4cm},
  grow=east, 
  edge from parent/.style={draw, thick}, 
  every node/.style={rectangle, draw, minimum size=6mm,fill=cyan!30}, 
  leaf/.style={rectangle, draw, fill=gray!30, minimum size=6mm}, 
  block/.style={rectangle, draw, fill=blue!20, minimum width=0.7cm, minimum height=5mm, rotate around={0:(0,0)}}, 
  gateX/.style={rectangle, draw, fill=green!20, minimum size=5mm, rotate around={0:(0,0)}} 
]

\node (root1) {$\hat{S}$}
  child {
    node {$\hat{S}$}
    child { 
      node[leaf] {$\boldsymbol{x}_3$} 
      edge from parent {}
    }
    child { 
      node[leaf] {$\boldsymbol{x}_2$} 
      edge from parent{
      }
    }
    edge from parent {
      node[midway,xshift=0.4cm, yshift=-0.4cm,  block, rotate=-45] {$\hat{U}^{(2)}$} 
    }
  }
  child {
    node {$\hat{S}$}
    child { 
      node[leaf] {$\boldsymbol{x}_1$} 
      edge from parent {}
    }
    child { 
      node[leaf] {$\boldsymbol{x}_0$} 
      edge from parent{ 
      }
    }
    edge from parent {
      node[midway, xshift=0.4cm,yshift=0.4cm, block, rotate=45] {$\hat{U}^{(2)}$} 
    }
  };
\node (root2) at (16,0) {$\hat{S}^{\dagger}$}
[grow=left]
  child {
    node {$\hat{S}^{\dagger}$}
    child { 
      node[leaf] {$\boldsymbol{x}_{11}$} 
      edge from parent {
      }
    }
    child { 
      node[leaf] {$\boldsymbol{x}_{10}$} 
      edge from parent {}
    }
    edge from parent {
      node(Ugate)[midway,xshift=-0.4cm,yshift=0.4cm, block, rotate=-45] {$\hat{U}^{(2)}$} 
    }
  }
  child {
    node {$\hat{S}^{\dagger}$}
    child { 
      node[leaf] {$\boldsymbol{x}_{01}$} 
      edge from parent {
      }
    }
    child { 
      node[leaf] {$\boldsymbol{x}_{00}$} 
      edge from parent {}
    }
    edge from parent {
      node[midway, xshift=-0.4cm, yshift=-0.4cm, block, rotate=45] {$\hat{U}^{(2)}$} 
    }
  };

\draw[thick] (-4, 0) -- (root1) node(Ugateroot)[midway, block, rotate=0] {$\hat{U}^{(1)}$};
\draw[thick] (root2) -- (20,0) node[midway, block] {$\hat{U}^{(1)}$};



\draw[thick, fill=red] (-3.6,0) circle (3pt);
\draw[thick, fill=white] (-3.3,0) circle (3pt);
\halfcoloredcircle{-3.3}{0}{white}{red}
\halfcoloredcircle{-3}{0}{white}{red}


\halfcoloredcircle{-1.4}{0}{red}{blue}
\halfcoloredcircle{-1.1}{0}{white}{blue}
\halfcoloredcircle{-0.8}{0}{white}{red}



\draw[thick, fill=white] (1.7,1.7) circle (3pt);
\draw[thick,fill=red] (1.5,1.5) circle (3pt);
\draw[thick,fill=red] (1.3,1.3) circle (3pt);

\draw[thick,fill=white] (3.4,3.4) circle (3pt);
\draw[thick,fill=red] (3.2,3.2) circle (3pt);
\draw[thick,fill=blue] (3,3) circle (3pt);


\draw[thick, fill=red] (1.7,-1.7) circle (3pt);
\draw[thick,fill=white] (1.5,-1.5) circle (3pt);
\draw[thick,fill=red] (1.3,-1.3) circle (3pt);

\draw[thick,fill=red] (3.4,-3.4) circle (3pt);
\draw[thick,fill=white] (3.2,-3.2) circle (3pt);
\draw[thick,fill=red] (3,-3) circle (3pt);


\draw[thick,fill=red] (6,5) circle (3pt);
\draw[thick,fill=white] (6.25,5.125) circle (3pt);
\draw[thick,fill=white] (6.5,5.25) circle (3pt);


\draw[thick, fill=white] (6,3) circle (3pt);
\draw[thick, fill=red] (6.25,2.875) circle (3pt);
\draw[thick,fill=white] (6.5,2.745) circle (3pt);




\draw[thick,fill=red] (6,-5) circle (3pt);
\draw[thick,fill=white] (6.25,-5.125) circle (3pt);
\draw[thick, fill=red] (6.5,-5.25) circle (3pt);


\draw[thick,fill=white] (9.5,5.25) circle (3pt);
\draw[thick,fill=white] (9.75,5.125) circle (3pt);
\draw[thick,fill=white] (10,5) circle (3pt);


\draw[thick, fill=white] (9.5,2.745) circle (3pt);
\draw[thick, fill=red] (9.75,2.875) circle (3pt);
\draw[thick,fill=white] (10,3) circle (3pt);




\draw[thick,fill=red] (9.5,-5.25) circle (3pt);
\draw[thick,fill=white] (9.75,-5.125) circle (3pt);
\draw[thick, fill=red] (10,-5) circle (3pt);


\draw[thick, fill=white] (14.3,1.7) circle (3pt);
\draw[thick,fill=red] (14.5,1.5) circle (3pt);
\draw[thick,fill=white] (14.7,1.3) circle (3pt);

\draw[thick, fill=white] (12.6,3.4) circle (3pt);
\draw[thick,fill=red] (12.8,3.2) circle (3pt);
\draw[thick,fill=white] (13,3) circle (3pt);


\draw[thick,fill=red] (14.3,-1.7) circle (3pt);
\draw[thick,fill=white] (14.5,-1.5) circle (3pt);
\draw[thick, fill=red] (14.7,-1.3) circle (3pt);

\draw[thick,fill=red] (12.6,-3.4) circle (3pt);
\draw[thick,fill=white] (12.8,-3.2) circle (3pt);
\draw[thick, fill=red] (13,-3) circle (3pt);


\halfcoloredcircle{19.6}{0}{white}{red}
\halfcoloredcircle{19.3}{0}{white}{red}
\halfcoloredcircle{19}{0}{red}{white}

\halfcoloredcircle{17.4}{0}{white}{red}
\halfcoloredcircle{17.1}{0}{white}{blue}
\halfcoloredcircle{16.8}{0}{red}{white}


\draw[dashed] (-2.7,6.5)--(-2.7,-6.5) node[fill=none,draw=none,left,anchor=east]
{};
\draw[dashed] (-1.65,6.5)--(-1.65,-6.5) node[fill=none,draw=none,left,anchor=east]
{};
\fill[gray!20, opacity=0.5] (-2.7,6.5) -- (-1.65,6.5)  -- (-1.65,-6.5) -- (-2.7,-6.5)  -- cycle;
\node[draw=none, fill=none] at (-2.1,-6.7){ Step 1};

\draw[->,-Stealth] (-1,7)-- (-1,6){};

\node[fill=none,draw=none, fill=none] at (-1,7.5){\Large$\ket{\psi_1}$};

\draw[dashed] (-0.5,6.5)--(-0.5,-6.5) node[fill=none,draw=none,left,anchor=east]
{};
\draw[dashed] (1.15,6.5)--(1.15,-6.5) node[fill=none,draw=none,left,anchor=east]
{};
\fill[gray!20, opacity=0.5] (-0.5,6.5)  --  (1.15,6.5)--(1.15,-6.5) -- (-0.5,-6.5) -- cycle;
\node[fill=none,draw=none] at (0.4,-6.7){Step 2};

\draw[->,-Stealth] (1.5,7)-- (1.5,6){};

\node[fill=none,draw=none] at (1.5,7.5){\Large$\ket{\psi_2}$};

\draw[dashed] (1.9,6.5)--(1.9,-6.5) node[fill=none,draw=none,left,anchor=east]{};
\draw[dashed] (2.85,6.5)--(2.85,-6.5) node[fill=none,draw=none,left,anchor= east]{};
\fill[gray!20, opacity=0.5] (1.9,6.5)  --  (2.85,6.5)--(2.85,-6.5) -- (1.9,-6.5) -- cycle;
\node[fill=none,draw=none] at (2.4,-6.7){Step 3};

\draw[->,-Stealth] (3.2,7)-- (3.2,6){};

\node[fill=none,draw=none] at (3.2,7.5){\Large$\ket{\psi_3}$};

\draw[dashed] (3.6,6.5)--(3.6,-6.5) node[fill=none,draw=none,left,anchor=east]
{};
\draw[dashed] (5.7,6.5)--(5.7,-6.5) node[fill=none,draw=none,left,anchor=east]
{};
\fill[gray!20, opacity=0.5] (3.6,6.5)  --  (5.7,6.5)--(5.7,-6.5) -- (3.6,-6.5) -- cycle;
\node[fill=none,draw=none] at (4.7,-6.7){Step 4};

\draw[->,-Stealth] (6.2,7)-- (6.2,6){};

\node[fill=none,draw=none] at (6.2,7.5){\Large$\ket{\psi_4}$};

\draw[dashed,] (6.8,6.5)--(6.8,-6.5) node[fill=none,draw=none,left,anchor=east]
{};
\draw[dashed,] (9.2,6.5)--(9.2,-6.5) node[fill=none,draw=none,left,anchor=east]
{};
\fill[gray!20, opacity=0.5] (6.8,6.5)  --  (9.2,6.5)--(9.2,-6.5) -- (6.8,-6.5) -- cycle;
\node[fill=none,draw=none] at (8,-6.7){Querying};

\draw[->,-Stealth] (9.8,7)-- (9.8,6){};

\node[fill=none,draw=none] at (9.8,7.5){\Large$\ket{\psi_Q}$};

\draw[dashed] (10.3,6.5)--(10.3,-6.5) node[fill=none,draw=none,left,anchor=east]
{};
\draw[dashed] (12.4,6.5)--(12.4,-6.5) node[fill=none,draw=none,left,anchor=east]
{};
\fill[gray!20, opacity=0.5] (10.3,6.5)  --  (12.4,6.5)--(12.4,-6.5) -- (10.3,-6.5) -- cycle;
\node[fill=none,draw=none] at (11.4,-6.7){Step 5};

\draw[->,-Stealth] (12.7,7)-- (12.7,6){};

\node[fill=none,draw=none] at (12.7,7.5){\Large$\ket{\psi_5}$};

\draw[dashed] (13.15,6.5)--(13.15,-6.5) node[fill=none,draw=none,left,anchor=east]{};
\draw[dashed] (14.1,6.5)--(14.1,-6.5) node[fill=none,draw=none,left,anchor=east]{};
\fill[gray!20, opacity=0.5] (13.15,6.5)  --  (14.1,6.5)--(14.1,-6.5) -- (13.15,-6.5) -- cycle;

\node[fill=none,draw=none] at (13.7,-6.7){Step 6};

\draw[->,-Stealth] (14.5,7)-- (14.5,6){};

\node[fill=none,draw=none] at (14.5,7.5){\Large$\ket{\psi_6}$};

\draw[dashed] (14.85,6.5)--(14.85,-6.5) node[fill=none,draw=none,left,anchor=east]
{};
\draw[dashed] (16.5,6.5)--(16.5,-6.5) node[fill=none,draw=none,left,anchor=east]
{};
\fill[gray!20, opacity=0.5] (14.85,6.5)  --  (16.5,6.5)--(16.5,-6.5) -- (14.85,-6.5) -- cycle;
\node[fill=none,draw=none] at (15.7,-6.7){Step 7};

\draw[->,-Stealth] (17,7)-- (17,6){};

\node[fill=none,draw=none] at (17,7.5){\Large$\ket{\psi_7}$};

\draw[dashed] (17.65,6.5)--(17.65,-6.5) node[fill=none,draw=none,left,anchor=east]
{};
\draw[dashed] (18.7,6.5)--(18.7,-6.5) node[fill=none,draw=none,left,anchor=east]
{};
\fill[gray!20, opacity=0.5] (17.65,6.5)  --  (18.7,6.5)--(18.7,-6.5) -- (17.65,-6.5) -- cycle;
\node[fill=none,draw=none] at (18.2,-6.7){Step 8};


\node[draw=none, fill=none] at (-3,-0.8){\scalebox{1.2}{$(d,l)=(1,1)$}};

\node[draw=none, fill=none, rotate=-45] at (2,-1){\scalebox{1.2}{$(2,1)$}};

\node[draw=none, fill=none, rotate=45] at (2,1){\scalebox{1.2}{$(2,2)$}};

\node[draw=none, fill=none,rotate=-22.5] at (5,-5){\scalebox{1.2}{$(3,1)$}};

\node[draw=none, fill=none, rotate=22.5] at (5,-2.9){\scalebox{1.2}{$(3,2)$}};

\node[draw=none, fill=none, rotate=-22.5] at (5,2.9){\scalebox{1.2}{$(3,3)$}};

\node[draw=none, fill=none, rotate=22.5] at (5,5){\scalebox{1.2}{$(3,4)$}};


\node[draw=none, fill=none,rotate=22.5] at (11,-5){\scalebox{1.2}{$(3',1)$}};

\node[draw=none, fill=none, rotate=-22.5] at (11,-2.9){\scalebox{1.2}{$(3',2)$}};

\node[draw=none, fill=none, rotate=22.5] at (11,2.9){\scalebox{1.2}{$(3',3)$}};

\node[draw=none, fill=none, rotate=-22.5] at (11,5){\scalebox{1.2}{$(3',4)$}};

\node[draw=none, fill=none, rotate=45] at (14,-1){\scalebox{1.2}{$(2',1)$}};

\node[draw=none, fill=none, rotate=-45] at (14,1){\scalebox{1.2}{$(2',2)$}};

\node[draw=none, fill=none] at (19,-0.8){\scalebox{1.2}{$(1',1)$}};

\draw[->, >={Stealth[length=6pt]}, very thick] (-4,0.5) -- (-3,0.5)node[fill=none,draw=none,midway,above] {\Large$\ket{\psi_{in}}$};
\draw[->, >={Stealth[length=6pt]}, very thick] (19,0.5) -- (20,0.5)node[fill=none,draw=none,midway,above] {\Large$\ket{\psi_{out}}$};

\end{tikzpicture}
\caption{
Step-by-step evolution of the qRAM protocol in the case of a maximally entangled superposition of addresses.
The figure illustrates the routing dynamics through successive applications of the gates $\hat{U}^{(d)}$ and $\hat{S}$, followed by the message retrieval and inverse routing phases.
The walker states are shown as they propagate from the root node to the memory cells and back.}

\label{rout_qram_ent}
\end{figure}

\subsubsection{Message copy}
To write the message contained in the cells $x_{00}$ and $x_{11}$ in $\ket{\psi_4}$, i.e. Eq. ~(\ref{exsuperposprecopy}), we use the copy method defined by Eq. ~(\ref{Ucopya}). Suppose that the memory cell $x_{00}$ contains the message $\boldsymbol{b}^{(00)}=1$ and the memory cell $x_{11}$ contains the message $\boldsymbol{b}^{(11)}=0$. Using the tools explained in Section ~\ref{SecMessCop} we obtain:
\begin{align}\nonumber
\ket{\psi_Q}&=\left(\hat{U}_{copy}^{(00)}\otimes I^{(01)}\otimes I^{(10)}\otimes\hat{U}_{copy}^{(11)}\right)\ket{\psi_4}\\ \nonumber&
=\left[\ket{R^{3,1}}_{D_0}\bra{R^{3,1}}\otimes \hat{U}_{copy}^{loc}\left(M_1^{3,1},D_1\right)\right] \otimes I^{(01)} \otimes I^{(10)} \otimes\left[\ket{R^{3,4}}_{D_0}\bra{R^{3,4}}\otimes \hat{U}_{copy}^{loc}\left(M^{3,4}_1,D_1\right)\right]\otimes \hat{I}_{A_2,A_1}\ket{\psi_4}\\\nonumber
&=\frac{\ket{R^{3',1}}_{D_1}\ket{R^{3',1}}_{D_0}\ket{\emptyset^{3'}}_{A_2}\ket{\emptyset^{3'}}_{A_1}+\ket{\emptyset^{3'}}_{D_1}\ket{R^{3',4}}_{D_0}\ket{R^{3',3}}_{A_2}\ket{R^{3',1}}_{A_1}}{\sqrt{2}}.
\end{align}
\subsubsection{Inverse routing}
We will now explain the steps to recollect the quantum walkers from the memory cells to the output node of the qRAM tree.
{\bf Steps 5 and 6:} We apply the gate $\hat{S}^{\dagger}$ to the state $\ket{\psi_Q}$, to route the particles from $d=3'$ to $d=2'$. Red particles with an odd leaf index $l$ are routed from level $d$ to level $d-1$ without any color change. In contrast, red particles with an even index are also routed from level $d$ to $d-1$, but their color changes from red to blue. For clarity, we will omit again the walker $D_0$ used during the copy procedure from the walker's state.  After the application of the gate we get:
\begin{equation*}    \ket{\psi_5}=\hat{S}^{\dagger}\ket{\psi_Q}=\frac{\ket{\red^{2',1}}_{D_1}\ket{\zero^{2'}}_{A_2}\ket{\zero^{2'}}_{A_1}+\ket{\zero^{2'}}_{D_1}\ket{\red^{2',2}}_{A_2}\ket{\red^{2',1}}_{A_1}}{\sqrt{2}}\hspace{0.5cm} d=2',\,\mbox{ post-}\hat S^\dagger \mbox{ gate, \,pre-}\hat U^{(2)}\mbox{ gate} \, .
\end{equation*}
Now we apply the gate $\hat{U}^{(2)}$ present at this depth. Since in the first component of the state $A_2$ is absent, no action will be performed on $D_1$. In the second part of the state $A_2$ is in color red, thus the gate will act on $D_1$, which, however, is absent. Thus, the overall effect makes the state stay the same:
\begin{equation*}    \ket{\psi_6}=\hat{U}^{(2)}\ket{\psi_5}=\frac{\ket{\red^{2',1}}_{D_1}\ket{\emptyset^{2'}}_{A_2}\ket{\zero^{2'}}_{A_1}+\ket{\zero^{2'}}_{D_1}\ket{R^{2',2}}_{A_2}\ket{\red^{2',1}}_{A_1}}{\sqrt{2}}\hspace{0.5cm} d=2',\,\mbox{ post-}\hat U^{(2)} \mbox{ gate, \,pre-}\hat S^\dagger\mbox{ gate} \, .
\end{equation*}
{\bf Steps 7 and 8}: The last gate $\hat{S}^{\dagger}$ routes the walkers from $d=2'$ to $d=1'$, changing the color of $A_2$ in the second component of the superposition:
\begin{equation*}   \ket{\psi_7}=\hat{S}^{\dagger}\ket{\psi_6}=\frac{\ket{\red^{1',1}}_{D_1}\ket{\emptyset^{1'}}_{A_2}\ket{\zero^{1'}}_{A_1}+\ket{\zero^{1'}}_{D_1}\ket{B^{1',1}}_{A_2}\ket{\red^{1',1}}_{A_1}}{\sqrt{2}} \hspace{0.5cm} d=1',\,\mbox{ post-}\hat S^\dagger \mbox{ gate, \,pre-}\hat U^{(1)}\mbox{ gate} \, . 
\end{equation*}
The final step is the application of $\hat{U}^{(1)}$, which uses the first address walker $A_1$ as a control. Once again, the first component of the superposition remains unaffected, as $A_1$ is absent. In the second component, however, $A_1$ is in the red state, triggering a Pauli $\hat{X}$ gate on both $A_2$ and $D_1$. We finally obtain the output state:
\begin{equation*}   \ket{\psi_{out}}=\frac{\ket{\red^{1',1}}_{D_1}\ket{\emptyset^{1'}}_{A_2}\ket{\emptyset^{1'}}_{A_1}+\ket{\zero^{1'}}_{D_1}\ket{R^{1',1}}_{A_2}\ket{R^{1',1}}_{A_1}}{\sqrt{2}}.
\end{equation*}

\section{Backup variant qRAM}\label{SecBackup}

We now describe the backup variant of our qRAM. The main difference of this variant is that each subsystem the qRAM registers described before, namely the address $A=\{A_1,A_2,\dots,A_n\}$ and the data $D=\{D_0,D_1,\dots,D_m\}$ registers, has a backup subsystem next to it. Hence, the backup of each $A_i$ is written as $\tilde A_i$ and similarly for the data particles. The address register used in this variant is  $A=\{A_1, \tilde A_1,A_2,\tilde A_2,\dots A_n,\tilde A_n\}$ and similarly the data register is $D=\{D_1,\tilde D_1,\dots, D_{m-1},\tilde D_{m-1},D_m\}$, where we have no $D_0$ and $D_m$ has no backup walker. 

A main feature of this variant is that each backup walker is initialized in the state $\ket{R}$. Hence, all the data and the backup walkers are initialized as red walkers, while the (non-backup) address walkers have the same encoding $a_i = 0 \rightarrow \ket{\emptyset}_{A_i}$, $a_i = 1 \rightarrow \ket{R}_{A_i}$ used before. In case of an $n=2$ and $m=2$ scenario, the initial state considered in this variant is:
\begin{equation}\label{psiinback}
    \ket{\psi_{in}} = \ket{R}_{D_2}\ket{R}_{\tilde D_1}\ket{R}_{D_1} \ket{R}_{\tilde A_2}
    \ket{a_2}_{A_2}
    \ket{R}_{\tilde A_1}
    \ket{a_1}_{A_1}
\end{equation}
Similarly to the previous scheme, we inject the carriers in the following sequence
 $\{D_m,\dots,\tilde D_1,D_1,\tilde A_n,A_n,\dots,\tilde A_1,A_1\}$, where $A_1$ ($D_m$) is the first (last) to be inserted.

The main reason to introduce this backup variant, which roughly doubles the total number of carriers, is to simplify the long-range gate $\hat U^{(d)}$ used by the standard method to distribute the address information.  
The backup scheme allows a decomposition of $\hat U^{(d)}$  into short-range blocks that can be applied repeatedly. Concerning the scattering gates $\hat S$ and $\hat S^\dagger$, no modifications are required.

{In order to implement this version of the algorithm, it is sufficient to consider the procedure presented in the previous section for the standard long-range qRAM while taking care of the modifications proposed below for $\hat U^{(d)}$ and the copy of the message into the data walkers. Therefore, we do not go through another step-by-step example and we focus on exposing the details of this alternative variant. }

\subsection{Decomposition of $\hat{U}^{(d)}$ in short-range building blocks}\label{BackupGates}

The main advantage of the backup protocol concerns the role of the gate $\hat{U}^{(d)}$, which, in the model presented above, was defined through long-range interactions. In this variant, we decompose its action
through an initial operation $\hat{U}_{in}$ applied once followed by a block gate $\hat{U}_B$ applied a number of times equal to half of the particles following $A_d$, namely $n+m-(d+1)$ times, as the particles flow through its sites. {We underline that, unlike all the other controlled gates of this work, $U_B$ is activated when the control particle is blue.}

First, we remind that that $\hat U^{(d)}$ used the $d$-th address walker as control to flip the color of all the following walkers. 
The first operation $\hat{U}_{in}$ is a control-NOT  operation which, in case $A_n$ is red, flips the color of  $\tilde A_n$ to blue. 
Thus, the gate takes the form (in standard circuit representation):

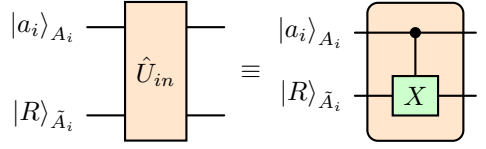
\begin{equation*}
 \begin{quantikz}
     \lstick{$\ket{a_i}_{A_i}$}& \gate[2, style={fill=orange!20}, background]{\hat{U}_{in}} & \\
     \lstick{$\ket{R}_{\tilde A_i}$}& & 
 \end{quantikz}
 \equiv
\begin{quantikz}
        \lstick{$\ket{a_i}_{A_i}$} &   \ctrl{1}\gategroup[wires=2,steps=1,style={rounded corners,inner sep=6pt, fill=orange!20},background]{} &\\
        \lstick{$\ket{R}_{\tilde A_i}$} &   \gate[1,style={fill=green!20},background]{X} &
\end{quantikz}.
\end{equation*}
where $a_i$ is the state given to $A_i$ by the encoding described above.

The second building block is the gate $\hat{U}_B$, which can be seen as a double-target controlled gate control-NOT-NOT, where the control is a backup walker $\ket{\tilde w_i}_{\tilde{W}_i}$ and the two targets are (i) the following (non-backup) walker $\ket{ w_i}_{W_{i+1}}$ and its corresponding backup $\ket{\tilde w_i}_{\tilde{W}_{i+1}}$, where $W_i\in\{A_1,\dots,A_n,D_1,\dots,D_{m}\}$, $\tilde W_i$ is its correspondent backup and $w_i$ and $\tilde w_i$ are their corresponding walker internal states.
The gate takes the form (in standard circuit representation):

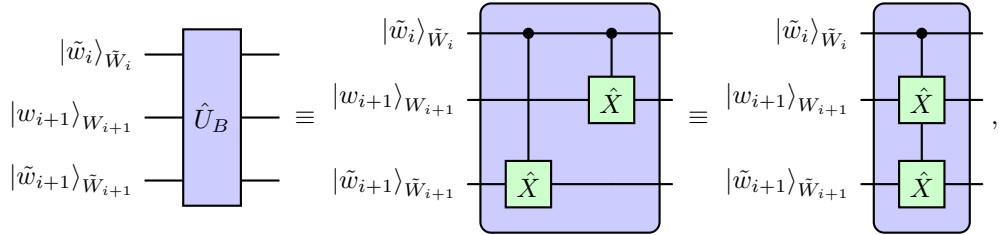
\begin{equation*}
 \begin{quantikz}[row sep =0.5cm]
     \lstick{$\ket{\tilde w_i}_{\tilde W_i}$}& \gate[3, style={fill=blue!20}, background]{\hat{U}_{B}} & \\
     \lstick{$\ket{w_{i+1}}_{W_{i+1}}$}& & \\
    \lstick{$\ket{\tilde w_{i+1}}_{\tilde W_{i+1}}$} & & 
 \end{quantikz}
 \equiv
    \begin{quantikz}[row sep=0.5cm]
        \lstick{$\ket{\tilde w_i}_{\tilde W_i}$} & \ctrl{2}\gategroup[wires=3,steps=2,style={rounded corners,inner sep=6pt, fill=blue!20},background]{} & \ctrl{1} & \\
        \lstick{$\ket{w_{i+1}}_{W_{i+1}}$} &  & \gate[1,style={fill=green!20},background]{\hat{X}} & \\
        \lstick{$\ket{\tilde w_{i+1}}_{\tilde W_{i+1}}$} & \gate[1,style={fill=green!20},background]{\hat{X}} & &  
    \end{quantikz}
    \equiv
\begin{quantikz}[row sep=0.5cm]
        \lstick{$\ket{\tilde w_i}_{\tilde W_i}$} & \ctrl{2}\gategroup[wires=3,steps=1,style={rounded corners,inner sep=6pt, fill=blue!20},background]{} & \\
        \lstick{$\ket{w_{i+1}}_{W_{i+1}}$} & \gate[1,style={fill=green!20},background]{\hat{X}} & \\
        \lstick{$\ket{\tilde w_{i+1}}_{\tilde W_{i+1}}$} & \gate[1,style={fill=green!20},background]{\hat{X}} &   
    \end{quantikz},
\end{equation*}
{where this controlled operation is activated if the backup control walker is colored blue.}

These two building blocks, namely the unitary gates $\hat U_{in}$ and $\hat U_B$, decompose the gate $\hat U^{(d)}$ that we find at depth $d$ as follows. First, we apply $\hat U_{in}$ between $A_d$ and $\tilde A_d$. In this way, we effectively transfer the path information from $A_i$ to its corresponding backup $\tilde A_i$. Then, we start applying $\hat U_B$ using $\tilde A_i$ as control. We keep applying $\hat U_B$ with the control on each backup walker, until all the particles, address and data ones, have received its action. The gate $\hat U_B$ effectively spreads the path information contained in $A_i $ to two more particles at each of its applications. 

The described combined action of $\hat U_{in}$ and $\hat U_B$, at depth $d=1$ for $n=2$ and $m=2$ reads:
\begin{eqnarray*}
\hat U_B^{\tilde D_1,D_2}\,
\hat U_B^{\tilde A_2,\tilde D_1}\,\hat U_B^{\tilde A_1,\tilde A_2}\left(\hat U_{in}  \ket{\psi_{in}} \right) &=& 
\hat U_B^{\tilde D_1,D_2}\,
\hat U_B^{\tilde A_2,\tilde D_1}\,\hat U_B^{\tilde A_1,\tilde A_2}\left(\hat U_{in}  \ket{R}_{D_2}\ket{R}_{\tilde D_1}\ket{R}_{D_1} \ket{R}_{\tilde A_2}
    \ket{a_2}_{A_2}
    \ket{R}_{\tilde A_1}
    \ket{a_1}_{A_1}\right) \\  
    &=& \hat U^{(1)}   \ket{R}_{D_2}\ket{R}_{\tilde D_1}\ket{R}_{D_1} \ket{R}_{\tilde A_2}
    \ket{a_2}_{A_2}
    \ket{R}_{\tilde A_1}
    \ket{ a_1}_{A_1} \, ,
\end{eqnarray*}
where $\ket{\bar{a}_1}=\hat{X}\ket{a_1}$ is the state of $A_1$ after the application of a NOT gate and $\hat U_B^{\tilde I_i, \tilde I_{i+1}}$ corresponds to $\hat U_B$ applied with $\tilde I_i$ as control and $I_{i+1}$ and $\tilde I_{i+1}$ as targets, where $I_i\in\{A_1,\dots,A_n,D_1,\dots,D_{m}\}$. We can prove this equality by direct inspection. For instance, for the address $\boldsymbol{a}=a_1 \,a_2=10$, we obtain:
\begin{eqnarray*}
\hat U_B^{\tilde D_1,D_2}\,
\hat U_B^{\tilde A_2,\tilde D_1}\,\hat U_B^{\tilde A_1,\tilde A_2}\left(\hat U_{in}  \ket{\psi_{in}} \right) = 
\hat U_B^{\tilde D_1,D_2}\,
\hat U_B^{\tilde A_2,\tilde D_1}\,\hat U_B^{\tilde A_1,\tilde A_2}\hat U_{in}  \ket{\red}_{D_2}\ket{\red}_{\tilde D_1}\ket{\red}_{D_1} \ket{\red}_{\tilde A_2}
    \ket{\emptyset}_{A_2}
    \ket{\red}_{\tilde A_1}
    \ket{\red}_{A_1} &&\\ 
    = \hat U_B^{\tilde D_1,D_2}\,
\hat U_B^{\tilde A_2,\tilde D_1}\,\hat U_B^{\tilde A_1,\tilde A_2}  \ket{\red}_{D_2}\ket{\red}_{\tilde D_1}\ket{\red}_{D_1} \ket{\red}_{\tilde A_2}
    \ket{\emptyset}_{A_2}
    \ket{\blue}_{\tilde A_1}
    \ket{\red}_{A_1}&& \\ 
    =
    \hat U_B^{\tilde D_1,D_2}\,
\hat U_B^{\tilde A_2,\tilde D_1} \ket{\red}_{D_2}\ket{\red}_{\tilde D_1}\ket{\red}_{D_1} \ket{\blue}_{\tilde A_2}
    \ket{\emptyset}_{A_2}
    \ket{\blue}_{\tilde A_1}
    \ket{\red}_{A_1}&&  \\ 
    =
    \hat U_B^{\tilde D_1,D_2}\,
  \ket{\red}_{D_2}\ket{\blue}_{\tilde D_1}\ket{\blue}_{D_1} \ket{\blue}_{\tilde A_2}
    \ket{\emptyset}_{A_2}
    \ket{\blue}_{\tilde A_1}
    \ket{\red}_{A_1} && \\ 
    =
    \
  \ket{\blue}_{D_2}\ket{\blue}_{\tilde D_1}\ket{\blue}_{D_1} \ket{\blue}_{\tilde A_2}
    \ket{\emptyset}_{A_2}
    \ket{\blue}_{\tilde A_1}
    \ket{\red}_{A_1} &&   
\end{eqnarray*}
which provides the same action of $\hat U^{(1)}$ (where $A_1$ is the control and $\tilde A_1,A_2,\dots,D_2$ are the targets), as shown in Eq.~(\ref{explU1}).

\subsection{Message copy}\label{BackupMessCopy}

\begin{figure}
    \centering
    \begin{tikzpicture}[x=1cm,y=1cm, scale=0.9]
  \def\w{8}
  \def\h{1.5}
  \def\dr{3pt}
  \def\dist{0.3}

  \begin{scope}[shift={(-2,2)}, scale=0.9, transform shape]
    \draw[rounded corners, thick, fill=gray!30] (1,0) rectangle ({\w-2},\h);

    \foreach \i in {1,...,3} {
      \pgfmathsetmacro{\x}{\i/5*\w}
      \draw[line width=1pt] (\x, \h/2) -- (\x, -\dist);
    }

    \foreach \i in {1,...,3} {
      \pgfmathsetmacro{\x}{\i/5*\w}
      \draw[line width=1pt] ({\x+0.8}, \h/2) -- ({\x+0.8}, -\dist);
    }

    \foreach \j in {1,3,5} {
      \pgfmathsetmacro{\xStart}{(\j+1)/5*(\w/2)}
      \pgfmathsetmacro{\xEnd}{(\j+2)/5*(\w/2)}
      \draw[line width=1pt] (\xStart, {\h/2}) -- (\xEnd, {\h/2});
    }
    
    \foreach[count=\k from 1] \col in {red, white, red} {
        \pgfmathsetmacro{\x}{\k/5*\w}  
        \filldraw[fill=\col, draw=black, thick] (\x, {\h/2}) circle [radius=\dr];
      }

    \draw[thick] (-1,-\dist) -- (\w,-\dist);

    \foreach[count=\k from 1] \col in {red,red,red,red,red,red} {
        \pgfmathsetmacro{\x}{(\k-3)/9*(\w/2)}
        \pgfmathsetmacro{\y}{-\dist}
        \filldraw[fill=\col, draw=black, thick] (\x, \y) circle [radius=\dr];
      }

    \foreach[count=\k from 1] \col in {black,black,black,black,black,black} {
        \pgfmathsetmacro{\x}{(\k+1)/5*(\w/2)}
        \pgfmathsetmacro{\y}{-\dist}
        \filldraw[fill=\col, draw=black, thick] (\x, \y) circle [radius=\dr-1];
      }

    \node at (1.7,1.2) {{\scalebox{0.8}{$M_{3}^{(\boldsymbol{a})}$}}};

    \node at (3.3,1.2) {{\scalebox{0.8}{$M_{2}^{(\boldsymbol{a})}$}}};
    
    \node at (4.9,1.2) {{\scalebox{0.8}{$M_{1}^{(\boldsymbol{a})}$}}};

      \node at (-0.9,-0.9) {{\scalebox{0.8}{$D_3$}}};

     \node at (-0.45,-0.9) {{\scalebox{0.8}{$\tilde{D}_2$}}};

    \node at (0,-0.9) {{\scalebox{0.8}{$D_2$}}};

    \node at (0.45,-0.9) {{\scalebox{0.8}{$\tilde{D}_1$}}};

    \node at (0.9,-0.9) {{\scalebox{0.8}{$D_1$}}};

    \node at (1.35,-0.9) {{\scalebox{0.8}{$\tilde{A}_n$}}};
    
  \end{scope}

\draw[dashed] (-3,{\h-0.8})--({\w-2.2},{\h-0.8});

  \begin{scope}[shift={(-2,-1)}, scale=0.9, transform shape]
    \draw[rounded corners, thick, fill=gray!30] (1,0) rectangle ({\w-2},\h);

   \node[draw, rounded corners, fill=blue!20, minimum width=17mm, minimum height=25mm, opacity=0.9] at (3.65,0.2) {};

    \foreach \i in {1,...,3} {
      \pgfmathsetmacro{\x}{\i/5*\w}
      \draw[line width=1pt] ({\x+0.8}, \h/2) -- ({\x+0.8}, -\dist);
    }

    \foreach \j in {1,3,5} {
      \pgfmathsetmacro{\xStart}{(\j+1)/5*(\w/2)}
      \pgfmathsetmacro{\xEnd}{(\j+2)/5*(\w/2)}
      \draw[line width=1pt] (\xStart, {\h/2}) -- (\xEnd, {\h/2});
    }

   \node[draw, dashed, rounded corners, fill=green!20, minimum width=9mm, minimum height=16mm, opacity=0.9] at (3.37,0.13) {};

    \foreach \i in {1,...,3} {
      \pgfmathsetmacro{\x}{\i/5*\w}
      \draw[line width=1pt] (\x, \h/2) -- (\x, -\dist);
    }

    \foreach[count=\k from 1] \col in {red, white, red} {
        \pgfmathsetmacro{\x}{\k/5*\w}  
        \filldraw[fill=\col, draw=black, thick] (\x, {\h/2}) circle [radius=\dr];
      }

    \draw[thick] (-1,-\dist) -- (\w,-\dist);

    \foreach[count=\k from 1] \col in { red, red,red,red,red, red} {
        \pgfmathsetmacro{\x}{(\k+1)/5*(\w/2)}
        \pgfmathsetmacro{\y}{-\dist}
        \filldraw[fill=\col, draw=black, thick] (\x, \y) circle [radius=\dr];
      }

    \node at (1.7,1.2) {{\scalebox{0.8}{$M_{3}^{(\boldsymbol{a})}$}}};

    \node at (3.3,1.2) {{\scalebox{0.8}{$M_{2}^{(\boldsymbol{a})}$}}};
    
    \node at (4.9,1.2) {{\scalebox{0.8}{$M_{1}^{(\boldsymbol{a})}$}}};


    \node at (1.65,-0.9) {{\scalebox{0.8}{$D_3$}}};

    \node at (2.45,-0.9) {{\scalebox{0.8}{$\tilde{D}_2$}}};

    \node at (3.25,-0.9) {{\scalebox{0.8}{$D_2$}}};

    \node at (4.05,-0.9) {{\scalebox{0.8}{$\tilde{D}_1$}}};

    \node at (4.85,-0.9) {{\scalebox{0.8}{$D_1$}}};

    \node at (5.65,-0.9) {{\scalebox{0.8}{$\tilde{A}_n$}}};
    

\node at (4.1,1.1) {{\scalebox{0.8}{$\hat{U}_{\text{copy}}$}}};

\node at (3.55,0) {{\scalebox{0.7}{$\hat{U}_{\text{copy}}^{\text{loc}}$}}};

  \end{scope}

\draw[thick](6.4,4)--(6.4,-2.5);

   \begin{scope}[shift={(8,2)}, scale=0.9, transform shape]
    \draw[rounded corners, thick, fill=gray!30] (1,0) rectangle ({\w-2},\h);

    \foreach \i in {1,...,3} {
      \pgfmathsetmacro{\x}{\i/5*\w}
      \draw[line width=1pt] ({\x+0.8}, \h/2) -- ({\x+0.8}, -\dist);
    }

    \foreach \j in {1,3,5} {
      \pgfmathsetmacro{\xStart}{(\j+1)/5*(\w/2)}
      \pgfmathsetmacro{\xEnd}{(\j+2)/5*(\w/2)}
      \draw[line width=1pt] (\xStart, {\h/2}) -- (\xEnd, {\h/2});
    }

    \foreach \i in {1,...,3} {
      \pgfmathsetmacro{\x}{\i/5*\w}
      \draw[line width=1pt] (\x, \h/2) -- (\x, -\dist);
    }

    \foreach[count=\k from 1] \col in {red, white, red} {
        \pgfmathsetmacro{\x}{\k/5*\w}  
        \filldraw[fill=\col, draw=black, thick] (\x, {\h/2}) circle [radius=\dr];
      }

    \draw[thick] (-1,-\dist) -- (\w+0.5,-\dist);

    \foreach[count=\k from 1] \col in { red, red,white,red,red, red} {
        \pgfmathsetmacro{\x}{(\k+1)/5*(\w/2)}
        \pgfmathsetmacro{\y}{-\dist}
        \filldraw[fill=\col, draw=black, thick] (\x, \y) circle [radius=\dr];
      }

    \node at (1.7,1.2) {{\scalebox{0.8}{$M_{3}^{(\boldsymbol{a})}$}}};

    \node at (3.3,1.2) {{\scalebox{0.8}{$M_{2}^{(\boldsymbol{a})}$}}};
    
    \node at (4.9,1.2) {{\scalebox{0.8}{$M_{1}^{(\boldsymbol{a})}$}}};


    \node at (1.65,-0.9) {{\scalebox{0.8}{$D_3$}}};

    \node at (2.45,-0.9) {{\scalebox{0.8}{$\tilde{D}_2$}}};

    \node at (3.25,-0.9) {{\scalebox{0.8}{$D_2$}}};

    \node at (4.05,-0.9) {{\scalebox{0.8}{$\tilde{D}_1$}}};

    \node at (4.85,-0.9) {{\scalebox{0.8}{$D_1$}}};

    \node at (5.65,-0.9) {{\scalebox{0.8}{$\tilde{A}_n$}}};
     
\end{scope}

\begin{scope}[shift={(8,-1)}, scale=0.9, transform shape]
    \draw[rounded corners, thick, fill=gray!30] (1,0) rectangle ({\w-2},\h);

    \foreach \i in {1,...,3} {
      \pgfmathsetmacro{\x}{\i/5*\w}
      \draw[line width=1pt] (\x, \h/2) -- (\x, -\dist);
    }

    \foreach \i in {1,...,3} {
      \pgfmathsetmacro{\x}{\i/5*\w}
      \draw[line width=1pt] ({\x+0.8}, \h/2) -- ({\x+0.8}, -\dist);
    }

    \foreach \j in {1,3,5} {
      \pgfmathsetmacro{\xStart}{(\j+1)/5*(\w/2)}
      \pgfmathsetmacro{\xEnd}{(\j+2)/5*(\w/2)}
      \draw[line width=1pt] (\xStart, {\h/2}) -- (\xEnd, {\h/2});
    }
    
    \foreach[count=\k from 1] \col in {red, white, red} {
        \pgfmathsetmacro{\x}{\k/5*\w}  
        \filldraw[fill=\col, draw=black, thick] (\x, {\h/2}) circle [radius=\dr];
      }

    \draw[thick] (-1,-\dist) -- (\w+0.5,-\dist);

    \foreach[count=\k from 1] \col in {red,red,white,red,red,red} {
        \pgfmathsetmacro{\x}{(\k+12.5)/9*(\w/2)}
        \pgfmathsetmacro{\y}{-\dist}
        \filldraw[fill=\col, draw=black, thick] (\x, \y) circle [radius=\dr];
      }

    \foreach[count=\k from 1] \col in {black,black,black,black,black,black} {
        \pgfmathsetmacro{\x}{(\k+1)/5*(\w/2)}
        \pgfmathsetmacro{\y}{-\dist}
        \filldraw[fill=\col, draw=black, thick] (\x, \y) circle [radius=\dr-1];
      }

    \node at (1.7,1.2) {{\scalebox{0.8}{$M_{3}^{(\boldsymbol{a})}$}}};

    \node at (3.3,1.2) {{\scalebox{0.8}{$M_{2}^{(\boldsymbol{a})}$}}};
    
    \node at (4.9,1.2) {{\scalebox{0.8}{$M_{1}^{(\boldsymbol{a})}$}}};
   
    \node at (6,-0.9) {{\scalebox{0.8}{$D_3$}}};

    \node at (6.45,-0.9) {{\scalebox{0.8}{$\tilde{D}_2$}}};

    \node at (6.9,-0.9) {{\scalebox{0.8}{$D_2$}}};

    \node at (7.35,-0.9) {{\scalebox{0.8}{$\tilde{D}_1$}}};

    \node at (7.8,-0.9) {{\scalebox{0.8}{$D_1$}}};

    \node at (8.25,-0.9) {{\scalebox{0.8}{$\tilde{A}_n$}}};
  \end{scope}

\draw[dashed] (7,{\h-0.8})--({\w+7.8},{\h-0.8});

\node at (-3,3.3){(a)};
\node at (-3,0.3){(b)};
\node at (7,3.3){(c)};
\node at (7,0.3){(d)};

\end{tikzpicture}
    \caption{Steps involved in the message copy process in the backup variant of the qRAM protocol. (a) The data particles, their corresponding backup particles, and the final address backup arrive at the memory cell; (b) since the backup particles $\tilde{D}_{i-1}$ are always red (the control particle necessary to write the message on $D_1$ is $\tilde A_n$), they serve as control particles for activating the local unitary gate defined in Eq. \ref{backupcopy}; (c) the message is copied into the data walkers; (d) finally the particles exit the memory cell and enter the inverse routing phase to the final output port.}
    \label{copy_backup}
\end{figure}
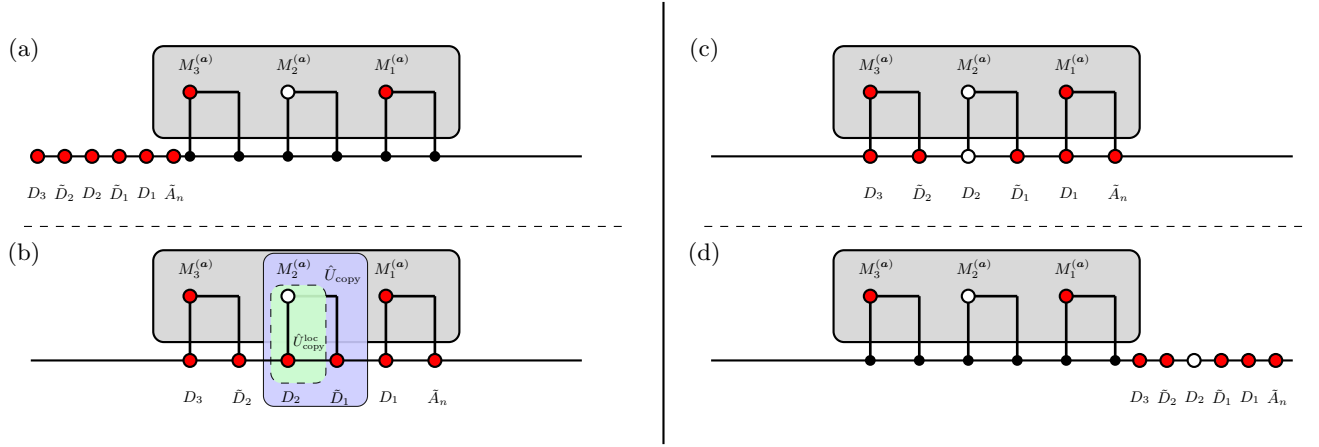

Although the control copy operation can be executed in the same way as the standard protocol, the presence of backup walkers makes it possible for a different short-range activation mechanism. More precisely, we conditionally copy the classical message $b_j^{(\boldsymbol{a})}$ contained in the $j$-th qubit of the memory cell $M^{(\boldsymbol{a})}$ into the $j$-th data walker $D_j$. Since all data particles are followed by a corresponding backup, we can define a controlled-copy gate where the backup of the $j-1$-th data particle plays the role of the control qubit. More precisely, if we want to copy the information carried by the message walker $M_j^{(\boldsymbol{a})}$ into $D_j$, we apply the copy of the former into the latter only if $\tilde D_{j-1}$ is present. The control on $\tilde D_{j-1}$ guarantees that no message is written on those cells not reached by any data walker. Notice that the particle that plays the control role for the writing of $M_1^{(\boldsymbol{a})}$ into $D_1$ is $\tilde A_n$. Indeed, by design, the final address backup $\tilde A_n$ arrives at the target memory cell in the red state $\ket R$ together with the data particles. The functioning of this mechanism is shown in Fig.~\ref{copy_backup}.
The mathematical expression that represents this controlled gate is:
\begin{equation}\label{backupcopy}
    \hat{U}_{copy}(\tilde{D}_{j-1};D_j,M_j^{(\boldsymbol{a})})=\ket{R}_{\tilde{D}_{j-1}}\!\!\bra{R}\otimes \hat{U}_{copy}^{loc}(M_j^{(\boldsymbol{a})},D_j)+\left(\ket{\emptyset}_{\tilde{D}_{j-1}}\!\!\bra{\emptyset}+\ket{B}_{\tilde{D}_{j-1}}\!\!\bra{B}\right)\otimes \hat{I}_{D_j,M_j^{(\boldsymbol{a})}}
\end{equation}
where $\hat{U}_{copy}^{loc}$ is the unitary defined in Eq.~(\ref{loc_copy}).

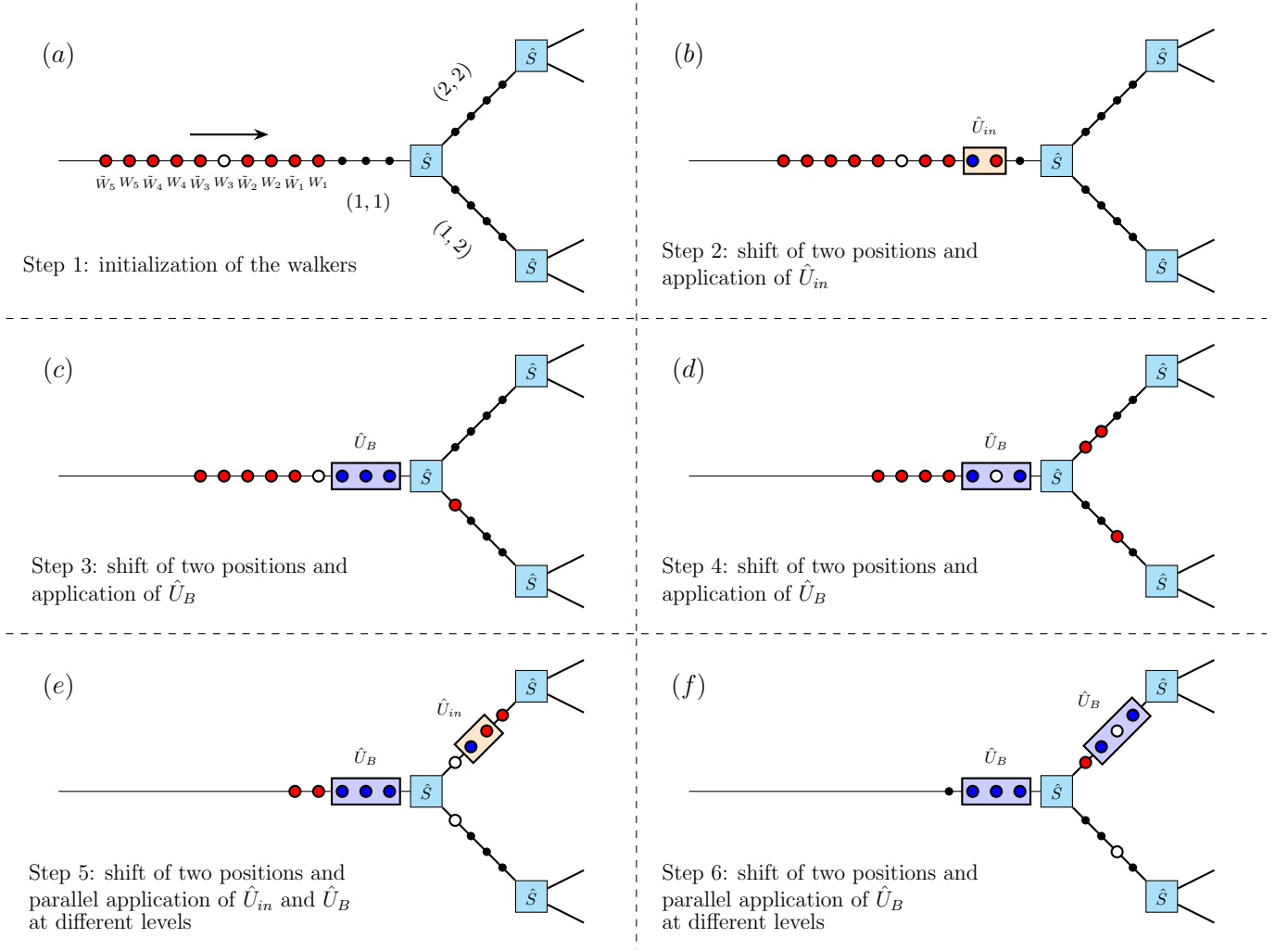
\begin{figure}
    \centering
\begin{tikzpicture}[scale=0.75, transform shape,
  level 1/.style={level distance=2cm, sibling distance=4cm},
  level 2/.style={level distance=1cm, sibling distance=1cm},
  grow=east, 
  edge from parent/.style={draw, thick}, 
  every node/.style={rectangle, draw, minimum size=6mm,fill=cyan!30}, 
  leaf/.style={rectangle, draw, fill=gray!30, minimum size=6mm}, 
  block/.style={rectangle, draw=none,dashed,fill=none, minimum width=1.8cm, minimum height=5mm, rotate around={0:(0,0)}}, 
  gateX/.style={rectangle, draw, fill=green!20, minimum size=5mm, rotate around={0:(0,0)}} 
]

\begin{scope}[shift={(0,0)}]

\node (root1) {$\hat{S}$}
  child {
    node {$\hat{S}$}
    child { 
      edge from parent {}
    }
    child { 
      edge from parent{
      }
    }
    edge from parent {
      node[midway,xshift=0cm, yshift=-0cm,  block, rotate=-45] {} 
    }
  }
  child {
    node {$\hat{S}$}
    child { 
    }
    child { 
      edge from parent{ 
      }
    }
    edge from parent {
      node[midway, xshift=0cm,yshift=0cm, block, rotate=45] {} 
    }
  };

\draw (-7, 0) -- (root1) node(Ugateroot)[xshift=-1.37cm, block, rotate=0] {};






\draw[fill=black] (-0.7,0) circle (2pt);
\draw[fill=black] (-1.15,0) circle (2pt);
\draw[fill=black] (-1.6,0) circle (2pt);
\draw[fill=black] (-2.05,0) circle (2pt);
\draw[thick, fill=red] (-2.05,0) circle (3pt) node[below=3pt, fill = none, draw=none]{\scriptsize$W_1$};
\draw[thick, fill=red] (-2.5,0) circle (3pt) node[below=3pt, fill = none, draw=none]{\scriptsize$\tilde{W}_1$};
\draw[thick, fill=red] (-2.95,0) circle (3pt)node[below=3pt, fill = none, draw=none]{\scriptsize$W_{2}$};
\draw[thick, fill=red] (-3.4,0) circle (3pt)node[below=3pt, fill = none, draw=none]{\scriptsize$\tilde{W}_{2}$};
\draw[thick, fill=white] (-3.85,0) circle (3pt)node[below=3pt, fill = none, draw=none]{\scriptsize$W_{3}$};
\draw[thick, fill=red] (-4.3,0) circle (3pt)node[below=3pt, fill = none, draw=none]{\scriptsize$\tilde{W}_{3}$};
\draw[thick, fill=red] (-4.75,0) circle (3pt)node[below=3pt, fill = none, draw=none]{\scriptsize$W_{4}$};
\draw[thick, fill=red] (-5.2,0) circle (3pt)node[below=3pt, fill = none, draw=none]{\scriptsize$\tilde{W}_{4}$};
\draw[thick, fill=red] (-5.65,0) circle (3pt)node[below=3pt, fill = none, draw=none]{\scriptsize$W_{5}$};
\draw[thick, fill=red] (-6.1,0) circle (3pt)node[below=3pt, fill = none, draw=none]{\scriptsize$\tilde{W}_{5}$};


\draw[fill=black] (1.45,1.45) circle (2pt);
\draw[fill=black] (1.15,1.15) circle (2pt);
\draw[fill=black] (0.85,0.85) circle (2pt);
\draw[fill=black] (0.55,0.55) circle (2pt);


\draw[ fill=black] (1.45,-1.45) circle (2pt);
\draw[fill=black] (1.15,-1.15) circle (2pt);
\draw[fill=black] (0.85,-0.85) circle (2pt);
\draw[fill=black] (0.55,-0.55) circle (2pt);

\node[draw=none, fill=none] at (-1.1,-0.8){\scalebox{1.2}{$(1,1)$}};

\node[draw=none, fill=none, rotate=-45] at (0.5,-1.5){\scalebox{1.2}{$(1,2)$}};

\node[draw=none, fill=none, rotate=45] at (0.5,1.5){\scalebox{1.2}{$(2,2)$}};


\draw[->, >={Stealth[length=6pt]}, thick] (-4.5,0.5) -- (-3,0.5)node[fill=none,draw=none,midway,above] {};

\node[draw=none, fill=none, align=left] at (-4.5,-2)
{\large Step 1: initialization of the walkers};

\node[draw=none, fill=none] at (-7,2){\Large $(a)$};

\end{scope}

\begin{scope}[shift={(12,0)}]

\node (root1) {$\hat{S}$}
  child {
    node {$\hat{S}$}
    child { 
      edge from parent {}
    }
    child { 
      edge from parent{
      }
    }
    edge from parent {
      node[midway,xshift=0cm, yshift=-0cm,  block, rotate=-45] {} 
    }
  }
  child {
    node {$\hat{S}$}
    child { 
    }
    child { 
      edge from parent{ 
      }
    }
    edge from parent {
      node[midway, xshift=0cm,yshift=0cm, block, rotate=45] {} 
    }
  };

\draw (-7, 0) -- (root1) node(Ugateroot)[xshift=-1.37cm, block, rotate=0] {};






\node[block,draw, thick,solid, minimum width = 0.8cm, fill=orange!20,label={[yshift=1mm]above:$\hat{U}_{in}$}] at (-1.37,0){};


\draw[fill=black] (-0.7,0) circle (2pt);
\draw[thick, fill=red] (-1.15,0) circle (3pt);
\draw[fill=blue,thick] (-1.6,0) circle (3pt);
\draw[fill=red,thick] (-2.05,0) circle (3pt);
\draw[thick, fill=red] (-2.5,0) circle (3pt);
\draw[thick, fill=white] (-2.95,0) circle (3pt);
\draw[thick, fill=red] (-3.4,0) circle (3pt);
\draw[thick, fill=red] (-3.85,0) circle (3pt);
\draw[thick, fill=red] (-4.3,0) circle (3pt);
\draw[thick, fill=red] (-4.75,0) circle (3pt);
\draw[thick, fill=red] (-5.2,0) circle (3pt);


\draw[fill=black] (1.45,1.45) circle (2pt);
\draw[fill=black] (1.15,1.15) circle (2pt);
\draw[fill=black] (0.85,0.85) circle (2pt);
\draw[fill=black] (0.55,0.55) circle (2pt);


\draw[ fill=black] (1.45,-1.45) circle (2pt);
\draw[fill=black] (1.15,-1.15) circle (2pt);
\draw[fill=black] (0.85,-0.85) circle (2pt);
\draw[fill=black] (0.55,-0.55) circle (2pt);

\node[draw=none, fill=none, align=left] at (-4.5,-2)
{\large Step 2: shift of two positions and \\ \large application of $\hat{U}_{in}$};

\node[draw=none, fill=none] at (-7,2){\Large $(b)$};
   
\end{scope}


\begin{scope}[shift={(0,-6)}]

\node (root1) {$\hat{S}$}
  child {
    node {$\hat{S}$}
    child { 
      edge from parent {}
    }
    child { 
      edge from parent{
      }
    }
    edge from parent {
      node[midway,xshift=0cm, yshift=-0cm,  block, rotate=-45] {} 
    }
  }
  child {
    node {$\hat{S}$}
    child { 
    }
    child { 
      edge from parent{ 
      }
    }
    edge from parent {
      node[midway, xshift=0cm,yshift=0cm, block, rotate=45] {} 
    }
  };

\draw (-7, 0) -- (root1) node(Ugateroot)[xshift=-1.37cm, block, rotate=0] {};






\node[block,draw,thick,minimum width = 1.3cm, solid, fill=blue!20,label={[yshift=1mm]above:$\hat{U}_B$}] at (-1.15,0){};


\draw[fill=blue, thick] (-0.7,0) circle (3pt);
\draw[fill=blue,thick] (-1.15,0) circle (3pt);
\draw[fill=blue,thick] (-1.6,0) circle (3pt);
\draw[fill=white,thick] (-2.05,0) circle (3pt);
\draw[thick, fill=red] (-2.5,0) circle (3pt);
\draw[thick, fill=red] (-2.95,0) circle (3pt);
\draw[thick, fill=red] (-3.4,0) circle (3pt);
\draw[thick, fill=red] (-3.85,0) circle (3pt);
\draw[thick, fill=red] (-4.3,0) circle (3pt);


\draw[fill=black] (1.45,1.45) circle (2pt);
\draw[fill=black] (1.15,1.15) circle (2pt);
\draw[fill=black] (0.85,0.85) circle (2pt);
\draw[fill=black] (0.55,0.55) circle (2pt);


\draw[ fill=black] (1.45,-1.45) circle (2pt);
\draw[fill=black] (1.15,-1.15) circle (2pt);
\draw[fill=black] (0.85,-0.85) circle (2pt);
\draw[fill=red,thick] (0.55,-0.55) circle (3pt);

\node[draw=none, fill=none, align=left] at (-4.5,-2)
{\large Step 3: shift of two positions and \\ \large application of $\hat{U}_{B}$};

\node[draw=none, fill=none] at (-7,2){\Large $(c)$};

\end{scope}


\begin{scope}[shift={(12,-6)}]

\node (root1) {$\hat{S}$}
  child {
    node {$\hat{S}$}
    child { 
      edge from parent {}
    }
    child { 
      edge from parent{
      }
    }
    edge from parent {
      node[midway,xshift=0cm, yshift=-0cm,  block, rotate=-45] {} 
    }
  }
  child {
    node {$\hat{S}$}
    child { 
    }
    child { 
      edge from parent{ 
      }
    }
    edge from parent {
      node[midway, xshift=0cm,yshift=0cm, block, rotate=45] {} 
    }
  };

\draw (-7, 0) -- (root1) node(Ugateroot)[xshift=-1.37cm, block, rotate=0] {};






\node[block,draw,thick,minimum width = 1.3cm, solid, fill=blue!20,label={[yshift=1mm]above:$\hat{U}_B$}] at (-1.15,0){};


\draw[fill=blue,thick] (-0.7,0) circle (3pt);
\draw[fill=white,thick] (-1.15,0) circle (3pt);
\draw[fill=blue,thick] (-1.6,0) circle (3pt);
\draw[fill=red,thick] (-2.05,0) circle (3pt);
\draw[thick, fill=red] (-2.5,0) circle (3pt);
\draw[thick, fill=red] (-2.95,0) circle (3pt);
\draw[thick, fill=red] (-3.4,0) circle (3pt);


\draw[fill=black] (1.45,1.45) circle (2pt);
\draw[fill=black] (1.15,1.15) circle (2pt);
\draw[fill=red,thick] (0.85,0.85) circle (3pt);
\draw[fill=red,thick] (0.55,0.55) circle (3pt);

\draw[ fill=black] (1.45,-1.45) circle (2pt);
\draw[fill=red,thick] (1.15,-1.15) circle (3pt);
\draw[fill=black] (0.85,-0.85) circle (2pt);
\draw[fill=black] (0.55,-0.55) circle (2pt);

\node[draw=none, fill=none, align=left] at (-4.5,-2)
{\large Step 4: shift of two positions and \\ \large application of $\hat{U}_{B}$};

\node[draw=none, fill=none] at (-7,2){\Large $(d)$};

\end{scope}


\begin{scope}[shift={(0,-12)}]

\node (root1) {$\hat{S}$}
  child {
    node {$\hat{S}$}
    child { 
      edge from parent {}
    }
    child { 
      edge from parent{
      }
    }
    edge from parent {
      node[midway,xshift=0cm, yshift=-0cm,  block, rotate=-45] {} 
    }
  }
  child {
    node {$\hat{S}$}
    child { 
    }
    child { 
      edge from parent{ 
      }
    }
    edge from parent {
      node[midway, xshift=0cm,yshift=0cm, block, rotate=45] {} 
    }
  };

\draw (-7, 0) -- (root1) node(Ugateroot)[xshift=-1.37cm, block, rotate=0] {};







\node[block,draw,thick,minimum width = 1.3cm, solid, fill=blue!20,label={[yshift=1mm]above:$\hat{U}_B$}] at (-1.15,0){};

\node[block,draw, thick,solid, minimum width = 0.8cm, fill=orange!20, rotate = 45,label={[yshift=1mm]above:$\hat{U}_{in}$}] at (1,1){};




\draw[fill=blue, thick] (-0.7,0) circle (3pt);
\draw[fill=blue, thick] (-1.15,0) circle (3pt);
\draw[fill=blue,thick] (-1.6,0) circle (3pt);
\draw[fill=red,thick] (-2.05,0) circle (3pt);
\draw[thick, fill=red] (-2.5,0) circle (3pt);



\draw[fill=red,thick] (1.45,1.45) circle (3pt);
\draw[fill=red,thick] (1.15,1.15) circle (3pt);
\draw[fill=blue,thick] (0.85,0.85) circle (3pt);
\draw[fill=white,thick] (0.55,0.55) circle (3pt);



\draw[ fill=black] (1.45,-1.45) circle (2pt);
\draw[fill=black] (1.15,-1.15) circle (2pt);
\draw[fill=black] (0.85,-0.85) circle (2pt);
\draw[fill=white,thick] (0.55,-0.55) circle (3pt);
















\node[draw=none, fill=none, align=left] at (-4.5,-2)
{\large Step 5: shift of two positions and \\ \large parallel application of $\hat{U}_{in}$ and $\hat{U}_B$\\\large  at different levels};

\node[draw=none, fill=none] at (-7,2){\Large $(e)$};

\end{scope}


\begin{scope}[shift={(12,-12)}]

\node (root1) {$\hat{S}$}
  child {
    node {$\hat{S}$}
    child { 
      edge from parent {}
    }
    child { 
      edge from parent{
      }
    }
    edge from parent {
      node[midway,xshift=0cm, yshift=-0cm,  block, rotate=-45] {} 
    }
  }
  child {
    node {$\hat{S}$}
    child { 
    }
    child { 
      edge from parent{ 
      }
    }
    edge from parent {
      node[midway, xshift=0cm,yshift=0cm, block, rotate=45] {} 
    }
  };

\draw (-7, 0) -- (root1) node(Ugateroot)[xshift=-1.37cm, block, rotate=0] {};






\node[block,draw,thick,minimum width = 1.3cm, solid, fill=blue!20,label={[yshift=1mm]above:$\hat{U}_B$}] at (-1.15,0){};

\node[block,draw,thick,minimum width = 1.3cm, solid, fill=blue!20, rotate = 45,label={[yshift=1mm]above:$\hat{U}_B$}] at (1.15,1.15){};


\draw[fill=blue,thick] (-0.7,0) circle (3pt);
\draw[fill=blue,thick] (-1.15,0) circle (3pt);
\draw[fill=blue,thick] (-1.6,0) circle (3pt);
\draw[fill=black] (-2.05,0) circle (2pt);


\draw[fill=blue,thick] (1.45,1.45) circle (3pt);
\draw[fill=white,thick] (1.15,1.15) circle (3pt);
\draw[fill=blue,thick] (0.85,0.85) circle (3pt);
\draw[fill=red,thick] (0.55,0.55) circle (3pt);



\draw[ fill=black] (1.45,-1.45) circle (2pt);
\draw[fill=white,thick] (1.15,-1.15) circle (3pt);
\draw[fill=black] (0.85,-0.85) circle (2pt);
\draw[fill=black] (0.55,-0.55) circle (2pt);

\node[draw=none, fill=none, align=left] at (-4.5,-2)
{\large Step 6: shift of two positions and \\ \large parallel application of $\hat{U}_{B}$ \\ \large at different levels};

\node[draw=none, fill=none] at (-7,2){\Large $(f)$};

\end{scope}

\draw[dashed] (4,3)--(4,-15);

\draw[dashed] (-8,-3)--(16,-3);

\draw[dashed] (-8,-9)--(16,-9);

\end{tikzpicture}
\caption{{Example of parallelization schedule, in the backup variant setting, between the first two levels of the binary tree: (a) the walkers are initialized at level $d=1$ of the tree; (b) after a two-site shift, the initialization gate $\hat{U}_{in}$ is applied to the pair $(\tilde W_1,W_1)$; (c) following another two-site shift, the block gate $\hat{U}_B$ acts on the triplet $(\tilde W_{2},W_{2},\tilde W_1)$; (d) after the next shift $\hat{U}_B$ is applied to the $(\tilde W_{3},W_{3},\tilde W_{2})$; (e) subsequent shifts allow the simultaneous application $\hat{U}_{B}$ at level $d=1$ and $\hat{U}_{in}$ at level $d=2$; (f) from this point onward the applications of gates $\hat{U}_B$ at levels $d=1$ and $d=2$ can be in parallel.}}
\label{parallelization}
\end{figure}

\subsection{Depth analysis}\label{DepthAnalysis}
In our backup scheme, the walkers interact through local range-2 operations implemented by the gate $\hat{U}_B$. In this section, we show how this locality can be exploited to reduce the overall circuit depth and physical length from quadratic to linear in the number of input walkers $n$. 
In case all gates are applied sequentially along each branch of the tree,
that is by first transforming all the target particles at a given depth $d$ through the action of $\hat U_{in}$ and $\hat U_B$ before routing the particles at the next depth through the application of $\hat S$,
the total execution time would scale quadratically with the number of walkers, namely $T_{\mathrm{serial}}\sim \mathcal{O}(n^2)$.

Notice, however, that the activation of the gates $\hat U_{in}$ and $\hat U_B$ at different depths can be parallelized thanks to the presence of backup walkers, which allow a gradual spreading of the address information contained in the address walkers.
To be more specific, at each routing step, the walkers are shifted by two positions and either $\hat{U}_{in}$ or $\hat{U}_B$ is applied. At level $d$, the protocol first applies $\hat{U}_{in}$ between the walker $A_d$ and its corresponding backup $\tilde{A}_d$. The walkers are then shifted by two positions, and the block operation $\hat{U}_B$ acts on the triple $(\tilde{A}_d,A_{d+1},\tilde{A}_{d+1})$. During this step, the first walker $A_d$ exits the active edge and no longer participates in subsequent routing operations. Shifting again by two positions, $\hat{U}_B$ is applied on the next triple $(\tilde{A}_{d+1},A_{d+2},\tilde{A}_{d+2})$, while the walkers $\tilde{A}_d$ and $A_{d+1}$ are routed to the next level of the tree, namely at depth $d+1$. 
After these first three steps, the execution of $\hat{U}_B$ at level $d$ and $\hat{U}_{in}$ at level $d+1$ can be parallelized, and from the next step onward, we can execute $\hat{U}_B$ in parallel on both levels $d$ and $d+1$.
This parallelization improves the overall circuit depth from quadratic to linear in the number of input walkers.  
The maximum achievable parallelization is reached at depth $d=\lfloor (n+m)/2 \rfloor$, after which the number of simultaneous operations decreases symmetrically as the particles start to exit the qRAM. We graphically represent our parallelization strategy in Fig.~\ref{parallelization}.

We conclude with an operational description of our scheme with the aim of providing an intuitive explanation of this linear scaling. We start by focusing on the following two observations: (i) the whole train of particles moves collectively along the train, which always moves two steps at time, and (ii) the total time of execution can be evaluated as the interval between the time when the first address walker is injected in the tree and the time when the last data walker exits. The length of each edge in the qRAM tree is fixed to 4 sites, thus being independent of $n$ and $m$. As a consequence, the time steps needed for $A_1$ to move from one of the available sites at one level to the corresponding spot at the next level are constant and equal to 4, regardless of the depth. 
Now, we explain why this is the case.  By looking at Fig.~\ref{parallelization}, we see that the operations implemented in our scheme alternate the parallel action of $\hat U_B$ and $\hat U_{in}$ gates at different levels and the synchronous movement of the particles of two sites. Hence, every two logical steps, the train of particles shifts by two sites.
Therefore, the total time needed for $A_1$ to reach the memory cells is $4n$. 
Once that $A_1$ reached the memory cells, depending on the different protocols used for the query phase, $A_1$ may take different times to travel through them. For instance, if we consider the strategy described in Fig.~\ref{copy_backup}, $A_1$ has to travel $2m$ sites, which can be covered in $2m$ time steps. Following this logic, the total time needed for $A_1$ to exit the three is $8n + 2m$. After $A_1$ exits the tree, the remaining $2(n+m-1)$ walkers are still inside the qRAM. In particular, the last data walker, $D_m$, is located $2(n+m)$ sites away from the output port. As the spatial propagation is fully parallelized, these remaining sites are covered in $2(n+m)$ additional time steps. Hence, the final time of execution is
\begin{equation*}
    T_{parallel}  = 10n + 4m + t_{\mathcal Q}\sim \mathcal{O}(n+m),
\end{equation*}
where $t_{\mathcal Q}$ is the \textit{extra} time needed to write the information on the data walkers during the query phase. For instance, the strategy described in Fig.~\ref{copy_backup} corresponds to $t_{\mathcal Q}=0$: the information is written on the data particles in parallel with the execution of the gates $U_B$ and $U_{in}$ executed during the inverse-routing phase.
\color{black}

\section{Alternative encodings}\label{SecAlternative}

Although the protocol was originally formulated with bosonic information carriers such as photons or phonons, its logical structure can be straightforwardly adapted to fermionic systems. 
This adaptability arises from the fact that the core mechanism of the protocol relies only on the presence or absence of a particle and its internal state, rather than on bosonic symmetrization or multi-occupancy properties. As a result, fermionic carriers, such as electrons in quantum dots or cold atoms in optical lattices, can implement the same logic gates provided proper exclusion constraints are respected. To accommodate the fermionic nature of the carriers, we propose encoding schemes that prevent double occupancy and ensure distinguishability of operational paths. In particular, we have outlined two concrete alternatives: one based on a dual-rail encoding where each logical site is represented by two spatial modes, and another using four-level qudit systems. Both schemes preserve the full functionality of the original bosonic protocol while adhering to fermionic statistics. These alternative encodings open the possibility of implementing the protocol on a broader class of hardware platforms, such as superconducting circuits with fermionic qubits, trapped fermionic atoms, or mesoscopic systems. Importantly, the modifications required by the fermionic implementations do not affect the scaling of resources or the overall depth of the protocol.

\subsubsection{Dual-rail approach}

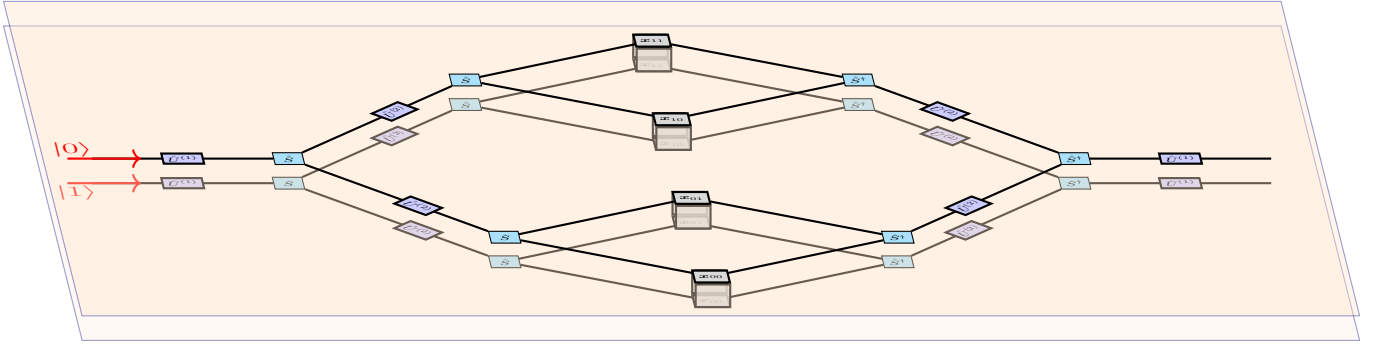
\begin{figure}[t]
\centering
\begin{tikzpicture}[scale=0.65, transform shape]

\def\dist{0.5} 

\begin{scope}[cm={1,0,-0.1,0.4,(0,0)},
  level 1/.style={level distance=4cm, sibling distance=8cm},
  level 2/.style={level distance=4cm, sibling distance=4cm},
  grow=east,
  edge from parent/.style={draw, thick},
   every node/.style={rectangle, draw, minimum size=6mm,fill=cyan!30},
  leaf/.style={rectangle, draw, fill=gray!30, minimum size=6mm},
  block/.style={rectangle, draw, fill=blue!20, minimum width=0.7cm, minimum height=5mm},
  gateX/.style={rectangle, draw, fill=green!20, minimum size=5mm}]

\fill[orange!10, draw=blue!50!black, opacity=0.4] (-5,-8) rectangle (21,8);

\node (root1) {$\hat{S}$}
  child {
    node {$\hat{S}$}
    child { 
      node[leaf] {$\boldsymbol{x}_3$} 
      edge from parent {}
    }
    child { 
      node[leaf] {$\boldsymbol{x}_2$} 
      edge from parent{
      }
    }
    edge from parent {
      node[midway,xshift=0.4cm, yshift=-0.4cm,  block, rotate=-45] {$\hat{U}^{(2)}$} 
    }
  }
  child {
    node {$\hat{S}$}
    child { 
      node[leaf] {$\boldsymbol{x}_1$} 
      edge from parent {}
    }
    child { 
      node[leaf] {$\boldsymbol{x}_0$} 
      edge from parent{ 
      }
    }
    edge from parent {
      node[midway, xshift=0.4cm,yshift=0.4cm, block, rotate=45] {$\hat{U}^{(2)}$} 
    }
  };

\node (root2) at (16,0) {$\hat{S}^{\dagger}$}
[grow=left]
  child {
    node {$\hat{S}^{\dagger}$}
    child { 
      node[leaf] {$\boldsymbol{x}_{11}$} 
      edge from parent {
      }
    }
    child { 
      node[leaf] {$\boldsymbol{x}_{10}$} 
      edge from parent {}
    }
    edge from parent {
      node(Ugate)[midway,xshift=-0.4cm,yshift=0.4cm, block, rotate=-45] {$\hat{U}^{(2)}$} 
    }
  }
  child {
    node {$\hat{S}^{\dagger}$}
    child { 
      node[leaf] {$\boldsymbol{x}_{01}$} 
      edge from parent {
      }
    }
    child { 
      node[leaf] {$\boldsymbol{x}_{00}$} 
      edge from parent {}
    }
    edge from parent {
      node[midway, xshift=-0.4cm, yshift=-0.4cm, block, rotate=45] {$\hat{U}^{(2)}$} 
    }
  };

\draw[thick] (-4, 0) -- (root1) node(Ugateroot)[midway, block, rotate=0] {$\hat{U}^{(1)}$};
\draw[thick] (root2) -- (20,0) node[midway, block] {$\hat{U}^{(1)}$};

\draw[->,thick, red] (-4.5,0)--(-3,0)node [below,pos=0.1, draw=none, fill=none] {\huge$\ket{1}$};

\node[thick,transform shape=false, fill=gray!30, rectangle, minimum height =8.5pt, minimum width=13pt, opacity=1] at (8.065,6.98) {}; 

\node[thick,transform shape=false, fill=gray!30, rectangle, minimum height =8.5pt, minimum width=13pt, opacity=1] at (8.065,2.98) {}; 

\node[thick,transform shape=false, fill=gray!30, rectangle, minimum height =8.5pt, minimum width=13pt, opacity=1] at (8.065,-1.02) {}; 

\node[thick, draw=black,transform shape=false, fill=gray!30, rectangle, minimum height =8.5pt, minimum width=13pt, opacity=1] at (8.065,-5.02) {}; 

\node[thick,transform shape=false, fill=gray!30, rectangle, minimum height =8.5pt, minimum width=13pt, fill opacity=0.8] at (8.065,6.27) {}; 

\node[thick,transform shape=false, fill=gray!30, rectangle, minimum height =8.5pt, minimum width=13pt, fill opacity=0.8] at (8.065,2.27) {}; 

\node[thick,transform shape=false, fill=gray!30, rectangle, minimum height =8.5pt, minimum width=13pt, fill opacity=0.8] at (8.065,-1.73) {}; 

\node[thick,transform shape=false, fill=gray!30, rectangle, minimum height =8.5pt, minimum width=13pt, fill opacity=0.8] at (8.065,-5.73) {};

\end{scope}

\begin{scope}[cm={1,0,-0.1,0.4,(0,\dist)},
  level 1/.style={level distance=4cm, sibling distance=8cm},
  level 2/.style={level distance=4cm, sibling distance=4cm},
  grow=east,
  edge from parent/.style={draw, thick},
every node/.style={rectangle, draw, minimum size=6mm,fill=cyan!30},
  leaf/.style={rectangle, draw, fill=gray!30, minimum size=6mm},
  block/.style={rectangle, draw, fill=blue!20, minimum width=0.7cm, minimum height=5mm},
  gateX/.style={rectangle, draw, fill=green!20, minimum size=5mm}]

\fill[orange!20, draw=blue!50!black, opacity=0.4] (-5,-8) rectangle (21,8);

\node (root1t) {$\hat{S}$}
  child {
    node {$\hat{S}$}
    child { 
      node[leaf] {$\boldsymbol{x}_3$} 
      edge from parent {}
    }
    child { 
      node[leaf] {$\boldsymbol{x}_2$} 
      edge from parent{
      }
    }
    edge from parent {
      node[midway,xshift=0.4cm, yshift=-0.4cm,  block, rotate=-45] {$\hat{U}^{(2)}$} 
    }
  }
  child {
    node {$\hat{S}$}
    child { 
      node[leaf] {$\boldsymbol{x}_1$} 
      edge from parent {}
    }
    child { 
      node[leaf] {$\boldsymbol{x}_0$} 
      edge from parent{ 
      }
    }
    edge from parent {
      node[midway, xshift=0.4cm,yshift=0.4cm, block, rotate=45] {$\hat{U}^{(2)}$} 
    }
  };

\node (root2t) at (16,0) {$\hat{S}^{\dagger}$}
[grow=left]
  child {
    node {$\hat{S}^{\dagger}$}
    child { 
      node[leaf] {$\boldsymbol{x}_{11}$} 
      edge from parent {
      }
    }
    child { 
      node[leaf] {$\boldsymbol{x}_{10}$} 
      edge from parent {}
    }
    edge from parent {
      node(Ugate)[midway,xshift=-0.4cm,yshift=0.4cm, block, rotate=-45] {$\hat{U}^{(2)}$} 
    }
  }
  child {
    node {$\hat{S}^{\dagger}$}
    child { 
      node[leaf] {$\boldsymbol{x}_{01}$} 
      edge from parent {
      }
    }
    child { 
      node[leaf] {$\boldsymbol{x}_{00}$} 
      edge from parent {}
    }
    edge from parent {
      node[midway, xshift=-0.4cm, yshift=-0.4cm, block, rotate=45] {$\hat{U}^{(2)}$} 
    }
  };

\draw[thick] (-4, 0) -- (root1t) node[midway, block, rotate=0] {$\hat{U}^{(1)}$};
\draw[thick] (root2t) -- (20,0) node[midway, block] {$\hat{U}^{(1)}$};

\draw[->,thick, red] (-4.5,0)--(-3,0)node [above,pos=0.1, draw=none, fill=none] {\huge$\ket{0}$};

\end{scope}

\end{tikzpicture}
\caption{Schematic representation of the qRAM protocol in the dual-rail fermionic encoding. Each rail evolves independently throughout the protocol, with no direct interaction between them, except during the control-COPY step. In this phase, the position of the data walker on the rails is coherently updated based on the classical information stored in the memory cells.}
\label{dualrailQRAM}
\end{figure}

A possible approach to generalizing the protocol to fermionic carriers involves adopting a dual-rail encoding for the address register, complemented by an internal degree of freedom used to control routing through the binary tree (see Fig.~\ref{dualrailQRAM}). In this encoding scheme, we define two spatial modes, referred to as the upper and lower rails, denoted respectively by the states $\ket{0}_R$ and $\ket{1}_R$. Each rail independently hosts a perfect binary tree of depth $d$, and both trees are ultimately connected to the same set of memory cells. Each fermionic particle on the rails possesses an internal "color" degree of freedom, which can take values in the set $\{\ket{R}_C, \ket{B}_C\}$. The full set of basis states for a particle is then given by $\{\ket{0,R}_{RC}, \ket{0,B}_{RC}, \ket{1,R}_{RC}, \ket{1,B}_{RC}\}$. The mapping from the bosonic address states to the fermionic encoding reads:
\begin{equation*}
\ket{\emptyset} \rightarrow \ket{0,R}_{RC}, \quad \ket{R} \rightarrow \ket{1,R}_{RC}.
\end{equation*}
At the start of the protocol, address particles are distributed across the two rails according to the binary string they encode, while message carriers are all initialized in the state $\ket{1,R}_{RC}$, i.e., on the lower rail with red color, replicating the role of data walkers in the bosonic version. The protocol then proceeds via the sequential application of unitary gates $\hat{U}^{i}$ on each level of the binary trees. These gates operate independently on each rail: if an address particle occupies the upper rail, the corresponding gate on the lower rail remains inactive, allowing message particles to continue straight along the tree (taking a right turn). Conversely, if the address particle is on the lower rail, the gate $\hat{U}^{i}$ acts non-trivially by flipping the internal state of all particles on that rail from $\ket{R}$ to $\ket{B}$, causing them to take a left turn. Through this mechanism, the routing of data particles is fully determined by the presence or absence of address fermions on specific rails, closely mirroring the logic of the bosonic protocol. In terms of resource requirements, the dual-rail scheme introduces a doubling in the number of controlled gates, but preserves the total number of information carriers as in the standard implementation.
The message retrieval stage, i.e. the controlled-COPY operation between the memory cell and the data walkers, can also be mapped onto this fermionic dual-rail framework. In this case, the COPY gates must act conditionally on both the rail and internal color of the walkers, preserving fermionic statistics and ensuring anti-symmetrization where necessary. More specifically, the same logical structure used in the bosonic version can be retained by identifying the “red” states as active information carriers and defining local gates that flip the data walker’s rail state depending on the content of the memory cell and the presence of a control (flag) particle. This guarantees the logical consistency of the message copy even in the fermionic encoding. We stress that while the physical implementation of such conditional gates may differ across platforms, the mapping preserves both the routing logic and the retrieval behavior of the original protocol, ensuring that the fermionic variant remains functionally equivalent.

\subsubsection{Qudit-based approach}

Another possible realization of the protocol employs multi-level quantum systems (qudits), and in particular four-level systems, instead of standard two-level qubits. In this setting, each particle is allowed to occupy one of four orthogonal states, defined as:
\begin{equation*}
    \{\ket{0},\ket{1},\ket{2},\ket{3}\} \equiv \{\ket{\redcircleopen},\ket{\bluecircleopen},\ket{\redcirclefill},\ket{\bluecirclefill}\},
\end{equation*}
which correspond one-to-one with the dual-rail encoding introduced earlier:
\begin{equation*}
    \ket{\redcircleopen} \equiv \ket{0,R}_{RC},\quad
    \ket{\bluecircleopen} \equiv \ket{0,B}_{RC},\quad
    \ket{\redcirclefill} \equiv \ket{1,R}_{RC},\quad
    \ket{\bluecirclefill} \equiv \ket{1,B}_{RC}.
\end{equation*}
In this qudit-based formulation, the entire routing procedure is again performed on a single binary tree. The routing logic is now fully encoded in the internal four-level structure of each particle, and the unitary operations are generalized to act directly on the full Hilbert space spanned by these four states. As in the previous protocols, each unitary gate $\hat{U}^{i}$ acts conditionally using the address walker $A_i$ as a control. If $A_i$ is in the state $\ket{\redcirclefill}$, then all controlled walkers in state $\ket{\redcirclefill}$ are flipped to $\ket{\bluecirclefill}$ and vice versa, while the states $\ket{\redcircleopen}$ and $\ket{\bluecircleopen}$ remain unaffected. Conversely, if the control walker is in the state $\ket{\redcircleopen}$, the transformation changes $\ket{\redcircleopen}$ to $\ket{\bluecircleopen}$ and vice versa, while leaving the other states unchanged. This design ensures that the gates $\hat U^i$ are activated whenever the internal state of the control walker is either $\ket{\redcirclefill}$  or $\ket{\redcircleopen}$.

The spatial routing along the binary tree is handled by an operation $\hat{S}$, whose action depends on the internal state of the incoming walker. Specifically, a walker in state $\ket{\redcirclefill}$ continues to the right without changing color. A walker in state $\ket{\bluecirclefill}$ is routed to the left and its internal state is flipped to $\ket{\redcirclefill}$. Similarly, a walker in state $\ket{\redcircleopen}$ is routed to the left with no internal change, while a walker in state $\ket{\bluecircleopen}$ is sent to the right and simultaneously flipped to $\ket{\redcircleopen}$.

The message retrieval stage in the qudit-based protocol can also be implemented in full analogy with the bosonic scheme. Since each data walker’s logical and internal degrees of freedom are jointly encoded in a single four-level system, the COPY operation can be performed via gates acting entirely within this qudit subspace. Specifically, the control-COPY mechanism can be defined as a conditional transformation that flips the state of a data qudit between $\ket{\redcirclefill}$ and $\ket{\redcircleopen}$, depending on the message content stored in the corresponding memory qudit. These operations can be structured to act only if the qudit is in a well-defined subspace, either $\text{span}\{\ket{\redcircleopen},\ket{\bluecircleopen}\}$ or $\text{span}\{\ket{\redcirclefill},\ket{\bluecirclefill}\}$, ensuring compatibility with the overall routing and retrieval logic. From a hardware perspective, this implementation may be advantageous in platforms where multi-level quantum systems arise naturally, such as trapped ions, superconducting qudits, or encoded fermionic states. Importantly, the qudit-based version does not require spatial rail separation, potentially simplifying the experimental layout while retaining full logical equivalence with the dual-rail scheme.

A possible reformulation of this qudit approach consists of replacing each four-level system with two independent qubits. Hence, each subsystem $A_i$ and $D_j$ is realized by two qubits $q_1$ and $q_2$: the first, which we call information qubit, stores the address/data information of $A_i$/$D_j$, while the second is called routing qubit and is exploited to spread the information among the subsystems in order to let the particles turn as desired. We write their internal states through the following symbolic notation:
$$
\ket{q_1,q_2}, \mbox{\,\,\,\,where\,\,\,} q_1 \in \mbox{Span}\{\circleopen,\circlefill\} \,\,\mbox{ and }\,\, q_2   \in \mbox{Span}\{\red,\blue\} \, . 
$$
For instance, 
$\ket{q_1,q_2} = \ket{\circlefill,\red}$ represents the  state $\ket{\redcirclefill}$ of the four-level qudit presented at the beginning of this section.
Using this reformulation, 
at level $i$, the information qubit $q_1$  of the $i$-th address subsystem $A_i$, via a unitary transformation, acts on the routing qubit $q_2$ of the same susbystem $A_i$.
Later, the repeated action of a unitary operation applied on adjacent routing qubits of different subsystems, having a similar role of $\hat U^B$, spreads the routing information of $A_i$ to the routing qubits $q_2$ of the following subsystems $A_{i+1}$, $A_{i+2}$, $\dots$, $D_1$, $\dots$, $D_m$, without altering their information qubits $q_1$. Finally, the scattering gate $\hat{S}$ conditionally routes the subsystems according to the state of $q_2$, resetting them to the red state. We remember that each subsystem is composed of an information and a routing qubit $q_1$ and $q_2$, respectively. 
Hence, at level $i$, if the subsystem $A_i$ is represented by $\ket{q_1,q_2}_{A_i} = \ket{\circleopen, \red}_{A_i}$, this protocol does not change the state of the routing qubits, which remains in the red state and $A_i$ and all its following subsystems turn left. Instead, if $\ket{q_1,q_2}_{A_i} = \ket{\circlefill, \red}_{A_i}$, our protocols (i) changes $A_i$'s routing qubit into blue, namely $\ket{q_1,q_2}_{A_i} = \ket{\circlefill, \red}_{A_i}\rightarrow \ket{\circlefill, \blue}_{A_i}$, (ii) the routing qubits of the following subsystems get a color change to blue, (iii) $A_i$ and its following subsystems turn right through $\hat S$ while getting a color reset (to red) for the routing qubits.
The rest of the scheme proceeds with the same logic as the other protocols discussed before, where the message (or any superposition) that has to be retrived from the memory cells is written into the information qubits of the data walkers. 
Crucially, unlike the other proposed protocols, this method transmits path information only through short‐range, nearest‐neighbor interactions, without the need for additional backups.

\end{document}